\documentclass[12pt]{article}
\usepackage[utf8]{inputenc}
\usepackage{amsmath}
\usepackage{amssymb}
\usepackage{mathtools}
\usepackage{graphicx}
\usepackage{courier}
\usepackage{scalerel}
\usepackage{cite}

\numberwithin{equation}{section}

\newcommand{\fdot}{ \scalerel*{\cdot}{\bigodot} }

\usepackage{listings}
\usepackage{color}

\usepackage[colorlinks=true,linkcolor=blue,citecolor=blue,urlcolor=blue,filecolor=black,linktocpage=true]{hyperref}

 \csname
@addtoreset\endcsname{equation}{section}
\newcommand{\q}[1]{``\textit{#1}''}

\newcommand{\bibname}[1]{\textit{#1},}
\newcommand{\arxivnew}[1]{\href{http://arxiv.org/abs/#1}{arXiv:#1 [hep-th]}.}
\newcommand{\arxivold}[1]{\href{http://arxiv.org/abs/hep-th/#1}{arXiv:#1 [hep-th]}.}

\usepackage{empheq}
\newcommand*\widefbox[1]{\fbox{\hspace{2em}#1\hspace{2em}}}

\newcommand{\uuline}[1]{\underline{\underline{#1}}}

\newcommand{\dint}{\int d^D x \, }

\definecolor{dkgreen}{rgb}{0,0.6,0}
\definecolor{gray}{rgb}{0.5,0.5,0.5}
\definecolor{mauve}{rgb}{0.58,0,0.82}

\begin{document}

\begin{titlepage}
	\begin{center}
		{\huge\bfseries Higher spin theory\\}
		\vspace{1.5cm}
		{\Large\bfseries Ivan Vuković}\\[5pt]
		\textit{Advisor:} Prof. Predrag Dominis Prester, PhD\\[14pt]
		\href{mailto:vukovic.ac@gmail.com}{vukovic.ac@gmail.com} \\[1cm]
		\textsc{\Large{{Master's Thesis}}} \\[5pt]
		\vspace{3cm}
		\emph{{University of Rijeka,}}\\
		\emph{Department of Physics,}\\
		\emph{Graduate Study Programme in Physics,}\\
		\emph{Astrophysics and Elementary Particle Physics}\\[2cm]
		{Rijeka,} \\{ September 2017}

	\end{center}
\end{titlepage}
\linespread{1.25}

\newpage

\begin{abstract} 
	\noindent We provide a review of gauge field theories with higher spin, focusing on the classical theory of massless bosons in flat Minkowski spacetime.	
	A brief introduction to the concept of spin is provided along with a historical review of some of the most important problems and accomplishments in higher spin theory, followed by a review of familiar lower spin theories.	
	Using a particularly elegant formalism, we examine the free higher spin theory and the theory coupled to a generic external current. For both of these theories, we review their constrained and the unconstrained formulation, focusing our attention on the latter.
	We also review the existing literature, confirming the results for the non-local unconstrained formulation and correcting some calculational errors.
	We briefly discuss the geometrical motivation behind the construction of the basic building blocks of the theory and we entertain the possibility of there being a single equation of motion for all bosonic higher spin fields.	
	Some explicit calculations obtained using computer-assisted methods are provided in the appendix, along with the relevant computer code.\\[14pt]
\end{abstract}

{\bf Keywords:}	Higher Spin, Fronsdal Equation, Gauge Theory, Field Theory in Flat Spacetime

\newpage

\tableofcontents

\newpage
\section{Introduction}\label{s1}

	The problem of constructing a consistent theory of interacting fields with higher spin is a long-standing one. Mainly, it was an interesting theoretical curiosity, but nowadays, thanks to advances in string field theory and attempts to construct a quantum theory of gravity, we see a renaissance of interest in higher spins, which may illuminate some fundamental aspects behind those theories.
	
	Before embarking on a journey into the yet uncharted territory of higher spin theory, in this section we review what we already know about spin.
	
	After a basic introduction, in \textbf{Section \ref{s2}} we discuss some of the main motivating factors that keep driving new research in this area.
	
	As history frequently teaches us important lessons, in \textbf{Section \ref{s3}} we go through a "folk" history of higher spin theory, having the obvious advantage of hindsight. One should keep in mind that this is far from a complete history, it is a story consciously biased towards the topics we will be dealing with here. This section also serves as an appropriate place to expose some of the most important ideas and developments in higher spin theory.\newline We review several theorems whose implications seem to render the investigation of higher spins pointless. 
	
	In \textbf{Section \ref{s5}}, as a warm-up exercise, we review the well-known theories of fields with spin $0, 1$ and $2$, carefully examining the role of spin and trying to catch a glimpse of the underlying pattern common to all spins.
	
	Finally, in \textbf{Section \ref{s6}}, we present the higher spin theory of massless bosons in flat spacetime using an elegant mathematical formalism that works for all spins in all dimensions. We discuss both free and interacting theory (with a generic external current) in their constrained and unconstrained forms. We also examine the issue of non-locality and we investigate the geometric formulation of higher spin theories. 
	
	In \textbf{Appendix \ref{a1}}, we provide some explicit calculational results in higher spin theory. These results were obtained and checked using a simple C\texttt{++} code written specifically for this purpose. The core snippets of the code are provided in \textbf{Appendix \ref{a2}}.

\newpage
\subsection{What is spin?}
	
	Spin is an intrinsic property of relativistic fields, which (after quantization) give rise to particles. It can be viewed as an additional degree of freedom unrelated to spatial degrees of freedom specified by position and momentum.	
	Its name comes from the fact that, mathematically, spin behaves like quantized angular momentum. Unlike orbital angular momentum, spin quantum numbers may have half-integer values and fundamental particles cannot be made to stop "spinning" or to spin faster or slower, spin can only change its orientation.	
	Particles (fields) with integral spin are called \textbf{bosons}, and those with half-integral spin are known as \textbf{fermions}.
		
	\textit{Spin-statistics theorem}, one of the rigorous results of axiomatic quantum field theory, tells us that bosons and fermions behave in a drastically different manner. While the former respect the \textbf{Bose-Einstein} statistics, the latter respect the \textbf{Fermi-Dirac} statistics, and as a consequence are subject to the \textit{Pauli exclusion principle}. Fermions form particles of matter, while bosons mediate the interactions between them.

\subsection{Where does spin come from?}
	
	The existence of spin is a direct consequence of the most fundamental mathematical properties of our universe, \textbf{spacetime symmetries}. This is why the true meaning of spin has to be discussed in the context of a fully Lorentz-invariant theory.	
	Quantum field theory, which is the underlying formalism describing the Standard Model of particle physics, is such a theory. We introduce a field for each fundamental particle species, which transforms \textit{nicely} under Lorentz transformations. Once we pick a particular representation of the Lorentz transformations, it specifies the spin. After quantizing the field, one finds that the field operator creates or annihilates particles of definite spin, which was of course, the spin associated with the classical field to begin with.
	
\subsubsection{Spin from irreducible representations in four-dimensional spacetime}\label{IRREPs}
	
	Instead of simply looking at Lorentz transformations, we have to look at the full class of spacetime isometries (i.e. isometries of \textit{Minkowski space} $\mathcal{M}^4$). They are locally described by the \textit{Poincaré group}
	\begin{equation}
	\mathcal{P} = \mathbb{R}^{(1,3)} \rtimes \mathrm{SO} (1,3) \, ,
	\end{equation} 
	a ten-dimensional noncompact Lie group, corresponding to ten independent symmetries (3 spatial translations, time translation, 3 spatial rotations and 3 Lorentz boosts).\newline	
	A closer look at continuous local symmetries leads us to analyze the identity component of $\mathcal{P}$, which acts as a stabilizer group of the origin, the \textit{proper orthochronous Lorentz group}, 
	\begin{equation}
	\mathrm{SO}^+ (1,3) = \left\{ \Lambda \in \mathrm{GL}  (4, \mathbb{R})  \, \big| \,  \Lambda^T \eta \Lambda = \eta, \, \eta = \mathbf{diag} (-1,1,1,1) \right\} \, .
	\end{equation}
	The fundamental physical fields must carry the irreducible representations of its Lie algebra,
	\begin{equation}
	\mathfrak{so}(1,3) \hookrightarrow \mathfrak{so}(1, 3)_\mathbb{C} \cong \mathfrak{su}(2)_\mathbb{C} \oplus \mathfrak{su}(2)_\mathbb{C} \cong \mathfrak{sl} (2, \mathbb{C}) \, ,
	\end{equation}
	which generates the covering Lie group of $\mathcal{P}$,
	\begin{equation}
	\mathbb{R}^{(1,3)} \rtimes \mathrm{SL} (2,\mathbb{C}) \hookleftarrow \mathcal{P}  \, .
	\end{equation} 	
	Elements of this group are generated by terms of the form
	\begin{equation}
	\exp(i a_\mu P^\mu) \exp\left(\frac{i}{2} \omega_{\mu\nu} M^{\mu\nu }\right) \, ,
	\end{equation}
	where $a_\mu$ parametrizes translations generated by $P^\mu$, and $\omega_{\mu\nu}$ parametrizes Lorentz transformations (rotations and boosts) generated by $M^{\mu\nu}$.
	This Lie algebra is defined by
	\begin{align}
	\left[ P_\mu, P_\nu \right] &= 0 \, , \\
	\left[ M_{\mu\nu}, P_\rho \right] &= i \, \eta_{\rho [ \mu} P_{\nu ]} \, , \\
	\left[ M_{\mu\nu}, M_{\rho\sigma} \right] &=   i \, \eta_{[ \mu \nu} M_{\rho \sigma]} \, ,
	\end{align}
	where $[\dots]$ stands for unweighted anti-symmetrization of indices with the minimal number of terms and $M_{\mu\nu}$ is defined in terms of the rotation generator $J_i$ and boost generator $K_\mu$ as
	\begin{align}
	J_i &= \frac{1}{2} \varepsilon_{ijk} M^{jk} \, , \\
	K_i &= M_{0i}.
	\end{align}
	The \textit{Casimir invariants} of the Poincaré group are
	\begin{equation}
	P_\mu P^\mu := m^2 \, ,
	\end{equation}
	where $m$ stands for mass and\footnotemark
	\footnotetext{This is true if $m \neq 0$, but $m=0$ does not imply $W^2=0$.}
	\begin{equation}
	W_\mu W^\mu = m^2 s (s+1),
	\end{equation}
	where $s$ stands for spin\footnotemark.\footnotetext{At this level of analysis, which is purely mathematical, \textit{mass} and \textit{spin} are simply names we give to these quantities. Their physical properties only become apparent after introducing the actual physics, i.e. equations of motion.}
	$W_\mu$ is the \textit{Pauli-Lubanski} pseudovector, defined as the Hodge dual of $\mathbf{J} \wedge \mathbf{P}$, i.e.
	\begin{equation}
	W_\mu := \frac{1}{2} \varepsilon_{\mu\nu\rho\sigma} J^{\nu\rho} P^\sigma.
	\end{equation}
	\newpage
	Next, we look at \textit{Wigner's little groups}, stabilizer subgroups of various mass states.
	\begin{itemize}
		\item $m>0$, stabilizer of $P=(m,0,0,0)$
		\newline$\implies$ massive states with mass $m$ and spin $s \in \mathbb{N}_0/2$
		\item $m=0$ and $P_0 > 0$, stabilizer of $P=(k,0,0,k)$
		\newline$\implies$ $s \in \mathbb{N}_0/2$ IRREPs and \textit{continuous spin} representation
		\item $m^2 < 0$, stabilizer of $P=(0,0,0,m)$
		\newline$\implies$ \textit{tachyons}\footnotemark \footnotetext{Fields with an imaginary mass, which propagate faster-than-light excitations and lead to theories with instabilities and violation of causality.}
		\item $m=0$ and $P^\mu=0$
		\newline$\implies$ trivial representation, the \textit{vacuum} state 
	\end{itemize}
	We can now classify the physically relevant\footnotemark\, finite-dimensional irreducible representations of the double cover of the Poincaré group by two numbers, $m \in \mathbb{R}^+$ and $s \in \frac{1}{2} \mathbb{Z}$. 
	\footnotetext{We follow the standard prescription of ignoring tachyons and \textit{continuous spin} representations.\newline The latter seem to give rise to fields whose excitations cannot be compactly localized, but are instead localized on semi-infinite spacelike strings (see \cite{infspin}) that we do not seem to find in nature.}	
	These irreducible representations are further classified by two numbers, $j_1$ and $j_2$ such that $j_1 + j_2 = s$ and are labeled as $(j_1, j_2)$ representations, summarized in the table below.\\
	
	\begin{tabular}{ r | c | c | c | c}
		\hline
		\textbf{spin}	& \textbf{representation}					& \textbf{field}				& \textbf{eq. of motion}		& \textbf{example}					\\
		\hline
		$0$ 			& $(0,0)$									& scalar						& \textit{Klein-Gordon}			& Higgs								\\ 
		$1/2$ 			& $(\frac{1}{2},0) \oplus (0, \frac{1}{2})$	& spinor						& \textit{Dirac}				& electron							\\
		$1$				& $(\frac{1}{2}, \frac{1}{2})$				& vector						& \textit{Proca}				& photon							\\
		$3/2$ 			& $(\frac{1}{2},1) \oplus (1, \frac{1}{2})$	& spinor-vector					& \textit{Rarita-Schwinger}		& gravitino\footnotemark			\\
		$2$				& $(1,1)$									& 2-tensor\footnotemark		& \textit{linearized Einstein}	& graviton\footnotemark		\\
		$\vdots$		& $\vdots$									& $\vdots$						& $\vdots$						& $\vdots$							\\
		\hline
	\end{tabular}\newline
	\footnotetext[5]{Graviton's fermionic superpartner in theories with Bose-Fermi symmetry (supersymmetry), specifically in supergravity.}
	\footnotetext[6]{Symmetric tensor of order two.}
	\footnotetext[7]{Hypothetical quantized excitation of the gravitational field. }
	
	The dots at the end of the table represent the fact that, from a mathematical point of view, nothing prevents the existence of higher spin fields at this level of analysis, nor does anything directly imply that we should expect them to behave differently from their lower spin counterparts.
	
	For a detailed group-theoretical analysis of spin, see, for example, \cite{weinbergQFT}, \cite{sternberg}, \cite{BBIRREPs} or \cite{schwichtenberg}.

\newpage
\subsubsection{Spin from an action principle}

	It has long been thought that spin cannot be formulated with an action principle within the framework of Lagrangian mechanics with particles.	
	This issue was particularly relevant with the introduction of Feynman's sum-over-histories approach to quantization.\newline Even Feynman himself wrote in his 1965 book \cite{feynmanpath}:\newline	
	\q{With regards to quantum mechanics, path integrals suffer most grievously from a serious defect. They do not permit a discussion of spin operators or other such operators in a simple and lucid way. ... It is a serious limitation that the half-integral spin of the electron does not find a simple and ready representation.}\newline	
	However, this is actually possible and relatively straightforward, but the appropriate phase space formulation was not fully obvious until several years later \cite{pathspin}.
	
	One begins by examining how spin behaves classically. It can be thought of as a spinning top, or a little arrow with fixed length, sticking out from the particle and pointing in a particular direction in three-dimensional physical space. Therefore, it is reasonable to assume that its phase space is a $2$-sphere $\mathcal{S}^2$, and its dynamical variables can be taken to be the polar and azimuthal angles $\theta$ and $\phi$.	
	This establishes the point particle as an entity described not only by its position and momentum, but also by the orientation of its spin "arrow". The rest is simply a matter of mathematical construction.	
	It is precisely this that caused so much confusion, because in order to write the action, one has to find the proper local geometric invariant on $\mathcal{S}^2$.
	The volume form on a $2$-sphere is
	\begin{equation}
	\mathbf{\omega} = \sin{\theta} \, d \phi \, \wedge \, d\theta \, ,
	\end{equation}
	and it is invariant under rotations. Since $\mathbf{\omega}$ is a closed form, we can write it locally as an exact form, i.e.
	\begin{equation}
	\mathbf{\omega} = d \mathbf{\chi} = d (\cos{\theta} \, d \phi) \, ,
	\end{equation}
	so the action for spin $J$ can be written as
	\begin{equation}
	S = J \int \chi = J \int d\phi \, \cos{\theta} = J \int dt \, \dot{\phi} \cos{\theta} \, . \label{int}
	\end{equation}
	A distinctive property of this system is that its phase space is compact (i.e. closed and bounded), which implies a finite-dimensional Hilbert space. Furthermore, invariance of $\mathcal{S}^2$ (as a manifold embedded in $\mathbb{R}^3$) under $\mathrm{SO}(3)$ guarantees the operators with correct commutation relations\footnotemark.
	\footnotetext{We will not derive this here. The interested reader is referred to \cite{condmatter}.}
	Let us demonstrate how quantized spin arises when we plug \eqref{int} into the Feynman integral. The spin term in the Feynman integral, with $J=j\hbar$ reads \footnotemark \footnotetext{Here and only here, we write $\hbar$ explicitly instead of working with natural units, where $\hbar=c=1$}
	\begin{equation}
	e^{i S / \hbar} = \exp\left(ij\int dt \, \dot{\phi} \cos{\theta}\right) \, .
	\end{equation}\newpage
	We proceed by using the Stokes' theorem\footnotemark\, on the integral, \footnotetext{$\int_{\partial \Omega} \omega = \int_{\Omega} d\omega$}
	\begin{equation}
	\int dt \, \dot{\phi} \cos{\theta} = \oint_C d \phi \cos{\theta} = \int_M d \phi d \theta \sin{\theta}
	\end{equation}
	where $C$ denotes a closed path on $\mathcal{S}^2$, bounding a $2$-surface $M$, i.e. $\partial M = C$. Since $\mathcal{S}^2$ is compact, the choice of $M$ is not unique, but the difference between two possible choices is simply the integral over entire $\mathcal{S}^2$. In other words, the difference between two possible choices for the action is
	\begin{equation}
	\Delta S = j \hbar \int_{\mathcal{S}^2} d \phi d \theta \sin{\theta} = 4 \pi  \hbar j \, .
	\end{equation}
	The path integral cannot be multivalued, which in turn means that $e^{iS/\hbar}$ has to be single-valued. Therefore,
	\begin{equation}
	e^{i \Delta S / \hbar} = 1 \quad \implies \quad 4 \pi j = 2 \pi N \quad ( \forall N \in \mathbb{Z}) \, ,
	\end{equation}
	i.e. spin $j$ can take only integer and half-integer values, the same conclusion we arrived at using group representation theory in \textbf{Section \ref{IRREPs}}.	
	Once again, nothing at the mathematical level of analysis prevents the existence of arbitrarily high spins nor does it indicate any sort of inconsistency.
\newpage
\subsection{What do we mean by higher spin?}
	
	Historically, the term \textit{higher spin}\footnotemark\, was used to refer to several different domains of theoretical constructs.
	\footnotetext{Often abbreviated as HS.}One of the reasons for including only spin $0$, $1/2$ and $1$ fields in the domain of \textit{lower spin} was the fact that only those result in renormalizable quantum field theories.
	Today, by \textit{higher spin}, we mean spin greater than \textbf{two}, i.e. spin-$5/2$ and higher for fermions, spin-$3$ and higher for bosons. This seems more appropriate since we \textit{do} have consistent \textit{classical} field theories for $s\leq2$, but all higher spins yield problematic constructions even before quantization.
	
	Constructing a consistent interacting theory of HS fields (sometimes referred to as \textit{higher-spin gravity} in the case of massless interacting fields) has been a long-standing problem in theoretical physics. So far, we only have a fully consistent interacting HS theory in $\mathrm{(A)dS}$ spacetimes, which has become known as \textit{Vasiliev's theory}. Similar attempts at constructing such theories in flat space have not been successful. Unfortunately, taking the flat-space limit of Vasiliev's theory in $\mathrm{(A)dS}$ in hope of recovering a theory in flat spacetime is by no means trivial and possibly not even well-defined.
		
	Interestingly, $\mathrm{(A)dS}$ spacetimes are highly symmetrical, and their symmetry group $\mathrm{SO}(1,4) \cong \mathbf{Sp}(2,2)$ reduces\footnotemark\, to the Poincaré group in the limiting case of infinite \newline(anti-)de Sitter radius, which may point to the $\mathrm{(A)dS}$ group as being more fundamental\cite{missed}.
	\footnotetext{This can be accomplished rigorously using the İnönü-Wigner group contraction.}

\newpage
\section{Why study HS theory?}\label{s2}

\subsection{String theory}\label{ST}

	String theory is a promising candidate for a consistent theory of quantum gravity. Its perturbative spectrum consists of states with arbitrarily high spins and masses.	
	One could say that higher spin gravity lies between supergravity and string theory, which makes it particularly interesting.
	
	In string theory, one finds an \textbf{infinite tower of massive string excitations} with increasing spin. The existence of this infinite tower of higher-spin fields is crucial for the absence of ultraviolet divergences, an extremely important feature of string theory.
	
	The only free parameter in string theory is the \textit{string constant}, denoted by $\alpha'$, which determines the characteristic length and mass scale of strings. In the $\alpha' \to 0$ limit, the theory reduces to supergravity, i.e. a theory with massless modes. On the other hand, in the $\alpha' \to \infty$ limit, all excitations become massless, and the theory resembles higher spin gravity.
	
	Furthermore, we know that it is possible to have a theory with only massless fields in its formal construction, which nevertheless produces no massless excitations after quantization. In other words, it is possible to begin with a Lagrangian with massless fields, which describes a quantum field theory without massless propagating degrees of freedom. There are at least two mechanisms, familiar from the Standard Model, that exhibit such behaviour. One is \textit{spontaneous symmetry breaking}, i.e. the \textit{Higgs mechanism} that gives mass to massive fundamental\footnotemark\footnotetext{Spontaneous symmetry breaking also occurs in non-fundamental descriptions, for example in the theory of superconductivity and superfluidity.} particles. The other one is \textit{color confinement}, the mechanism responsible for clumping of gluons and quarks into colorless hadrons. It is possible that a similar mechanism underlies the generation of massive states in string theory, in which case we would have to know how to construct a massless higher spin theory. There are strong indications that symmetries of string theory form a very large group, much larger than what can be seen using the perturbative approach, which spontaneously breaks down to a smaller group, giving mass to higher-spin excitations.
		
	Therefore, it is plausible that a firm understanding of HS theory could shed some light on the underlying mathematical structure of string theory. In particular, we would like to know what symmetries the theory possesses and what is the notion of spacetime geometry in string theory and HS gravity.
\newpage
\subsection{AdS/CFT correspondence}

	The $\mathrm{AdS/CFT}$ (\textit{anti-de Sitter/conformal field theory}) correspondence\footnotemark \footnotetext{Also known as \textit{Maldacena duality} or \textit{gauge/gravity duality}.}, in its strictest formulation, posits an equivalence between the theory of quantum gravity in anti-de Sitter spacetimes, as formulated in string theory, and conformal field theory on its boundary. However, although it seems to be valid generally, it is still technically a conjecture and its rigorous construction has not yet been completed.
	
	There are reasons to believe that studying HS theory could help us not only to understand string theory but also to elucidate this conjectured correspondence, particularly in the prominent example of type IIB string theory (a theory on $AdS_5 \times \mathcal{S}^5$ with five spacetime and five compact dimensions) and $\mathcal{N}=4$ supersymmetric Yang-Mills theory on its four-dimensional boundary.

\subsection{Why not?}

	Of course, from the viewpoint of pure mathematics, one needs no justification for studying anything. But in the case of HS theory, there is more to it than just curiosity. 
	
	We know that a rich mathematical structure emerges from the fully consistent (necessarily non-linear!) theory of spin-$2$ fields, i.e. \textbf{pseudo-Riemannian geometry}. Some say that Einstein's general theory of relativity was, in a sense, discovered prematurely, and it was only Einstein's deep geometric intuition that allowed him to make the leap to a fully geometrized description of gravity. Had we persisted on building the theory in a bottom-up way from a linear spin-$2$ theory, we might have not ended up with such an elegant theory years before the dawn of quantum field theory.
	
	It is not known at the time of writing whether fully consistent spin-$3$ and higher spin theories give rise to some exciting new connections between physics and mathematics, perhaps even hitherto unknown mathematical structures, but it doesn't seem so unlikely that they might.
	
\newpage
\section{Folk history of HS theory}\label{s3}
	
	One cannot fully appreciate the struggle to understand higher spins without its history. For that purpose, we review here the most important steps forward\footnotemark\footnotetext{Like in every area of research, some steps that did not quite lead forward have been made, which was not understood at the time. Today, we understand more, so we can only pick those results that lead somewhere, hence the title \textit{folk} history.} in understanding higher spins, and we use this opportunity to expose the very basics of the theory. We restrict our attention to fields with integral spin, since those are the focus of this thesis. 

\subsection{Fierz-Pauli equations (1939)} \label{fpe}
	
	Fierz and Pauli constructed a consistent set of equations describing free massive fields of arbitrary spin \cite{fierzpauli}.	
	They start with the Klein-Gordon equation,
	\begin{equation}
	(\Box - M^2) \phi = 0 \, ,
	\end{equation}
	which describes spin-$0$ fields, and generalize it directly to higher spins, imposing additional consistency constraints.	
	The Fierz-Pauli equations describing a spin-$s$ field are
	\begin{align}
	\phi_{\mu_1 \cdots \mu_s} &= \phi_{(\mu_1 \cdots \mu_s)}\, ,	\label{FP1}	\\
	(\Box - M^2) \phi_{\mu_1 \cdots \mu_s} &= 0 \, ,				\label{FP2}	\\
	\partial^{\mu_1} \phi_{\mu_1 \cdots \mu_s} &= 0 \, ,			\label{FP3}	\\
	\eta^{\mu_1 \mu_2} \phi_{\mu_1 \cdots \mu_s} &= 0 \, ,			\label{FP4}
	\end{align}
	where $(\dots)$ denotes the normalized symmetrization of indices. Equation \eqref{FP1} establishes the field as a fully symmetric tensor of order $s$. This condition ensures that the field transforms in accordance with the desired spin representation.	
	To ensure that it is an irreducible representation, it must be traceless, which is guaranteed by \eqref{FP4}.	
	Finally, \eqref{FP3} imposes the transversality condition, needed for the field to propagate the correct number of degrees of freedom, which we calculate in the following segment. 
	
	From group-theoretical considerations, we expect all\footnotemark\, massless bosons in four-dimensional Minkowski spacetime to propagate exactly \textbf{two} independent degrees of freedom.\footnotetext{Except scalar bosons, which always have a single degree of freedom. More precisely, \textit{all} massless bosons have a single degree of freedom, but it gets doubled due to parity transformations, except in the case of scalars, whose irreducible representations are one-dimensional.} In general, a spin-$s$ field in $D$-dimensional spacetime should propagate\footnotemark
	\begin{equation}
	\#(D-2,s) - \#(D-2, s-2) = {{D+s-3} \choose {s}} - {{D+s-5} \choose {s-2}} 
	\end{equation}
	independent degrees of freedom, as can be seen for example, using Wigner's classification. $\#(D,s)$ denotes the number of independent components of a fully symmetric tensor of order $s$ in $D$-dimensional spacetime. It is a simple exercise in combinatorics to check that $\#(D,s)={{D+s-1} \choose {s}}$.
	A spin-$s$ field is described by a totally symmetric doubly traceless tensor of order $s$, which contains 
	\begin{equation}
	\#(D, s) - \#(D, s-4) = \underbrace{{{D+s-1} \choose {s}}}_{\parbox{2cm}{\tiny{components of a\\ symmetric tensor}}} - \underbrace{{{D+s-5} \choose {s-4}}}_{\parbox{1.8cm}{\tiny{components of \\ its second trace}}}
	\end{equation}
	independent components.
	Gauge invariance eliminates the propagation of spurious degrees of freedom ($2 s^2$ components in $D=4$), leaving the correct number of remaining degrees of freedom.
	For a detailed calculation, see, for example, \cite{weinbergQFT} or \cite{BBIRREPs}.
	
\subsection{Singh-Hagen Lagrangian (1974) }
	
	It was not until 35 years later that the proper Lagrangian formulation of Fierz-Pauli equations was constructed, by Singh and Hagen \cite{singhhagen}.	
	The fundamental obstacle lay in the need for auxiliary non-dynamical fields of spins $s-2, s-3, \dots$, along with the spin-$s$ field.	
	Let us motivate their construction by starting with the trivial example of $s=1$, and then proceeding to the first non-trivial case of $s=2$.

\subsubsection{Spin-1: no auxiliary fields}
	
	A spin-$1$ field is described by $\phi_\mu(x)$, a Lorentz-tensor of order one (i.e. a 4-vector). Since it only has a single index, we do not have to worry about equation \eqref{FP4}, nor do we have to worry about the symmetry condition \eqref{FP1}. The Lagrangian for $s=1$ is the Proca Lagrangian
	\begin{equation}
	\mathcal{L}_{(1)} = -\frac{1}{2} (\partial_\mu \phi_\nu)^2 + \frac{1}{2} (\partial \cdot \phi )^2 - \frac{M^2}{2} \phi^2 \label{SH1}
	\end{equation}
	which produces the Proca equation of motion,
	\begin{equation}
	\Box \phi_\mu - \partial_\mu (\partial \cdot \phi) - M^2 \phi_\mu = 0 \, . \label{proca}
	\end{equation}
	At first glance, this is not equal to \eqref{FP2} for $s=1$, but a single divergence of \eqref{proca} gives
	\begin{equation}
	\partial^\mu \phi_\mu = 0 \, ,
	\end{equation}
	which gives the transversality condition \eqref{FP3}. Putting this back into \eqref{proca}, we are indeed left with the spin-$1$ version of equation \eqref{FP2},
	\begin{equation}
	(\Box - M^2) \phi_\mu = 0 \, .
	\end{equation}
\newpage
\subsubsection{Spin-2: scalar auxiliary field}
	
	Following \eqref{FP1} and \eqref{FP4}, a spin-$2$ field is described by $\phi_{\mu \nu}(x)$, a symmetric traceless Lorentz-tensor of order two. Instead of directly generalizing \eqref{SH1} to the spin-$2$ case by using a tensor of order two, we write the Lagrangian with an undetermined real parameter $\alpha$ in place of $1$,
	\begin{equation}
	\mathcal{L}_{(2)} = -\frac{1}{2} (\partial_\mu \phi_{\nu\rho})^2 + \frac{\alpha}{2} (\partial \cdot \phi_\mu )^2 - \frac{M^2}{2} \phi^2 \, . \label{SH2}
	\end{equation}
	The reason for doing so will become apparent soon.
	The corresponding equation of motion\footnotemark\, is found to be
	\footnotetext{$\mathcal{L}_{(2)}$ is varied taking into consideration the symmetry and the tracelessness of $\phi_{\mu \nu}$}
	\begin{equation}
	\Box \phi_{\mu \nu} - \frac{\alpha}{2} \left( \partial_\mu \partial \cdot \phi_\nu + \partial_\nu \partial \cdot \phi_\mu - \frac{2}{D} \eta_{\mu \nu} \partial^2 \cdot \phi \right) - M^2 \phi_{\mu \nu} = 0 \, , \label{sheom}
	\end{equation}
	where $D$ is the dimension of spacetime. A single divergence of \eqref{sheom} gives
	\begin{equation}
	\left( 1 - \frac{\alpha}{2} \right) \Box \partial \cdot \phi_\mu + \alpha \left( \frac{1}{D} - \frac{1}{2} \right) \partial_\mu \partial^2 \cdot \phi - M^2 \partial \cdot \phi_\mu = 0 \, . \label{shtrouble}
	\end{equation}
	We seem to be in trouble, because (assuming $D>2$) we can only partially restore the transversality condition $\eqref{FP3}$ by setting $\alpha=2$, which eliminates the first term in \eqref{shtrouble}. Had we generalized \eqref{SH1} directly, instead of leaving $\alpha$ undetermined, we would not have been able to eliminate it.
	
	To eliminate the second term in \eqref{shtrouble}, we introduce the auxiliary scalar field $\pi(x)$ by adding to $\mathcal{L}_{(2)}$ (with $\alpha=2$) additional terms with two undetermined real parameters, $c_1$ and $c_2$,
	\begin{equation}
	\mathcal{L}_\pi = \pi \partial^2 \cdot \phi + c_1 (\partial_\mu \pi)^2 + c_2 \pi^2 \, .
	\end{equation}
	The corresponding equations of motion for $\mathcal{L} = \mathcal{L}_{(2)}\big\rvert_{\alpha=2} + \mathcal{L}_\pi$ are found to be
	\begin{align}
	\phi :& \quad \Box \phi_{\mu \nu} - \partial_\mu \partial \cdot \phi_\nu - \partial_\nu \partial \cdot \phi_\mu + \frac{2}{D} \eta_{\mu \nu} \partial^2 \cdot \phi - M^2 \phi_{\mu\nu} + \partial_\mu \partial_\nu \pi - \frac{1}{D} \eta_{\mu \nu} \Box \pi = 0 \, , \label{phieq} \\
	\pi :& \quad \partial^2 \cdot \phi + 2(c_2 - c_1 \Box) \pi = 0 \, . \label{sheom2pi}
	\end{align}
	Taking twice the divergence of \eqref{phieq}, i.e. contracting it by $\partial_\mu \partial_\nu$, and multiplying it by $D$, yields
	\begin{equation}
	\left( (2-D) \Box - D M^2 \right) \partial^2 \cdot \phi + (D-1) \Box^2 \pi = 0 \, .\label{sheom2phi}
	\end{equation}
	The two equations, \eqref{sheom2pi} and \eqref{sheom2phi}, can be seen as a linear homogeneous system in $\partial^2 \cdot \phi$ and $\pi$. The system is solved by requiring that its determinant be non-vanishing and purely algebraic (without $\Box$ operators). Fortunately, this is possible if we choose
	\begin{align}
	c_1 &= \frac{D-1}{2(D-2)} \, , \\
	c_2 &= \frac{D(D-1)M^2}{2(D-2)^2} \, .
	\end{align}
	This way, the only solution of the linear system \eqref{sheom2pi}-\eqref{sheom2phi} is $\pi = 0$ and $\partial^2 \cdot \phi = 0$, which we plug into \eqref{shtrouble} with $\alpha=2$ to obtain the Fierz-Pauli transversality condition \eqref{FP3} for $s=2$,
	\begin{equation}
	\partial^\mu \phi_{\mu\nu} = 0 \, .
	\end{equation}
	Finally, plugging the transversality condition and the solution $\partial^2 \cdot \phi = 0$ into \eqref{sheom}, we get the Fierz-Pauli equation of motion \eqref{FP2} for $s=2$,
	\begin{equation}
		(\Box - M^2) \phi_{\mu\nu} = 0 \, .
	\end{equation}
	A similar procedure with $s-1$ auxiliary fields was shown to yield the correct Lagrangian for spin-$s$ fields, equivalent to the Fierz-Pauli equations \cite{singhhagen}.

\subsection{Fronsdal equation (1978)}\label{fronsdaleq}
	
	Soon after Singh and Hagen, Fronsdal investigated the massless case, taking the $M \to 0$ limit of their Lagrangian formulation \cite{fronsdal}.	
	In this limit, only the spin-$s$ and the first auxiliary spin-$s-2$ field survive, while all the lower spin auxiliary fields decouple. Furthermore, the remaining two fields can be neatly packed into a single field, with additional consistency constraints. Let us demonstrate here what happens in the spin-$2$ case.

\subsubsection{Spin-$2$ Fronsdal equation}
	
	We start from the $M \to 0$ limit of the Singh-Hagen Lagrangian for $s=2$,
	\begin{equation}
	\mathcal{L} = -\frac{1}{2} (\partial_\mu \phi_{\nu\rho})^2 + (\partial \cdot \phi_\mu)^2 + \pi \partial^2 \cdot \phi  + \frac{D-1}{2(D-2)} (\partial_\mu \pi)^2 \, .
	\end{equation}
	Next, we redefine $\pi$ and $\phi_{\mu \nu}$ into a new field $\varphi_{\mu\nu}$,
	\begin{equation}
	\varphi_{\mu\nu} := \phi_{\mu \nu} + \frac{1}{D-2} \eta_{\mu \nu} \pi \, ,
	\end{equation}
	which is no longer traceless.	
	The resulting Lagrangian is
	\begin{equation}
	\mathcal{L} = -\frac{1}{2} (\partial_\mu \varphi_{\nu\rho})^2 + (\partial \cdot \varphi_\mu)^2 + \frac{1}{2} (\partial_\mu \varphi)^2 + \varphi \partial^2 \cdot \varphi \, .
	\end{equation}
	Note that this is exactly the linearized Einstein-Hilbert Lagrangian, an important fact to which we will return later.	
	The equation of motion that follows from this Lagrangian is
	\begin{equation}
	\Box \varphi_{\mu\nu} - (\partial_\mu \partial \cdot \varphi_\nu + \partial_\nu \partial \cdot \varphi_\mu) + \partial_\mu \partial_\nu \varphi + \eta_{\mu \nu} \left( \partial^2 \cdot \varphi - \Box \varphi \right) = 0 \, , \label{f2eom}
	\end{equation}
	which is precisely the free linearized Einstein equation,
	\begin{equation}
	G_{\mu\nu}^{(lin)} = R_{\mu\nu}^{(lin)} - \frac{1}{2} \eta_{\mu \nu} R^{(lin)} = 0 \, ,
	\end{equation}
	but more on this in \textbf{Section \ref{spin2}}.
	
	We define $\mathcal{F}_{\mu\nu}$ as the \textit{Fronsdal tensor} of order two,
	\begin{equation}
	\mathcal{F}_{\mu\nu} = \Box \varphi_{\mu\nu} - (\partial_\mu \partial \cdot \varphi_\nu + \partial_\nu \partial \cdot \varphi_\mu) + \partial_\mu \partial_\nu \varphi \, ,
	\end{equation}	
	which is equal to $R_{\mu\nu}^{(lin)}$. Note that we can now write \eqref{f2eom} as
	\begin{equation}
	\mathcal{F}_{\mu\nu} - \frac{1}{2} \eta_{\mu \nu} \mathcal{F} = 0 \, , \label{fgeom1}
	\end{equation}
	which simply reduces to
	\begin{equation}
	\mathcal{F}_{\mu \nu} = \Box \varphi_{\mu\nu} - (\partial_\mu \partial \cdot \varphi_\nu + \partial_\nu \partial \cdot \varphi_\mu) + \partial_\mu \partial_\nu \varphi = 0 \, . \label{fgeom2}
	\end{equation}
	This is the \textbf{Fronsdal equation} for spin-$2$ fields. Note that \eqref{fgeom1} can reduce to \eqref{fgeom2} only because the theory is \textit{free}, analogous to the reduction of Einstein field equations to the vanishing of $R_{\mu\nu}$ in vacuum.
	
	Fronsdal equation is invariant under the gauge transformation
	\begin{equation}
	\delta \varphi_{\mu \nu} = \partial_\mu \Lambda_\nu + \partial_\nu \Lambda_\mu \, .
	\end{equation}
	This fact will be particularly important when we begin investigating the theory in detail.

\subsubsection{Spin-$3$ Fronsdal equation}
	
	We can try to generalize the spin-$2$ case directly on a totally symmetric (but not traceless!) tensor of order three\footnotemark, \footnotetext{Here, we begin to use a prime to denote a trace, e.g. $\varphi'_\nu := \varphi^\mu{}_{\mu\nu}$.}
	\begin{align}
	\mathcal{F}_{\mu\nu\sigma} &= \Box \varphi_{\mu\nu\sigma} - (\partial_\mu \partial \cdot \varphi_{\nu \sigma} + \partial_\nu \partial \cdot \varphi_{\sigma \mu} +  \partial_\sigma \partial \cdot \varphi_{\mu \nu}) \\ \nonumber &+ \partial_\mu \partial_\nu \varphi'_\sigma + \partial_\nu \partial_\sigma \varphi'_\mu + \partial_\nu \partial_\sigma \varphi'_\mu = 0\, .
	\end{align}
	The corresponding generalized gauge transformations reads
	\begin{equation}
	\delta \varphi_{\mu\nu\sigma} = \partial_\mu \Lambda_{\nu\sigma} + \partial_\nu \Lambda_{\sigma\mu} + \partial_\sigma \Lambda_{\mu\nu} \, ,
	\end{equation}
	but unlike in the case of $s=2$, now we do not have a fully gauge-invariant Fronsdal tensor. Instead, 
	\begin{equation}
	\delta \mathcal{F}_{\mu\nu\sigma} = 3 \partial_\mu \partial_\nu \partial_\sigma \Lambda' \, .
	\end{equation}
	In his original formulation\cite{fronsdal}, Fronsdal circumvents this problem by simply restricting the space of gauge parameters to ones that are traceless, i.e. by imposing the unusual constraint
	\begin{equation}
	\Lambda' = 0 \, .
	\end{equation}
	This amounts to restricting ourselves to a subclass of gauge transformations, instead of having fully unrestricted gauge invariance.

\subsubsection{Spin-$s$ Fronsdal equation}
	
	The traceless $\Lambda$ constraint leaves us with a fully consistent gauge-invariant theory of free higher spin fields obeying the spin-$s$ \textit{Fronsdal equation}\footnotemark,
	\begin{equation}
	\mathcal{F}_{\mu_1 \cdots \mu_s} = \Box \varphi_{\mu_1 \cdots \mu_s} - (\partial_{\underline{\mu_1}} \partial \cdot \varphi_{\underline{\mu_2 \cdots \mu_s}}) + \partial_{\underline{\mu_1}} \partial_{\underline{\mu_2}} \varphi'_{\underline{\mu_3 \cdots \mu_{s} }} = 0 \, .
	\end{equation}\footnotetext{Underlined indices stand for unweighted symmetrization with the minimal number of terms.}	
\subsubsection{Fronsdal Lagrangian}

	Fronsdal started with the Singh-Hagen Lagrangian formulation and naturally, he wanted to describe his theory using an action principle. 
	The Lagrangian that makes this possible is
	\begin{equation}
	\mathcal{L}_{\mathcal{F}} = \frac{1}{2} \varphi^{\mu_1 \cdots \mu_s} \left( \mathcal{F}_{\mu_1 \cdots \mu_s} - \frac{1}{2} \eta_{\underline{\mu_1 \mu_2}} \mathcal{F}'_{\underline{\mu_3 \cdots \mu_s}} \right) \, . \label{flang}
	\end{equation}
	As we show through explicit calculation in \textbf{Section \ref{freefronsdal}}, where we switch to a simpler notation, \eqref{flang} indeed yields the Fronsdal equation for $s<4$. For spins higher than four, we have to impose another unusual constraint,
	\begin{equation}
	\varphi'' = 0
	\end{equation}
	if we are to arrive at the Fronsdal equation of motion $\mathcal{F}=0$.

\subsection{Vasiliev's equations (1990) }
	
	M.A.Vasiliev successfully constructed a fully consistent non-linear theory of interacting higher spin fields in (anti-)de Sitter spacetimes\cite{vasiliev90}. The equations are notoriously complicated and since we will be dealing with massless bosonic fields in flat spacetime, we will not reproduce them here.
	
	It suffices to quote \cite{vasiliev}:\newline
	\q{The shortest route to Vasiliev equations covers 40 pages.}\newline
	\q{It is a sort of conventional wisdom that Vasiliev equations cannot be derived...}
	
	Similarly to string theory, Vasiliev's theory in spacetime dimensions four and higher can be consistent only if it contains an infinite tower of higher-spin fields. Only in dimensions three and lower can it be consistent with an upper limit on spin.

\subsection{No-go theorems}\label{s4}
	
	Throughout the history of HS theory, several important results have been obtained that severely constrain the properties of would-be interacting theories of higher spin fields. Vasiliev's theory\cite{vasiliev90}\cite{vasiliev} shows that the class of such theories is not empty, but we have yet to arrive at other theories of this kind.	
	We list here some of the most important \textit{no-go} theorems. For a more detailed discussion, see \cite{nogo},\cite{nogo2} and references therein.	
\paragraph{No long-range HS interactions\newline}
	Using the \textit{S-matrix} approach, Weinberg proved in 1964 that there are no consistent \textbf{long-range} interactions by massless bosons with spin greater than two\cite{weinberg}.

\paragraph{No local Lagrangians in HS theories\newline}
	Using the local \textit{Lagrangian formalism} and working in the \textit{soft limit}, Aragone and Deser proved in 1979 \cite{AragoneDeserNoGo1} (see also \cite{AragoneDeserNoGo2}) that HS fields cannot consistently interact with gravity. Since gravitational interaction is universal, this implies that there can be no consistent interacting HS fields.

\paragraph{No massless HS interactions in flat spacetime\newline}
	The Weinberg-Witten theorem\cite{WeinbergWitten} from 1980 states that no massless HS field can consistently interact with gravity in \textbf{flat spacetime}.\newline
	
	It is important to keep in mind that all no-go theorems start with some underlying assumptions that are not obviously satisfied in all physically possible cases. Therefore, the effort to construct consistent interacting HS theories might not be a fool's errand after all.

\newpage
\section{Review of lower spin theories}\label{s5}

	Instead of jumping head-first into some deeper problems of higher spin theory, let us review the familiar territory of lower spin bosonic theories.

\subsection{Spin-0 theory}

	Fields of spin $0$ are described by Lorentz scalars.	
	The general Lagrangian for these fields is
	\begin{equation}
	\mathcal{L}_{0} [\phi]= \frac{1}{2} ( \partial_\mu \phi  )^2 - \frac{m^2}{2} \phi^2 \, ,
	\end{equation}
	and it produces the equation of motion for scalar fields, the \textbf{Klein-Gordon equation},
	\begin{equation}
	(\Box + m^2) \phi = 0 \, .
	\end{equation}

\subsubsection{Example: Higgs boson}

	\textbf{Higgs field} is a well-known example of a scalar field, and it is the only scalar fundamental field in the Standard Model.
	
	It is a \textit{complex} scalar field, described by the Lagrangian
	\begin{equation}
	\mathcal{L}_H = \left| \partial_\mu \phi \right|^2 - V(\phi) \, .
	\end{equation}

\subsection{Spin-1 theory}

	Fields of spin $1$ are described by Lorentz vectors.	
	The general Lagrangian for these fields is
	\begin{equation}
	\mathcal{L}_1 [A^\mu]= -\frac{1}{2} F_{\mu \nu} F^{\mu \nu} + \frac{m^2}{2} A_\mu A^\mu \, ,
	\end{equation}
	with $F_{\mu \nu} = \partial_\mu A_\nu - \partial_\nu A_\mu$, and it produces the \textbf{Proca equation},
	\begin{equation}
	\Box A^\nu - \partial^\nu ( \partial_\mu A^\mu ) + m^2 A^\nu = 0 \, .
	\end{equation}

\subsubsection{Example: Maxwell's electrodynamics}

	In the Standard Model, four vector bosons take part in the \textbf{electroweak} interaction, the \textbf{photon} and three intermediate bosons, $W^\pm$ and $Z^0$.
	The free massive intermediate boson fields satisfy the Proca equation while the massless photon field satisfies \textbf{Maxwell's equations},
	\begin{equation}
	\partial_\mu F^{\mu \nu} = j^\nu.
	\end{equation}
	The strong force is also mediated by vector bosons, described by the massless \textbf{gluon} field.
	Before addressing the spin-2 theory, let us briefly discuss the issue of gauge invariance.

\paragraph{Spin-1 gauge invariance:}

	A massless spin-$1$ field $A^\nu (x)$ has 4 components, and it satisfies the equation of motion
	\begin{equation}
	\Box A^\nu - \partial^\nu (\partial_\lambda A^\lambda ) = 0.
	\end{equation}
	This theory is invariant under the Abelian gauge transformation 
	\begin{equation}
	\delta A_\mu (x) = \partial_\mu \Lambda (x) \, .
	\end{equation}
	The equation of motion can be cast into a simple wave equation form, 
	\begin{equation}
	\Box A_\mu (x) = 0 \, ,
	\end{equation}
	by choosing the Lorentz-invariant \textit{Lorenz gauge},
	\begin{equation}
	\partial_\mu A^\mu (x) = 0.
	\end{equation}
	This choice is a scalar constraint, which eliminates one of two spurious degrees of freedom, but there is a degree of gauge freedom left, i.e.
	\begin{equation}
	\delta (\partial_\mu A^\mu) = \Box \Lambda = 0.
	\end{equation}
	This is also a scalar constraint, so we are indeed left with two propagating degrees of freedom.

\subsection{Spin-2 theory}\label{spin2}

	Fields of spin $2$ are described by symmetric Lorentz tensors of order two. The general Lagrangian for these fields is
	\begin{equation}
	\mathcal{L}_2 [h^{\mu\nu}] = -\frac{1}{2} (\partial_\sigma h_{\mu\nu})^2 + \partial_\sigma h_{\mu\nu} \partial^\mu h^{\nu\sigma} - \partial \cdot h_\nu \partial^\nu h + \frac{1}{2} (\partial_\mu h)^2 \, , \label{Ls2}
	\end{equation}
	and it produces the equation of motion
	\begin{equation}
	\Box h_{\mu\nu} - \partial_\mu \partial \cdot h_\nu -  \partial_\nu \partial \cdot h_\mu + \partial_\mu \partial_\nu h + \eta_{\mu\nu} \partial^2 \cdot h - \eta_{\mu\nu} \Box h = 0 \, . \label{Es2}
	\end{equation}

\subsubsection{Example: General Relativity}
	
	\textbf{Einstein's General Relativity} is the archetypal example of a spin-2 theory. It describes gravitation, and it is the only spin-$2$ theory found in nature. 
	In its full form, general relativity is highly nonlinear, and it is described by \textbf{Einstein field equations},
	\begin{equation}
	G_{\mu \nu} \equiv R_{\mu \nu} - \frac{1}{2} g_{\mu \nu} R = 8 \pi T_{\mu \nu} ,
	\end{equation}
	where $G_{\mu \nu}$ is the \textbf{Einstein tensor}, $R_{\mu \nu}$ is the \textbf{Ricci tensor}, $R$ is the \textbf{Ricci scalar} and $T_{\mu \nu}$ is the matter \textbf{energy-momentum} tensor.
	The \textbf{Riemann curvature tensor} can be defined as
	\begin{equation}
	R^\rho{}_{\sigma\mu\nu} = \partial_\mu \Gamma^\rho{}_{\nu \sigma} - \partial_\nu \Gamma^\rho{}_{\mu \sigma} + \Gamma^\rho{}_{\mu \lambda} \Gamma^\lambda{}_{\nu \sigma} - \Gamma^\rho{}_{\nu \lambda} \Gamma^\lambda{}_{\mu \sigma} \, ,
	\end{equation} 
	using the \textit{torsionless connection} $\Gamma^\rho{}_{\mu\nu} = \Gamma^\rho{}_{\nu\mu}$,
	\begin{equation}
	\Gamma^\rho{}_{\mu\nu} = \frac{1}{2} g^{\rho\lambda} (\partial_\mu g_{\nu \lambda} + \partial_\nu g_{\mu \lambda} - \partial_\lambda g_{\mu \nu}) \, .
	\end{equation}
	Ricci tensor and scalar are simply given by
	\begin{align}
	R_{\mu \nu} &= R^\rho{}_{\mu \rho \nu} \, ,\\
	R &= R^\lambda{}_\lambda \, .
	\end{align}

\subsubsection{Example: Linearized Gravity}
	
	By considering small metric perturbations from the flat Minkowski spacetime, we can construct a linear theory of a dynamical spin-2 field in a static flat background. Explicitly, we decompose the metric so that
	\begin{equation}
	g_{\mu \nu} (x) = \eta_{\mu \nu} + h_{\mu \nu} (x) + \mathcal{O}(h^2) \, ,
	\end{equation}
	and we truncate the expansion to first order in $h_{\mu\nu}$, assuming $\| h(x) \| \ll 1$.
	The resulting theory is what we call \textbf{linearized gravity}, and it is described by \textit{linearized} Einstein field equations,
	\begin{equation}
	G^{(lin)}_{\mu \nu} = R^{(lin)}_{\mu \nu} - \frac{1}{2} \eta_{\mu \nu} R^{(lin)} = T^{(lin)}_{\mu\nu}.
	\end{equation}
	Or, using the metric perturbation field explicitly,
	\begin{align}
	R^{(lin)}_{\mu \nu} &= \Box h_{\mu \nu} - \partial_\mu (\partial^\lambda h_{\lambda \nu}) - \partial_\nu (\partial^\lambda h_{\lambda \mu}) + \partial_\mu \partial_\nu h^\lambda{}_\lambda \, , \\
	R^{(lin)} &= 2 \Box h^\lambda{}_\lambda -2 \partial^\lambda \partial^\sigma h_{\lambda \sigma} \, , \\
	G^{(lin)}_{\mu \nu} &= \Box h_{\mu \nu} - \partial_\mu (\partial^\lambda h_{\lambda \nu}) - \partial_\nu (\partial^\lambda h_{\lambda \mu}) + \partial_\mu \partial_\nu h^\lambda{}_\lambda + \eta_{\mu \nu} \partial^\lambda \partial^\sigma h_{\lambda \sigma} - \eta_{\mu\nu}\Box h^\lambda{}_\lambda  \, .
	\end{align}
	Note that $G^{(lin)}_{\mu \nu} = 0$ corresponds to \eqref{Es2}. This is no coincidence, since \eqref{Ls2} precisely describes the Lagrangian for linearized gravity in the absence of sources, i.e. with \newline$T^{(lin)}_{\mu\nu}=0$. 

\paragraph{Spin-2 gauge invariance:}
	
	The spin-$2$ field is described by a doubly traceless tensor $h_{\mu\nu} (x)$ of rank two and therefore has $\mathbf{10}$ independent components.
	In free theory, it satisfies the equation of motion
	\begin{equation}
	R^{(lin)}_{\mu\nu} = \Box h_{\mu\nu} - \partial_\mu (\partial^\lambda h_{\lambda\nu}) - \partial_\nu (\partial^\lambda h_{\mu\lambda}) + \partial_\mu \partial_\nu h^\lambda{}_\lambda = 0 \, .
	\end{equation}
	This theory is invariant under the Abelian gauge transformation
	\begin{equation}
	\delta h_{\mu \nu} (x) = \partial_\mu \xi_\nu (x) + \partial_\nu \xi_\mu (x) \, ,
	\end{equation}
	which allows us to cast the above equation into a simple wave equation form,
	\begin{equation}
	\Box h_{\mu \nu} (x) = 0 \, ,
	\end{equation}
	by choosing the Lorentz-invariant \textit{de Donder Gauge}\footnotemark,
	\begin{equation}
	\mathcal{D}_\mu (x) \equiv \partial^\lambda h_{\lambda \mu}  - \frac{1}{2} \partial_\mu h^\lambda{}_\lambda = 0.
	\end{equation}
	\textit{De Donder tensor} $\mathcal{D}_\mu$ is a $4$-vector, so we are left with $10 - 4 = \mathbf{6}$ degrees of freedom.
	Fixing the gauge in this way does not eliminate the gauge freedom completely. This can be seen from the \textit{de Donder} gauge condition, since
	\footnotetext{Also known as the \textit{harmonic gauge}, \textit{Lorentz gauge}, \textit{Einstein gauge}, \textit{Hilbert gauge} or \textit{Fock gauge}.}
	\begin{equation}
	\delta \mathcal{D}_\mu (x) = \Box \xi_\mu (x) = 0.
	\end{equation}
	This too is a $4$-vector constraint, which eliminates the remaining $4$ spurious degrees of freedom, leaving us with $6 - 4 = \mathbf{2}$ propagating degrees of freedom, as expected.

\newpage
\section{Higher spin theory of massless bosons}\label{s6}
\subsection{Francia-Sagnotti formalism}

	There exists an elegant formalism\footnotemark\, developed by D.Francia and A.Sagnotti\cite{fs1} \cite{fs2} \cite{st} \cite{introfree} \cite{fs3} \cite{fms} \cite{dariomass} \cite{dariopropm} \cite{dariocrete} that makes it easy to express and manipulate most of mathematical objects of HS theory in the linear approximation.\footnotetext{To be fair, it would be more precise to call it \textit{notation}, but as Feynman said\cite{feynmannotation}: \q{We could, of course, use any notation we want; do not laugh at notations; invent them, they are powerful. In fact, mathematics is, to a large extent, invention of better notations.}} This formalism is suitable for higher spin theory since the tensorial indices and spin are left implicit, but are easily recovered.
	
	A spin-$s$ field is simply written as
	\begin{equation}
	\phi_{\mu_1 \cdots \mu_s} \equiv \phi \, .
	\end{equation}
	The $n$-th gradient of $\phi$ is written as $\partial^n \phi$, the $n$-th divergence\footnotemark\, as $\partial^n \cdot \phi$ and the $n$-th trace as $\phi^{[n]}$. Lower traces are simply written with a prime, e.g. $\phi''$ for the second trace.
	\footnotetext{Where Francia and Sagnotti would use (for example) $\partial \cdot \partial \cdot \partial \cdot \varphi$, here we use $\partial^3 \cdot \varphi $ instead. This simplification seems to produce no ambiguities, as the reader is welcome to check.}	
	All indices are implicitly symmetrized, without weight factors, using the minimal number of terms.	
	For example, if $s=2$,
	\begin{align}
	\partial^2 \phi &\equiv \partial_\mu \partial_\nu \phi_{\sigma \rho} + \partial_\mu \partial_\sigma \phi_{\nu \rho} + \partial_\mu \partial_\rho \phi_{\sigma \nu} + \partial_\nu \partial_\sigma \phi_{\mu \rho} + \partial_\nu \partial_\rho \phi_{\sigma \mu} + \partial_\sigma \partial_\rho \phi_{\mu \nu} \, , \label{ex1} \\
	\partial (\partial \cdot \phi) &\equiv \partial_\nu (\partial^\lambda \phi_{\mu \lambda}) + \partial_\mu (\partial^\lambda \phi_{\lambda \nu}) \, , \label{ex2} \\
	\eta \partial^2 \cdot \phi &\equiv \eta_{\mu\nu} \partial^\rho \partial^\sigma \phi_{\rho \sigma} \, . \label{ex3}
	\end{align}
	The formalism implies the following set of rules:
	\begin{align}
	( \partial^p \phi )' &= \Box \partial^{p-2} \phi + 2 \partial^{p-1} \left( \partial \cdot \phi \right) + \partial^p \phi' \label{fs1} \\
	\partial \cdot (\partial^p \phi) &= \Box \partial^{p-1} \phi + \partial^p \left( \partial \cdot \phi \right)  \label{fs2} \\
	\left( \eta^k T_{(s)} \right)' &= [D + 2(s+k-1)] \eta^{k-1} T_{(s)} + \eta^k T_{(s)}' \label{traces} \\
	\partial^p \partial^q &= {{p+q}\choose{q}} \partial^{p+q} \\
	\eta^p \eta^q &= {{p+q}\choose{q}} \eta^{p+q} \\
	\partial \cdot \eta^{p} &= \eta^{p-1} \partial
	\end{align}
	\eqref{fs1} and \eqref{fs2} can further be generalized to:
	\begin{align}
	( \partial^n \phi )^{[p]} &= \sum_{k=0}^{p} \sum_{l=0}^{k} {p \choose k} {k \choose l} 2^l \Box^{p-k} \partial^{n - 2p + 2k -l} \left( \partial^l \cdot \phi^{[k-l]} \right) \\
	\partial^n \cdot ( \partial^p \phi ) &= \sum_{k=0}^{n} {n \choose k} \Box^{n-k} \partial^{p-n+k} \left( \partial^k \cdot \phi \right) \label{gen2}
	\end{align}
	The relations \eqref{fs1}-\eqref{gen2} will prove to be useful in simplifying our calculations.
	
	Note that this formalism is also implicit in the dimension of spacetime, as long as the relevant expressionts do not include traces of terms containing the metric tensor, as implied by \eqref{traces}.
	
	We introduce "$\fdot$" to denote maximal contraction between two tensors\footnotemark.\footnotetext{Francia and Sagnotti do not use this notation. Instead, such contractions are left implicit, which may look confusing to the untrained eye.} For tensors $\varphi$ of order $s$ and $\chi$ of order $r$, with $s>r$, the contraction is defined as
	\begin{equation}
	\varphi \fdot \chi \equiv \varphi_{\mu_1 \cdots \mu_r \mu_{r+1} \cdots \mu_s} \chi^{\mu_1 \cdots \mu_r} \, ,
	\end{equation}
	where both tensors are assumed to be symmetrized with the minimal number of unweighted terms, before contraction.
	For example, if $\varphi$ is a tensor of order three and $\chi$ is a tensor of order two,
	\begin{align}
	\varphi \fdot \chi &\equiv \varphi_{\mu\nu\sigma} \chi^{\mu \nu}  \, , \\
	\varphi \fdot \partial \chi &\equiv \varphi_{\mu\nu\sigma} \left( \partial^\mu \chi^{\nu\sigma} + \partial^\nu \chi^{\sigma\mu} + \partial^\sigma \chi^{\mu\nu} \right) = 3 \varphi_{\mu\nu\sigma} \partial^\mu \chi^{\nu\sigma} \, , \\
	\partial \varphi \fdot \eta \chi &\equiv \left( \partial_\mu \varphi_{\nu\sigma\rho} + \partial_\nu \varphi_{\sigma\rho\mu} + \partial_\sigma \varphi_{\rho\mu\nu} + \partial_\rho \varphi_{\mu\nu\sigma} \right) \\ \nonumber &\left( \eta^{\mu\nu} \chi^{\sigma\rho} + \eta^{\mu\sigma} \chi^{\nu\rho} + \eta^{\mu\rho} \chi^{\sigma\nu} + \eta^{\nu\sigma} \chi^{\mu\rho} + \eta^{\nu\rho} \chi^{\mu\sigma} + \eta^{\sigma\rho} \chi^{\mu\nu} \right) \\
	\nonumber &= 12 \partial^\sigma \varphi_{\sigma \mu \nu} \chi^{\mu\nu} + 12  \chi^{\mu\nu} \eta^{\sigma\rho} \partial_\mu \varphi_{\nu\sigma\rho} \, .
	\end{align}
	In other words, \textit{first} we symmetrize the tensors as in examples \eqref{ex1}-\eqref{ex3}, and \textit{then} we contract them. This notation will prove to be particularly useful in the analysis of actions and their variations. In the following segments, when we vary a Lagrangian, we will always vary it under the integral sign, as a variation of the action, i.e.
	\begin{equation}
	\delta \mathcal{S} [\varphi(x)] = \delta \int d^D x \, \mathcal{L}[\varphi(x)] = \int d^D x \, \delta \mathcal{L}[\varphi(x)] \, .
	\end{equation}
	When calculating such variations, we will often encounter terms of the form
	\begin{equation}
	\int d^D x \, A(x) \delta (\partial_\mu B(x)) \, ,
	\end{equation}
	where we perform partial integration to obtain
	\begin{equation}
	- \int d^D x \, \partial_\mu A(x) \delta B(x) + \text{(boundary terms)} \, .
	\end{equation}
	The boundary terms vanish due to the standard assumption that all fields vanish at infinity and that there are no non-trivial topological features of spacetime. This allows us to use the following relation:
	\begin{equation}
	\int d^D x \, A(x) \delta (\partial_\mu B(x)) = - \int d^D x \, \partial_\mu A(x) \delta B(x) \, .
	\end{equation}
	In the Francia-Sagnotti formalism, one should be careful when performing partial integration, since this operation might produce additional symmetry factors. For example, if $\varphi$ is a symmetric tensor of order $s$ and $\Lambda$ is a symmetric tensor of order $s-1$,
	\begin{align}
	\int d^D x \, \partial \Lambda \fdot \varphi &\equiv \int d^D x \, \left( \underbrace{\partial_{\mu_1} \Lambda_{\mu_2 \cdots \mu_{s}} + \dots +  \partial_{\mu_s} \Lambda_{\mu_1 \cdots \mu_{s-1}}}_{s \, \text{terms}} \right) \varphi^{\mu_1 \cdots \mu_s} \\	
	&= s \int d^D x \, \partial_{\mu_1}  \Lambda_{\mu_2 \cdots \mu_{s}} \varphi^{\mu_1 \cdots \mu_s} \\
	&= -s \int d^D x \, \Lambda_{\mu_1 \cdots \mu_{s-1}} \partial_{\mu_s} \varphi^{\mu_1 \cdots \mu_s} \\
	&\equiv -s \int d^D x \, \Lambda \fdot \partial \cdot \varphi \, .
	\end{align}
	Similarly, one should be careful when writing terms of the form $\varphi \fdot \eta \varphi'$ as terms of the form $\varphi' \fdot \varphi'$, because 
	\begin{align}
	\varphi \fdot \eta \varphi' &\equiv \varphi_{\mu_1 \cdots \mu_s} \eta_{\nu \sigma} \left( \underbrace{ \eta^{\mu_1 \mu_2} \varphi^{\mu_3 \cdots \mu_s \nu \sigma} + \dots + \eta^{\mu_{s-1} \mu_s} \varphi^{\mu_1 \cdots \mu_{s-2} \nu \sigma}}_{{s \choose 2} \, \text{terms}}  \right) \\
	&= {s \choose 2} \eta^{\mu_1 \mu_2} \varphi_{\mu_3 \cdots \mu_s} \eta_{\mu_1 \mu_2} \varphi^{\mu_3 \cdots \mu_s} \\
	&\equiv {s \choose 2} \varphi' \fdot \varphi' \, .
	\end{align}
	To drive the point home, we provide two additional examples that we will encounter in our calculations:
	\begin{align}
	\int d^D x \, \Lambda \fdot \partial^3 \varphi'' &=  - {{s-1} \choose 3} \int d^D x \, \partial^3 \cdot \Lambda \fdot \varphi'' \\
	\Lambda \fdot \eta \partial \cdot \varphi' &= {s-1 \choose 2} \Lambda' \fdot \partial \cdot \varphi'
	\end{align}
\subsection{Fronsdal's constrained theory}\label{fronsdalconstrained}
\subsubsection{Free theory}\label{freefronsdal}

	As we have already seen in \textbf{Section \ref{fronsdaleq}}, Fronsdal's HS theory, in the absence of sources, consists of the Fronsdal equation along with two unusual constraints, i.e.
	\begin{align}
	\mathcal{F} &= \Box \varphi - \partial (\partial \cdot \varphi) + \partial^2 \varphi' = 0 \, , \label{freq} \\
	\mathcal{L}_{\mathcal{F}} &= \frac{1}{2} \varphi \fdot \left( \mathcal{F} - \frac{1}{2} \eta \mathcal{F}' \right) \, , \label{lagrangianF} \\
	\delta \varphi &= \partial \Lambda \, , \label{gauge} \\
	\Lambda' &= 0 \, \label{constr1},\\
	\varphi'' &= 0 \, \label{constr2}.
	\end{align}
	
	Let us show how \eqref{lagrangianF} produces \eqref{freq} as the equation of motion in the absence of sources. Varying the action gives	
	\begin{align}
	\delta \mathcal{S}_\mathcal{F} = \dint  \delta \mathcal{L}_\mathcal{F} &= \frac{1}{2} \dint \left[ \delta \varphi \fdot \left( \mathcal{F} - \frac{1}{2} \eta \mathcal{F}' \right) +  \varphi \fdot  \left( \delta (\mathcal{F}) - \frac{1}{2} \eta (\delta \mathcal{F}') \right) \right] \\
	&= \frac{1}{2} \dint \Big\{ \delta \varphi \fdot \left( \Box \varphi - \partial (\partial \cdot \varphi) + \partial^2 \varphi' + \eta \partial^2 \cdot \varphi - \eta \Box \varphi \right) \\
	&+ 	 \varphi \fdot \left[ \delta (\Box \varphi) - \delta (\partial (\partial \cdot \varphi)) + \delta (\partial^2 \varphi') + \delta (\eta \partial^2 \cdot \varphi) - \delta (\eta \Box \varphi ) \right] \Big\}  \nonumber \\
	&= \dint \left( \Box \varphi - \partial (\partial \cdot \varphi) + \partial^2 \varphi' + \eta \partial^2 \cdot \varphi - \eta \Box \varphi \right)  \fdot \delta \varphi \\
	&= \dint \left( \mathcal{F} - \frac{1}{2} \eta \mathcal{F}' \right) \fdot \delta \varphi \, ,
	\end{align}
	so the equation of motion reads
	\begin{equation}
	\mathcal{F} - \frac{1}{2} \eta \mathcal{F}' = 0 \, , \label{ffeq}
	\end{equation}
	which indeed reduces to
	\begin{equation}
	\mathcal{F} = 0 \, ,
	\end{equation}	
	since there are no sources on the right-hand side of \eqref{ffeq}.
	Armed with the powerful formalism, let us now take a closer look at the two constraints \eqref{constr1} and \eqref{constr2}. We would like to find where exactly they come from so that we can construct an equivalent \textit{unconstrained} theory.
	
\paragraph{Why traceless $\Lambda$?\newline}
	The Fronsdal equation \eqref{freq} transforms under the gauge variation \eqref{gauge} as
	\begin{equation}
	\delta \mathcal{F} = 3 \partial^3 \Lambda' \label{FVar} \, ,
	\end{equation}
	which is why we demand that the gauge parameter be traceless.	
	If we could find an appropriate linear combination of fully gauge-invariant terms, we could formulate a theory without imposing this constraint.	
	One way of getting around this would be through a differential constraint,
	\begin{equation}
	\partial^3 \Lambda' \, (x) = 0 \, ,
	\end{equation}
	without directly constraining $\Lambda'$. If the gauge parameter $\Lambda(x)$ vanishes at infinity, the only solution would indeed be $\Lambda'=0$.	
	Another way to dispense with this constraint is to introduce a non-dynamical spin-$(s-3)$ \textit{compensator} field $\alpha(x)$, which transforms under the gauge variation as
	\begin{equation}
	\delta \alpha = \Lambda' \, ,
	\end{equation}
	and modify the equation of motion to
	\begin{equation}
	\mathcal{F} - 3 \partial^3 \alpha = 0 \, .
	\end{equation}
	If we introduce a second non-dynamical spin-$(s-4)$ field, this theory can be described by a Lagrangian, as we will explain in \textbf{Section \ref{localunconstrained}}.	
	The third way to avoid the traceless $\Lambda$ is to work within a manifestly gauge-invariant geometric framework. Unfortunately, as we will see in \textbf{Section \ref{nonlocal}}, this forces us to abandon locality and instead work with non-local or higher-order (in derivatives) terms.
	
\paragraph{Why doubly-traceless $\varphi$?\newline}

	The relation that lies at the heart of this constraint is the so-called \textit{anomalous\footnotemark Bianchi identity},\footnotetext{It is called \textit{anomalous} because it does not vanish. If the right-hand side vanishes, it is simply the \textit{Bianchi identity}.}
	\begin{equation}
	\partial \cdot \mathcal{F} - \frac{1}{2} \partial \mathcal{F}' = - \frac{3}{2} \partial^3 \varphi'' \, . \label{bianchi}
	\end{equation}
	We would like the Fronsdal action to be gauge invariant, so let us see what its gauge variation\footnotemark\, produces. Using \eqref{bianchi}, one obtains
	\footnotetext{We use $\delta_{\Lambda}$ to avoid confusing this variation with the usual functional variation $\delta$.}
	\begin{equation}
	\delta_{\Lambda} \mathcal{S}_\mathcal{F} = \dint \delta_{\Lambda} \mathcal{L}_{\mathcal{F}} = \frac{1}{2} \dint \left[ \partial \Lambda \fdot \left( \mathcal{F} - \frac{1}{2} \eta \mathcal{F}' \right) +  \varphi \fdot \delta_{\Lambda} \left(  - \frac{3}{2} \partial^3 \varphi'' \right) \right] \, .
	\end{equation}
	The second term under the integral vanishes if we impose $\Lambda' = 0$, so we have
	\begin{align}
	\delta_{\Lambda} \mathcal{S}_\mathcal{F} = \dint \delta_{\Lambda} \mathcal{L}_{\mathcal{F}} &= \frac{1}{2} \dint \partial \Lambda \fdot \left( \mathcal{F} - \frac{1}{2} \eta \mathcal{F}' \right) \\ 
	&= -\frac{s}{2} \dint \Lambda \fdot \partial \cdot \left( \mathcal{F} - \frac{1}{2} \eta \mathcal{F}' \right) \\
	&= -\frac{s}{2} \dint \Lambda \fdot \left( \underbrace{\partial \cdot \mathcal{F} - \frac{1}{2} \partial \mathcal{F}'}_{\eqref{bianchi}} - \frac{1}{2} \eta \partial \cdot \mathcal{F}' \right) \\
	&= -\frac{s}{2} \dint \left[ \Lambda \fdot \left( -\frac{3}{2} \partial^3 \varphi'' \right) - \frac{1}{2} \Lambda \fdot \eta \partial \cdot \mathcal{F}' \right] \\
	&= -3 \dint \left[ {s \choose 4} \partial^3 \cdot \Lambda \fdot \varphi'' - \frac{1}{4} {s \choose 3} \Lambda'\fdot \partial \cdot \mathcal{F}'  \right] \, .
	\end{align}
	Once again, the second term under the integral vanishes if we impose $\Lambda' = 0$. It follows that it is necessary to impose the additional constraint $\varphi'' = 0$ for the action to be gauge-invariant.	
	
	An alternative way to get around the double-tracelessness constraint is to work within a geometric framework, where we generalize the Fronsdal tensor into an equivalent object satisfying generalized Bianchi identities. As previously mentioned, the price to pay for the elegant geometric theory is higher-order terms or non-locality.
	
\paragraph{Counting degrees of freedom$\newline$}

	Let us show that the constrained Fronsdal equation propagates the correct number of degrees of freedom. In case of a massless spin-$s$ bosonic field, arguments from representation theory (as discussed in \textbf{Section \ref{fpe}} and specifically in relation to the Fronsdal equation in \cite{verybasics}) show that the correct number is
	\begin{equation}
	\#(D-2,s) - \#(D-2, s-2) = {{D+s-3} \choose {s}} - {{D+s-5} \choose {s-2}} \, . \label{dof}
	\end{equation}
	We begin by counting the number of independent components of $\varphi$. It is a fully symmetric doubly-traceless $D-$dimensional tensor of order $s$, so that number is
	\begin{equation}
	\#(D,s) - \#(D,s-4) = {{D+s-1} \choose {s}} - {{D+s-5} \choose {s-4}} \, .
	\end{equation}
	We proceed by partially fixing the gauge, imposing the de Donder gauge condition,
	\begin{equation}
	\mathcal{D} = \partial \cdot \varphi - \frac{1}{2} \partial \varphi' = 0 \, ,
	\end{equation}
	which reduces the Fronsdal equation to a wave equation,
	\begin{equation}
	\Box \varphi = 0 \, .
	\end{equation}
	Since $\mathcal{D}$ is traceless and of order $s-1$, fixing the de Donder tensor corresponds to eliminating
	\begin{equation}
	\#(D,s-1) - \#(D,s-3) = {{D+s-2} \choose {s-1}} - {{D+s-4} \choose {s-3}}
	\end{equation}
	independent components.
	However, fixing $\mathcal{D}$ does not fully fix the gauge, since
	\begin{equation}
	\delta \mathcal{D} = \Box \Lambda \, .
	\end{equation}
	Fixing this residual gauge freedom also corresponds to eliminating $\#(D,s-1) - \#(D,s-3)$ independent components.
	In total, this leaves us with
	\begin{align}
	&\#(D,s) - \#(D,s-4) - 2\left\{\#(D,s-1) - \#(D,s-3)\right\} = \nonumber \\ &\#(D-2,s) - \#(D-2, s-2) \, ,
	\end{align}
	which is the same as \eqref{dof}.	
	Note that we used the double tracelessness of $\varphi$ to count the propagating degrees of freedom. However, this is merely a \textit{sufficient} condition for the correct number, not a \textit{necessary} one.

\subsubsection{Interacting theory with an external current}\label{interacting constrained}
	Let us begin the analysis of the HS gauge field coupled to an external current within the framework of Fronsdal's constrained theory by defining the \textit{Fronsdal-Einstein} tensor $\mathcal{G}$,
	\begin{equation}
	\mathcal{G} := \mathcal{F} - \frac{1}{2} \eta \mathcal{F}' \, .
	\end{equation}
	We showed in \textbf{Section \ref{freefronsdal}} that this is precisely the left-hand side of the equation of motion, as obtained from \eqref{lagrangianF}.
	An interaction term in the Lagrangian, for some \textit{generic} totally symmetric external current $J$ can be written as
	\begin{equation}
	\mathcal{L}_{int} = -\frac{1}{2} \varphi \fdot J \, , \label{current}
	\end{equation}
	so the total action reads
	\begin{equation}
	\mathcal{S} [\varphi, J] = \dint \mathcal{L} = \dint \left( \mathcal{L}_{\mathcal{F}} + \mathcal{L}_{int} \right) = \mathcal{S}_\mathcal{F} - \frac{1}{2} \dint \varphi \fdot J \, . \label{constrainedS}
	\end{equation}
	As demonstrated in \textbf{Section \ref{freefronsdal}}, 
	\begin{equation}
	\delta_{\Lambda} \mathcal{S}_{\mathcal{F}} = 0 \, ,
	\end{equation}
	so we need to investigate the effect of the interaction term $\mathcal{L}_{int}$, since it need not be gauge-invariant.
	The equation of motion obtained by varying \eqref{constrainedS} reads
	\begin{equation}
	\mathcal{G} = J \, . \label{el}
	\end{equation}	
	Taking the trace of \eqref{el} yields
	\begin{equation}
	\mathcal{F}' = \frac{-2}{D+2(s-3)} J' \, , \label{fj}
	\end{equation}
	which in turn implies
	\begin{equation}
	J'' = 0 \, , \label{jpp0}
	\end{equation}
	since $\mathcal{F}''=0$ when $\varphi''=0$.
	We can now rewrite \eqref{el} as
	\begin{equation}
	\mathcal{F} = J - \frac{1}{D+2(s-3)} \eta J' \, . \label{fronsINT}
	\end{equation}
	Taking the divergence of \eqref{el} and using \eqref{bianchi}, we get
	\begin{align}
	\partial \cdot J = -\frac{1}{2} \eta \partial \cdot \mathcal{F}' \label{Whencefore It Cometh?}
	\end{align}
	Substituting \eqref{fj} into \eqref{Whencefore It Cometh?} yields
	\begin{equation}
	\partial \cdot J - \frac{1}{D+2(s-3)} \eta \partial \cdot \ J'  = 0 \, . \label{divtracelessJ}
	\end{equation}
	The left-hand side of \eqref{divtracelessJ} is actually the traceless part of $\partial \cdot J$. In general, the traceless part of a fully symmetric tensor $\chi$ of order $s$ in $D$-dimensional spactime is\footnotemark\footnotetext{To compactify the notation, here we begin to use the \textbf{falling factorial} function, defined as \newline $n^{\underline{k}} = \frac{n!}{(n-k)!}$, and we define the \textbf{falling double factorial} function $n^{\uuline{k}} = \frac{n!!}{(n-k)!!}$.}
	\begin{equation}
	\mathcal{T}_D[\chi] = \sum_{k=0}^{[s/2]} \frac{(-1)^k}{[D+2(s-2)]^{\uuline{k}}} \eta^k \chi^{[k]} :=  \sum_{k=0}^{[s/2]} \rho_k (D,s) \eta^k \chi^{[k]} \, , \label{traceless}
	\end{equation}
	which is easily checked by direct computation. In \eqref{traceless}, we define coefficients $\rho_k (D,s)$ for later convenience. Using \eqref{traceless} and \eqref{jpp0}, we see that indeed
	\begin{equation}
	\mathcal{T}_D[\partial \cdot J] = \partial \cdot J - \frac{1}{D+2(s-3)} \eta \partial \cdot \ J' \, . \label{jtraceless}
	\end{equation}
	Also,
	\begin{equation}
	\mathcal{F} = \mathcal{T}_{D-2} [J] \, .
	\end{equation}
	Therefore, in general, only the traceless part of the divergence of $J$ vanishes.
	\bigskip\par\centerline{*\,*\,*}\medskip\par
	To understand the physical meaning of \eqref{jtraceless}, we need to introduce the concept of \textit{current exchange}. Let us motivate the idea on a familiar case of spin-$1$ fields, i.e. Maxwell's theory of electrodynamics.
	
	In the manifestly Lorentz-covariant formalism, Maxwell's equations coupled to an external current $J^\mu$ read
	\begin{equation}
	\Box A^\mu - \partial^\mu (\partial \cdot A) = J^\mu \, ,
	\end{equation}
	where consistency demands that the current be conserved, i.e.
	\begin{equation}
	\partial_\mu j^\mu = 0 \, .
	\end{equation}
	In the momentum space, this translates to
	\begin{align}
	(p^2 \eta_{\mu\nu} - p_\mu p_\nu) A^\nu &= J_\mu \, , \\
	p^\mu J_\mu &= 0 \, .
	\end{align}
	It follows that, for a current-current interaction,
	\begin{equation}
	p^2 A_\mu J^\mu = J_\mu J^\mu \, .
	\end{equation}
    By current exchange, we mean the exchange between the degrees of freedom that take part in this interaction. As we know from electrodynamics, interactions mediated by photons only respond to the transverse part of the current, since the photon has no longitudinal degrees of freedom. Therefore, instead of considering the full Lorentzian product $J_\mu J^\mu$, we can project an on-shell current (i.e. a current satisfying the equation of motion)  $J_\mu (p)$ to its transverse part using the projection operator $\Pi$
    \begin{equation}
    \Pi_{\mu\nu} = \eta_{\mu\nu} - p_\mu \bar{p}_\nu - p_\nu \bar{p}_\mu \, , \label{projektor}
    \end{equation}
    where $p$ is the exchanged on-shell momentum, satisfying $p^2 = 0$, and $\bar{p}$ is a vector that satisfies $\bar{p}^2 = 0$ and $p_\mu \bar{p}^\mu = 1$.
    One can check by direct computation that, indeed,
    \begin{equation}
	p^\mu \Pi_{\mu\nu} J^\nu = 0 \, ,
    \end{equation}
    and
    \begin{equation}
    J_\mu J^\mu =  J^\mu \Pi_{\mu\nu} J^\nu \, . \label{jpj}
    \end{equation}
    Now, since 
    \begin{equation}
    \eta_{\mu \nu} \Pi^{\mu\nu} = D - 2 \, ,
    \end{equation}
    it follows\footnotemark\footnotetext{Equation \eqref{jpj} is basically an eigenvalue problem. The trace of a linear operator equals the sum of its eigenvalues. Since the eigenvalues of a projection operator equal $0$ or $1$, its trace equals the dimension of the subspace to which it projects.}  that the number of degrees of freedom taking part in the interaction is $D-2$.
    
    Instead of working in the manifestly Lorentz-covariant formulation, we can repeat the procedure in the light-cone formulation, where we have two null coordinates (i.e. coordinates on the light cone),
    \begin{align}
    x^+ &= \frac{t+x}{\sqrt{2}} \, , \\
    x^- &= \frac{t-x}{\sqrt{2}} \, ,
    \end{align}
    and the remaining $D-2$ coordinates are spatial. This allows us to work in the light-cone gauge,
    \begin{equation}
    A^+ = 0 \, ,
    \end{equation}
    which eliminates all unphysical degrees of freedom.
    In this formulation, Maxwell's equations in the momentum space take a simple form that involves only the spatial coordinates,
    \begin{equation}
    p^2 A_i = j_i \, ,
    \end{equation}
    where latin indices denote the components of ($D-2$)-dimensional Euclidean vectors. The current-current interaction becomes
    \begin{equation}
    p^2 j_i A^i = j_i j^i \, .
    \end{equation}
    Since all components are physical\footnotemark\footnotetext{Because the light-cone formulation corresponds to working in the "reference frame" of a massless particle, where all degrees of freedom are particle's proper degrees of freedom. This is why the light-cone frame is sometimes referred to as the \textit{infinite momentum frame.}}, we simply count the number of components of $j_i$, which is $D-2$, in agreement with our previous conclusion. 
    
    The general idea is to check whether
    \begin{equation}
    J_{\mu_1 \cdots \mu_s} \mathcal{P}^{\mu_1 \cdots \mu_s \nu_1 \cdots \nu_s} J_{\nu_1 \cdots \nu_s} = j_{a_1 \cdots a_s} j^{a_1 \cdots a_s} \label{currx}
    \end{equation}
    holds for spin-$s$ current exchanges, where $\mathcal{P}$ denotes the proper analogue of the projection operator \eqref{projektor}. The right-hand side of \eqref{currx} implies that the proper number of degrees of freedom in the current exchange is equal to the number of independent components of the current in the light-cone gauge. As we saw in \textbf{Section \ref{fpe}}, this number is equal to the number of independent components of a traceless fully symmetric tensor of order $s$ in $D-2$ dimensions.
    
    This means that $\mathcal{P}$ should be an operator that projects the current to its transverse part and then extracts its traceless part. Using \eqref{traceless} and \eqref{projektor}, we see that $\mathcal{P}$ has to be
    \begin{equation}
    \mathcal{P}^{(\mu)(\nu)} J_{(\nu)} = \mathcal{T}_{D-2} [\Pi \cdot J] \, , \label{conserved}
    \end{equation}
    where we write $(\mu)$ and $(\nu)$ to indicate a totally symmetric set of $s$ indices.
	\bigskip\par\centerline{*\,*\,*}\medskip\par
	Coming back to the interacting theory with an external current in the constrained formulation, we see that \eqref{jtraceless} determines the currrent exchange. It implies that
	\begin{align}
	J_{(\mu)} \mathcal{P}^{(\mu)(\nu)} J_{(\nu)} &= J \fdot \left( J - \frac{1}{D+2(s-3)} \eta J'\right) \, , \\
	&= J \fdot J - \frac{1}{D+2(s-3)}  J \fdot \eta J' \, , \\
	&= J \fdot J - \frac{s(s-1)}{2[D+2(s-3)]}  J' \fdot J' \, \\
	&= J \fdot J + \rho_1 (D-2,s) {s \choose 2} J' \fdot J' . \label{constrained exchange}
	\end{align}
	We will return to this result to compare it with the analogous result in the unconstrained formulation.
\subsection{Local unconstrained theory}\label{localunconstrained}

	Let us demonstrate how we can rewrite Fronsdal's theory without the usual
	\begin{equation}
	\Lambda' = 0 \quad \& \quad \varphi'' = 0
	\end{equation}
	constraints. This is accomplished here by introducing two compensator fields.
	
\subsubsection{Free theory}

	We begin by considering the Fronsdal tensor $\mathcal{F}$ and its gauge transformation \eqref{FVar}. From $\mathcal{F}$, one can build a fully gauge-invariant tensor,
	\begin{equation}
	\mathcal{A} := \mathcal{F} - 3 \partial^3 \alpha \, , \label{Acomp}
	\end{equation}
	where we introduce the field $\alpha(x)$ as a spin-$(s-3)$ \textit{compensator}, which transforms as
	\begin{equation}
	\delta_\Lambda \alpha = \Lambda'
	\end{equation}
	under the gauge transformation \eqref{FVar}.	
	The Bianchi identity for $\mathcal{A}$ reads
	\begin{equation}
	\partial \cdot \mathcal{A} - \frac{1}{2} \partial \mathcal{A}' = - \frac{3}{2} \partial^3 \left( \varphi'' - 4 \partial \cdot \alpha - \partial \alpha'  \right) =: -\frac{3}{2} \partial^3 \mathcal{C} \, , \label{ABianchi}
	\end{equation}
	where we have identified a gauge-invariant tensor, which we denote by $\mathcal{C}$, i.e.
	\begin{equation}
	\mathcal{C} = \varphi'' - 4 \partial \cdot \alpha - \partial \alpha' \, .
	\end{equation}
	In analogy with \eqref{lagrangianF}, we write the Lagrangian
	\begin{equation}
	\mathcal{L}_0 = \frac{1}{2} \varphi \fdot \left( \mathcal{A} - \frac{1}{2} \eta \mathcal{A}' \right) \, .
	\end{equation}
	Varying the action, we get
	\begin{align}
	\delta_{\Lambda} \mathcal{S}_0 = \dint \delta_{\Lambda} \mathcal{L}_0 &= \frac{1}{2} \dint \partial \Lambda \fdot \left( \mathcal{A} - \frac{1}{2} \eta \mathcal{A}' \right) \\ 
	&= - \frac{s}{2} \dint \Lambda \fdot \left( \underbrace{\partial \cdot \mathcal{A} - \frac{1}{2} \partial \mathcal{A}'}_{\eqref{ABianchi}} - \frac{1}{2} \eta \partial \cdot \mathcal{A}' \right) \\
	&= -3 \dint \left[ {s \choose 4} \partial^3 \cdot \Lambda \fdot \mathcal{C} - \frac{1}{4} {s \choose 3} \Lambda' \fdot \partial \cdot \mathcal{A}' \right] \, .
	\end{align}
	We can make all the terms under the integral vanish by adding to $\mathcal{L}_0$ 
	\begin{equation}
	\mathcal{L}_1 = -\frac{3}{4} {s \choose 3} \alpha \fdot \partial \cdot \mathcal{A}' + 3 {s \choose 4} \beta \fdot \mathcal{C} \, ,
	\end{equation}
	where we introduce the second compensator\footnotemark \footnotetext{Technically, it is just a Lagrange multiplier.}, a spin-$(s-4)$ field denoted by $\beta$ that transforms as
	\begin{equation}
	\delta_\Lambda \beta = \partial^3 \cdot \Lambda
	\end{equation}
	under the gauge transformation \eqref{FVar}.	
	Finally, we can write the fully gauge-invariant Lagrangian for the unconstrained local theory as
	\begin{equation}
	\mathcal{L} = \frac{1}{2} \varphi \fdot \left( \mathcal{A} - \frac{1}{2} \eta \mathcal{A}' \right) -\frac{3}{4} {s \choose 3} \alpha \fdot \partial \cdot \mathcal{A}' +  3 {s \choose 4} \beta \fdot \mathcal{C} \, . \label{abplagrangian}
	\end{equation}
	We can introduce the third gauge-invariant tensor $\mathcal{B}$,
	\begin{equation}
	\mathcal{B} := \beta  + \Box \partial \cdot \alpha + \frac{1}{2}  \partial (\partial^2 \cdot \alpha) - \frac{1}{2} \partial^2 \cdot \varphi' \, \label{Bcomp}
	\end{equation}
	and note that \eqref{abplagrangian} may be generalized to
	\begin{equation}
	\mathcal{L}_k = \frac{1}{2} \varphi \fdot \left( \mathcal{A} - \frac{1}{2} \eta \mathcal{A}' \right) -\frac{3}{4} {s \choose 3} \alpha \fdot \partial \cdot \mathcal{A}' +  3 {s \choose 4} \left( \beta - k \mathcal{B} \right) \fdot \mathcal{C} \, , \label{abplagrangianK}
	\end{equation}
	without affecting the equations of motion, so that \eqref{abplagrangian} corresponds to $k=0$.
	
	As shown in \cite{fms}, a more general analysis reveals that adding quadratic terms in $\mathcal{A}$, $\mathcal{B}$ and $\mathcal{C}$ to \eqref{abplagrangianK} does not produce any terms that would lead to different equations of motion.
	Therefore, the free unconstrained local theory is parametrized by a real parameter $k$ and the gauge-invariant field equations read
	\begin{align}
	E_\varphi (k) &:= \mathcal{A} - \frac{1}{2} \eta \mathcal{A}' + \frac{1+k}{4} \eta \partial^2 \mathcal{C} + (1-k) \eta^2 \mathcal{B} = 0 \, , \label{Evarphi} \\
	E_\alpha (k) &:= -\frac{3}{2} {s \choose 3} \left[ \partial \cdot \mathcal{A}'  -\frac{1+k}{2} \left( \partial \Box  + \partial^2 \partial \cdot \right) \mathcal{C} + (k-1) \left(  2 \partial + \eta \partial \cdot  \right) \mathcal{B}  \right] = 0 \, , \label{Ealpha} \\
	E_\beta (k) &:= 3 {s \choose 4} (1-k) \mathcal{C} = 0 \, . \label{Ebeta}
	\end{align}
	We can use these three tensors to write the final Lagrangian in a particularly elegant form,
	\begin{equation}
	\mathcal{L}_k = \frac{1}{2} \varphi \fdot E_\varphi (k) + \frac{1}{2} \alpha \fdot E_\alpha (k) + \frac{1}{2} \beta \fdot E_\beta (k) \, .
 	\end{equation}
    From equations \eqref{Evarphi}-\eqref{Ebeta}, if $k\neq1$, it follows that
	\begin{align}
	\mathcal{A} &\equiv \mathcal{F} - 3 \partial^3 \alpha = 0 \, , \\
	\mathcal{C} &\equiv \varphi'' - 4 \partial \cdot \alpha - \partial \alpha' = 0 \, .
	\end{align}
	After fixing the gauge to $\Lambda' = 0$ we are left with
	\begin{align}
	\mathcal{F} &= 0 \, , \\
	\varphi'' &= 0 \, ,
	\end{align}
	which is exactly equivalent to Fronsdal's constrained formulation.
	
\subsubsection{Interacting theory with an external current}
	In the unconstrained formulation described in the previous subsection, setting $k=0$, coupling to an external source \eqref{current} is described by 
	\begin{equation}
	\mathcal{A} - \frac{1}{2} \eta \mathcal{A}' + \eta^2 \mathcal{B} = J \label{abj}
	\end{equation}	
	We can define a quantity $\mathcal{K}$,
	\begin{equation}
	\mathcal{K} := J - \eta^2 \mathcal{B} \, , \label{kdef}
	\end{equation}
	and write the equation of motion as
	\begin{equation}
	\mathcal{A} - \frac{1}{2} \eta \mathcal{A}' = \mathcal{K} \, , \label{unconseom}
	\end{equation}
	so that, formally, $\mathcal{A}$ and $\mathcal{K}$ play the same role as $\mathcal{F}$ and $J$ play in the constrained formalism of \textbf{Section \ref{interacting constrained}}.
	Note that
	\begin{equation}
	\mathcal{A}'' = 3 \Box \mathcal{C} + 3 \partial (\partial \cdot \mathcal{C}) + \partial^2 \mathcal{C}' = 0 \, , \label{app}
	\end{equation}
	since $\mathcal{C}$ vanishes as a result of \eqref{Ebeta}. Since $\mathcal{A}''$ vanishes when the equations of motion are satisfied, we can write
	\begin{equation}
	\mathcal{B} = \sum_{k=2}^{n+1} \sigma_k \eta^k J^{[k]} \, ,
	\end{equation}
	where $n=\left[ \frac{s-1}{2} \right]$ and we can determine the coefficients $\sigma_k$ from the condition $\mathcal{K}'' = 0$. A direct computations yields
	\begin{equation}
	\mathcal{B} = \sum_{k=2}^{n+1} (1-n) \, \rho_n(D-2,s) \eta^k J^{[k]} \, , \label{bcons}
	\end{equation}
	which allows us to rewrite \eqref{abj} as
	\begin{equation}
	\mathcal{A} - \frac{1}{2} \eta \mathcal{A}' = J - \sum_{k=2}^{n+1} (1-n) \, \rho_n(D-2,s) \eta^k J^{[k]}
	\end{equation}	
	 We can use the formal correspondence between $\mathcal{A}$ and $\mathcal{K}$, and $\mathcal{F}$ and $J$, to skip the explicit calculation and immediately write
	\begin{equation}
	\mathcal{A} = \mathcal{K} - \frac{1}{D + 2(s-3)} \eta \mathcal{K}' \, , \label{acons}
	\end{equation}
	in analogy with \eqref{fronsINT}. Using \eqref{bcons} and \eqref{kdef}, we arrive at
	\begin{equation}
	\mathcal{A} = \sum_{k=0}^{n+1} \rho_k (D-2,s) \eta^k J^{[k]}
	\end{equation}
	The current exchange is thus
	\begin{align}
	J_{(\mu)} \mathcal{P}^{(\mu)(\nu)} J_{(\nu)} &= \sum_{k=0}^{n+1} \rho_k (D-2,s) J \fdot \eta^k J^{[k]}  \\
	&= \sum_{k=0}^{n+1} \rho_k (D-2,s) {{s-2k} \choose k} J^{[k]} \fdot J^{[k]} \, , \label{unconstrained exchange}
	\end{align}
	which agrees with \eqref{constrained exchange}, as we can see by expanding the first two terms,
	\begin{align}
	J_{(\mu)} \mathcal{P}^{(\mu)(\nu)} J_{(\nu)} & = J \fdot J + \rho_1 (D-2,s) {s \choose 2} J' \fdot J' \\ \nonumber &\quad + \sum_{k=2}^{n+1} \rho_k (D-2,s) J \fdot \eta^k J^{[k]} \, .
	\end{align}

\newpage
\subsection{Non-local unconstrained theory}\label{nonlocal}

	Instead of introducing compensator fields $\alpha$ and $\beta$ and formulating the theory in terms of $\mathcal{A}$, $\mathcal{C}$ and $\mathcal{B}$ tensors, we can construct it using only the gauge field $\varphi$ if we allow non-local operators, i.e. powers of $\frac{1}{\Box}$ \footnotemark.\footnotetext{Alternatively, we could multiply the equations with the appropriate power of $\Box$ and have a higher-order derivative theory instead. However, it is not clear if the higher-order formulation of the theory is equivalent to the non-local formulation.}	
	Let us show here how to construct the theory in this manner.

\subsubsection{Free theory}

	One begins by building a non-local tensor $\mathcal{H}$ that satisfies
	\begin{equation}
	\delta_{\Lambda} \mathcal{H} = 3 \Lambda' \, ,
	\end{equation}
	so that $\mathcal{F} - \partial^3 \mathcal{H}$ becomes gauge-invariant without any additional constraints or compensator fields.	
	As shown in \cite{fs1}, inspired by HS generalizations of metric connections from general relativity (developed in \cite{dwf} and later explained in more detail in \textbf{Section \ref{geometric}}), we can try to construct a generalized Fronsdal tensor $\mathcal{F}_n$ that transforms as
	\begin{equation}
	\delta_\Lambda \mathcal{F}_{n} = (2n+1) \frac{\partial^{2n+1}}{\Box^{n-1}} \Lambda^{[n]} \, \label{gaugenonlocal}
	\end{equation}
	under the gauge variation \eqref{gauge}. This way, for high enough $n$, $\mathcal{F}_n$ becomes gauge-invariant. Since we also want the action to be gauge invariant, we require that $\mathcal{F}_n$ satisfies a generalization of the Bianchi identity,
	\begin{equation}
	\partial \cdot \mathcal{F}_{n} - \frac{1}{2n} \partial \mathcal{F}_{n}{}' = - \left( 1 + \frac{1}{2n} \right) \frac{\partial^{2n+1}}{\Box^{n-1}} \varphi^{[n+1]} \, , \label{bianchinonlocal}
	\end{equation}
	which also vanishes for high enough $n$. The generalized Fronsdal tensor $\mathcal{F}_n$ that satisfies all these requirements reads	
	\begin{equation}
	\mathcal{F}_{n+1} = \mathcal{F}_{n} - \frac{1}{n+1} \frac{\partial}{\Box} \left( \partial \cdot \mathcal{F}_{n} \right) +  \frac{1}{(n+1)(2n+1)} \frac{\partial^2}{\Box} \mathcal{F}_{n}{}' \, ,
	\end{equation}
	where $\mathcal{F}_{1} = \mathcal{F}$ (or equivalently, $\mathcal{F}_0 = \Box \varphi$), as one can easily check through direct computation and using simple inductive arguments.

	To construct a spin-$s$ theory, we use $\mathcal{F}_{n+1}$ with $n=\left[ \frac{s-1}{2} \right]$, the minimal value for which the gauge variation and the Bianchi identity \eqref{bianchinonlocal} both vanish.	
	The corresponding generalized Einstein-like tensor reads
	\begin{equation}
	\mathcal{G}_{n} = \sum_{k=0}^{n+1} \frac{(-1)^k}{2^k (n+1)^{\underline{k}}} \eta^k \mathcal{F}_{n+1}^{[k]} \, .
	\end{equation}
	One can check that $\mathcal{G}_{n}$ is indeed divergenceless, as required by the gauge-invariance of the action, using the traces of \eqref{bianchinonlocal}, which satisfy
	\begin{equation}
	\partial \cdot \mathcal{F}_{n+1}^{[k]} - \frac{1}{2(n-k+1)} \partial \mathcal{F}_{n+1}^{[k+1]} = 0 \, , \quad (k \leq n) \label{pbianchinonlocal}
	\end{equation}
	and applying it successively to terms in $\partial \cdot \mathcal{G}_{n}$.
	For clarity, let us show how all the pieces fit together to make the action gauge-invariant.
	\begin{align}
	\delta_\Lambda \mathcal{S}_n &= \dint \delta_\Lambda \mathcal{L}_n \\ &= \frac{1}{2} \dint \left( \partial \Lambda \fdot \mathcal{G}_n + \varphi \fdot \delta_\Lambda \mathcal{G}_n \right) \\
	&= \frac{1}{2} \sum_{k=0}^{n+1} \frac{(-1)^k}{2^k (n+1)^{\underline{k}}} \dint \left( \partial \Lambda \fdot \eta^k \mathcal{F}_{n+1}^{[k]} +  \varphi \fdot \eta^k \underbrace{\delta_\Lambda \mathcal{F}_{n+1}^{[k]}}_{=0, \eqref{gaugenonlocal}} \right) \\
	&= -\frac{s}{2} \sum_{k=0}^{n+1} \frac{(-1)^k}{2^k (n+1)^{\underline{k}}} \dint \left[ \Lambda \fdot \partial \cdot \left( \eta^k \mathcal{F}_{n+1}^{[k]} \right) \right] \\
	&= -\frac{s}{2} \dint \Lambda \fdot \partial \cdot \mathcal{F}_{n+1} -\frac{s}{2} \sum_{k=1}^{n+1} \frac{(-1)^k}{2^k (n+1)^{\underline{k}}} \dint \Lambda \fdot  \eta^k \partial \cdot \mathcal{F}_{n+1}^{[k]} \\ \nonumber &\quad+s \sum_{k=0}^{n} \frac{(-1)^k}{2^k (n+1)^{\underline{k+1}}} \dint \Lambda \fdot  \eta^{k} \partial \mathcal{F}_{n+1}^{[k+1]} \\
	&= -\frac{s}{2}  \sum_{k=0}^{n} \frac{(-1)^k}{2^k (n+1)^{\underline{k}}} \dint \eta^k \left( \underbrace{ \partial \cdot \mathcal{F}_{n+1}^{[k]} - \frac{1}{2(n-k+1)} \partial \mathcal{F}_{n+1}^{[k+1]} }_{=0, \eqref{pbianchinonlocal}} \right) \\
	&= 0
	\end{align}

\subsubsection{Interacting theory with an external current}

	If the system is coupled to a generic totally symmetric external current $\mathcal{J}$, a natural starting point would be to write the Lagrangian as
	\begin{equation}
	\mathcal{L} = \frac{1}{2} \varphi \fdot \left( \mathcal{G}_{n} - \mathcal{J}  \right)
	\end{equation}
	and the field equations read
	\begin{equation}
	\mathcal{G}_{n} \equiv  \sum_{k=0}^{n+1} \frac{(-1)^k}{2^k (n+1)^{\underline{k}}} \eta^k \mathcal{F}_{n+1}^{[k]} = \mathcal{J} \, . \label{nonlocal current}
	\end{equation}
	We proceed like in all previous cases, inverting \eqref{nonlocal current} to extract the current exchange. Taking successive traces of \eqref{nonlocal current} and multiplying both sides with metric tensors to obtain a tensor of order $s$, one finds a general relation,
	\begin{equation}
	\rho_k (D-2n,s-1) \eta^k \mathcal{J}^{[k]} = (-1)^k   \sum_{p=k}^{n+1}  \frac{(-1)^{p}}{2^{p} (n+1)^{\underline{p}}} {p \choose k} \eta^p \mathcal{F}_{n+1}^{[p]} \, . \label{JFrelation}
	\end{equation}
	Summing both sides of \eqref{JFrelation} over $k$ ($0 \leq k \leq n+1$), one finds that the factor $(-1)^k {p \choose k}$ cancels all the terms over $p$ on the right-hand side, except $\mathcal{F}_{n+1}$, i.e.
	\begin{equation}
	\mathcal{F}_{n+1} = \sum_{k=0}^{n+1} \rho_k (D-2n,s-1)  \eta^k \mathcal{J}^{[k]} \, .
	\end{equation}
	Therefore, the current exchange is described by
	\begin{align}
	\mathcal{J}_{(\mu)} \mathcal{P}^{(\mu)(\nu)} \mathcal{J}_{(\nu)} &=  \sum_{k=0}^{n+1} \rho_k (D-2n,s-1)  \mathcal{J} \fdot \eta^k \mathcal{J}^{[k]} \\
	&=  \sum_{k=0}^{n+1} \rho_k (D-2n,s-1) {{s-2k} \choose 2} \mathcal{J}^{[k]} \fdot \mathcal{J}^{[k]} \label{thexchangesum}
	\end{align}
	Expanding the first two terms of the current exchange, we see that
	\begin{align}
    \mathcal{J}_{(\mu)} \mathcal{P}^{(\mu)(\nu)} \mathcal{J}_{(\nu)} &= \mathcal{J} \fdot \mathcal{J} + \frac{1}{2} \rho_1 (D-2n,s-1)  \mathcal{J}' \fdot \mathcal{J}'\\ \nonumber &\quad + \sum_{k=2}^{n+1} \rho_k (D-2n,s-1)  {{s-2k} \choose 2} \mathcal{J}^{[k]} \fdot \mathcal{J}^{[k]} \, ,
	\end{align}
	which, due to the presence of an additional $-2n$ in the denominator, clearly \textit{disagrees} with the constrained case \eqref{constrained exchange} and the unconstrained case \eqref{unconstrained exchange}, except in the case of lower spins, i.e. $s\leq2$. We explore the implications of this disagreement in the following segment.

\subsubsection{HS theory with proper current exchange}
	
	So far, we have seen three different ways to formulate a higher spin theory of massless bosons in flat spacetime. One is \textit{Fronsdal's constrained formulation}, explored in \textbf{Section \ref{freefronsdal}}, in which the gauge-invariance of the action is enforced by restricting the gauge parameter $\Lambda$ to a traceless tensor and restricting the gauge field $\varphi$ to a doubly-traceless tensor. The second one is the \textit{local unconstrained formulation}, explored in \textbf{Section \ref{localunconstrained}}, which requires additional non-dynamical fields $\alpha$ and $\beta$ to ensure a fully gauge-invariant action. The third one is the \textit{non-local unconstrained formulation}, explored in this section, which allows for a fully gauge-invariant theory without any additional fields, at the cost of having to use non-local operators $\frac{1}{\Box}$. As we concluded in the previous segment, current exchanges in the non-local unconstrained theory seem to disagree with other two formulations. Since the non-local formulation is based on simple geometric arguments, without imposing \textit{ad-hoc} constraints or adding additional fields to the theory, it is natural to take it as a starting point, try to understand the disagreement and try to formulate it in a way that naturally reduces to other two formulations. As we shall see, this leads us to a \textit{unique} form of the theory for each spin.
	
	Equation \eqref{unconstrained exchange} suggests that, in the constrained formulation, the operator $\mathcal{P}_c$, as defined in \eqref{currx}, is
	\begin{equation}
	\mathcal{P}_c \fdot J= \sum_{k=0}^{n+1} \rho_k (D-2,s)  \eta^k J^{[k]} \, .
	\end{equation}
	A direct computation shows that
	\begin{equation}
	(\mathcal{P}_c \fdot J)' = 2 \sum_{k=0}^{n+1} \rho_{k+1} (D-2,s)  \eta^k J^{[k]} \, ,
	\end{equation}
	and
	\begin{equation}
	(\mathcal{P}_c \fdot J)'' = 0 \, . \label{p''}
	\end{equation}
	Note that $\mathcal{P}_c$ precisely corresponds to \eqref{currx} if the current is conserved, since $\Pi$ effectively gets replaced by $\eta$.
	Thus, if we want to build an unconstrained theory with proper current exchanges, in analogy with \eqref{abj}, we postulate the non-local Einstein tensor $\mathcal{E}$ of form\footnotemark\footnotetext{We put $\varphi$ in the subscript to stress the fact that these quantities are to be built using only the gauge field $\varphi$.}
	\begin{equation}
	\mathcal{E} = \mathcal{A}_\varphi - \frac{1}{2} \eta \mathcal{A}_\varphi ' + \eta^2 \mathcal{B}_\varphi \, ,
	\end{equation}
	requiring that $\mathcal{A}_\varphi'' = 0$, reflecting \eqref{p''}, and $\partial \cdot \mathcal{E} = 0$, reflecting the fact that $\mathcal{P}_c$ corresponds to the generalized projection operator for a \textit{conserved} external current.
	We construct $\mathcal{A}_\varphi$ using gauge-invariant building blocks, which can all be expressed in terms of $\mathcal{F}_{n+1}$\footnotemark\footnotetext{Alternatively, we could have used $\mathcal{G}_n$ as the main building block.}, where $n=\left[ \frac{s-1}{2} \right]$. Since
	\begin{equation}
	\partial \cdot \mathcal{F}^{[k]}_{n+1} = \frac{1}{2(n-k+1)} \partial \mathcal{F}^{[k+1]}_{n+1} \, ,
	\end{equation}
	all divergences can be expressed in terms of traces, which means that our gauge-invariant building blocks are
	\begin{equation}
	\mathcal{F}_{n+1}, \, \mathcal{F}_{n+1}' , \, \dots \, , \mathcal{F}^{[N]}_{n+1}
	\end{equation}
	where $N=\left[ \frac{s}{2} \right]$. The general linear combination thus reads
	\begin{equation}
	\mathcal{A}_\varphi = \sum_{k=0}^{N} a_k \frac{\partial^{2k}}{\Box^k} \mathcal{F}^{[k]}_{n+1} \, .
	\end{equation}
	Demanding that $\mathcal{A}_\varphi$ satisfies the Bianchi identity (because we want to write the Lagrangian as $\mathcal{L}_\varphi = \frac{1}{2} \varphi \fdot \mathcal{E}$),
	\begin{equation}
	\partial \cdot \mathcal{A}_\varphi - \frac{1}{2} \partial \mathcal{A}_\varphi ' = 0 \,  \label{bianchnl}
	\end{equation}
	implies
	\begin{equation}
	a_k = (-1)^{k+1} (2k-1) \prod_{j=0}^{k-1} \frac{n+j}{n-j+1} \, .
	\end{equation}
	We can write $\mathcal{B}_\varphi$ as
	\begin{equation}
	\mathcal{B}_\varphi = \sum_{k=0}^{N-2} \eta^k \mathcal{B}_k \, ,
	\end{equation}
	where $\mathcal{B}_k$ terms contain no metric tensors.
	We solve for $\mathcal{B}_k$ by demanding that $\mathcal{E}$ be divergenceless, i.e.
	\begin{align}
	\partial \cdot & \left\{  \mathcal{A}_\varphi - \frac{1}{2} \eta \mathcal{A}_\varphi ' + \eta^2 \mathcal{B}_\varphi \right\} = 0 \,  \\
	\implies  & \partial \cdot \mathcal{A}_\varphi ' = 2 \partial \mathcal{B}_\varphi + \eta \partial \cdot \mathcal{B}_\varphi \, .
	\end{align}
	This in turn implies that $\mathcal{B}_0$ is pure gradient
	\begin{equation}
	\partial \cdot \mathcal{A}_\varphi' = 2 \partial \mathcal{B}_0 \, , \label{AB}
	\end{equation}
	and $\mathcal{B}_k$ tensors can therefore be expressed as traces of $\mathcal{B}_0$, i.e.
	\begin{equation}
	\mathcal{B}_\varphi = \sum_{k=0}^{N-2} \frac{1}{2^{k-1} (k+2)!} \eta^k \mathcal{B}^{[k]}_0 \, .
	\end{equation}
	Solving \eqref{AB} for $\mathcal{B}_0$ gives
	\begin{equation}
	\mathcal{B}_0 = \sum_{k=0}^{N} b_k \frac{\partial^{2k}}{\Box^k} \mathcal{F}^{[k+2]}_{n+1} \, ,
	\end{equation}
	where\footnotemark \footnotetext{This calculation was worked out with an error in \cite{fms} and later corrected in \cite{dariomass}, which unfortunately also seems to contain an error.}
	\begin{equation}
	b_k = \frac{a_k}{4(n-k)(n-k+1)} \frac{1-4n^2}{1-4k^2} \, .
	\end{equation}
	Note that the denominator of $b_k$ can be equal to zero, but that is not a problem, since $\mathcal{F}^{[k+2]}_{n+1}$ vanishes for those values.

\subsection{Geometric theory}\label{geometric}

	It is instructive to formulate the theory purely in terms of generalized geometric objects. We start by defining \textit{higher-spin curvatures}, which leads us to the construction of generalized Riemann and Einstein tensors.
	
\subsubsection{Higher-spin curvature}
	As explained in \cite{dwf}, HS curvatures are essentially a hierarchy of generalized (linearized!) \textit{Christoffel connections} $\Gamma$ built from derivatives of the gauge field $\varphi$.	
	The $m$-th connection reads
	\begin{equation}
		\Gamma^{(m)}_{\mu_1 \cdots \mu_m ; \nu_1 \cdots \nu_s} \equiv  \Gamma^{(m)} = \sum_{k=0}^{m} \frac{(-1)^k}{{m \choose k}} \partial_{(\nu)}^{m-k} \partial_{(\mu)}^k \varphi \, , \label{gamma}
	\end{equation}
	where we write $(\mu)$ and $(\nu)$ in the subscript of $\partial$ to denote that $m-k$ derivatives carry one set of symmetric indices $(\nu_1 \cdots \nu_s)$, whereas $k$ derivatives carry the other set of symmetric indices $(\mu_1 \cdots \mu_m)$.
	For example, if $s=2$ the first connection is
	\begin{align}
	\Gamma^{(1)} &=  \partial_{(\nu)} \varphi - \partial_{(\mu)} \varphi  \\
	& \equiv  \partial_{\nu_1} \varphi_{\nu_2 \mu} + \partial_{\nu_2} \varphi_{\nu_1 \mu} - \partial_\mu \varphi_{\nu_1 \nu_2} 
	\end{align}
	which is exactly the linearized first connection as we know it from linearized general relativity, up to a multiplicative constant.
	The gauge variation of $\Gamma^{(m)}$ is
	\begin{align}
		\delta \Gamma^{(m)} &=  \sum_{k=0}^{m} \frac{(-1)^k}{{m \choose k}} \partial_{(\nu)}^{m-k} \partial_{(\mu)}^k (\partial_{(\nu)} \Lambda + \partial_{(\mu)} \Lambda) \\
		&=  \sum_{k=0}^{m} \frac{(-1)^k}{{m \choose k}} \left[ (m-k+1) \partial_{(\nu)}^{m-k+1} \partial_{(\mu)}^k \Lambda + (k+1) \partial_{(\nu)}^{m-k} \partial_{(\mu)}^{k+1} \Lambda \right] \\
		& = (m+1) \partial_{(\nu)}^{m+1} \Lambda \,.
	\end{align}
	Since $\partial$ should carry $m+1$ $\nu$-indices, $\Lambda$ should carry $s-1$ $\mu$-indices and there are $m+s$ indices in total, the gauge variation vanishes for $m \geq s$. For this reason, we define the generalized Riemann tensor as
	\begin{align}
	\mathcal{R} := \Gamma^{(s)} \, .
	\end{align}

\subsubsection{Relationship between $\mathcal{R}$ and $\mathcal{F}$}
	
	Following \cite{fms}, we can relate the Fronsdal tensor $\mathcal{F}$ to the generalized Riemann tensor $\mathcal{R}$ using the generalized Fronsdal tensor $\mathcal{F}_n$,
	\begin{equation}
	\mathcal{F}_{n+1} = \frac{1}{\Box^n} \partial^{s - 2N} \cdot \mathcal{R}^{[N]} \, , \label{geometry}
	\end{equation}
	where, as before, $n=\left[ \frac{s-1}{2} \right]$, $N=\left[ \frac{s}{2} \right]$ and the contraction is performed on $\mu$-indices, i.e. the first set of indices in the $s$-th connection.	
	The correspondence \eqref{geometry} thus allows us to reformulate the theory using only (linearized!) geometric objects, which was one of the motivating factors that led to the construction of $\mathcal{F}_{n+1}$, as mentioned in \textbf{Section \ref{nonlocal}}
	
\newpage
\section{Discussion}\label{s7}
	Having gone through the analysis of massless bosonic massless fields, we are now in a position to concisely define the proper theory and show how it reduces to some interesting special cases.
	
	The full set of equations reads
	\begin{empheq}[box=\widefbox]{align}
	\mathcal{L} &= \frac{1}{2} \varphi \left( \mathcal{E} - \mathcal{J}  \right) \\
	\mathcal{E} &= \mathcal{A}_\varphi - \frac{1}{2} \eta \mathcal{A}_\varphi ' + \eta^2 \mathcal{B}_\varphi\\
	\mathcal{A}_\varphi &= \sum_{k=0}^{N} a_k \frac{\partial^{2k}}{\Box^k} \mathcal{F}^{[k]}_{n+1} \\
	\mathcal{B}_\varphi &= \sum_{k=0}^{N-2} \frac{1}{2^{k-1} (k+2)!} \eta^k \mathcal{B}_0^{[k]} \label{B} \\
	\mathcal{B}_0 &= \sum_{k=0}^{N} b_k \frac{\partial^{2k}}{\Box^k} \mathcal{F}^{[k+2]}_{n+1} \label{D} \\
	a_k &= (-1)^{k+1} (2k-1) \prod_{j=0}^{k-1} \frac{n+j}{n-j+1} \label{ak} \\
	b_k &= \frac{a_k}{4(n-k)(n-k+1)} \frac{1-4n^2}{1-4k^2} \label{bk} \\
	\mathcal{F}_{n+1} &= \mathcal{F}_n - \frac{1}{n+1} \frac{\partial}{\Box} \partial \cdot \mathcal{F}_n + \frac{1}{(n+1)(2n+1)} \frac{\partial^2}{\Box} \mathcal{F}_n' \\
	\mathcal{F}_0 &= \Box \varphi  \\
	n &= \left[ \frac{s-1}{2} \right],  \quad N = \left[ \frac{s}{2}  \right]
	\end{empheq}

	If the theory is free, $\mathcal{E} = 0$  reduces to $\mathcal{A}_\varphi = 0$ which in turn reduces to $\mathcal{F}_{n+1} = 0$.	
	If we want to cast the theory in a local unconstrained form, we express $\mathcal{A}$ in terms of $\varphi$ and $\alpha$ as described in \eqref{Acomp}, and we express $\mathcal{B}$ in terms of $\varphi$ and $\beta$ as described in \eqref{Bcomp}.	
	Finally, to go full circle and arrive at Fronsdal's constrained theory, we simply dispense with compensator fields $\alpha$ and $\beta$ and we impose $\Lambda'=0$ and $\varphi''=0$.

\newpage
\subsection{A single equation?}\label{disc}

	The Einstein-like tensor $\mathcal{E}$ reveals an interesting peculiarity when expressed explicitly in terms of $\varphi$. As you can see in \textbf{Appendix \ref{E}}, its general form for the spin-$s$ tensor $\mathcal{E}_s$ seems to be
	\begin{equation}
	\mathcal{E}_{s} [\varphi_s]= \mathcal{E}_{s-1}[\varphi_s] + \Delta_{s}[\varphi_s] \, ,
	\end{equation}
	where $\Delta_s$ contains only those terms that become non-vanishing for spin $s$, i.e.
	\begin{equation}
	\Delta_s [\varphi_{s'}] = 0 \quad \text{for} \quad s' < s \, .
	\end{equation}
	In other words, for spin $s$, all tensors $\mathcal{E}_k$ where $k\geq s$ are equally valid, since they trivially reduce to $\mathcal{E}_s$.	
	What this seems to imply is not only that there is a unique Einstein-like tensor that leads to a valid theory \textbf{for each spin}, but that there is a \textbf{single tensor}	$\mathcal{E}_\infty$ that leads to a valid theory \textbf{for all spins}.	
	Note that we have only evaluated $\mathcal{E}_s$ up to $s=15$ using computer-assisted methods described in \textbf{Appendix \ref{program}}, but it certainly seems natural that this pattern holds for general spin $s$.	
	This conjecture remains to be proved, and the possibility of an explicit construction of $\mathcal{E}_\infty$ also remains an open question.
	
\section{Conclusion}\label{s8}
	We have shown the proper form of equations for a theory of massless higher-spin bosons interacting with a generic external current. As it turns out, to construct a consistent unconstrained local theory, we either have to introduce non-local operators or high derivatives.
	This construction leads to a \textit{unique} theory for each spin, perhaps even a unique theory for \textit{all spins}, as discussed in \textbf{Section \ref{disc}}.
	
	Putting the $\mathrm{(A)dS}$ and the fermionic theory aside, an interesting step forward would perhaps be to find the proper HS gauge corresponding to the \textit{de Donder} gauge for the spin-$2$ theory. Some interesting results regarding generalized \textit{de Donder} gauges can be found in \cite{fs2}, but they certainly deserve further investigation.	
	Note that in spacetimes with more than four dimensions, fully symmetric tensors do not exhaust all the available possibilities and one should also consider mixed-symmetry tensors. This was considered, for example, in \cite{mix1}, \cite{mix2} or \cite{mix3}.

\appendix
\newpage
\section{Some useful results}\label{a1}

	For reader's convenience, we list some useful results that are commonly used throughout HS calculations.
	
\subsection{Fronsdal's tensor}

	All results listed here are valid for every integer spin $s$ and for every dimensionality $D$. All technically nonsensical terms are to be interpreted as non-existent, e.g. Fronsdal's equation for $s=1$ becomes simply $\Box \varphi$.
	
	All terms that vanish in Fronsdal's constrained theory (i.e. with $\Lambda'=0$ and $\varphi''=0$)  are \textbf{\underline{included}}. Furthermore, all terms with the same field are ordered by the lowest spin for which they do not vanish trivially.	

\subsubsection{$\mathcal{F}$-tensor}

\paragraph{Definition}

	\begin{equation}
	\boxed{\mathcal{F} = \Box \varphi - \partial (\partial \cdot \varphi) + \partial^2 \varphi' }
	\end{equation}
	
	\textbf{Important properties:}
	\begin{itemize}
		\item tensor of order $s$
		\item divergenceless for $s<2$
		\item $\Lambda$-gauge-invariant for $s<3$
	\end{itemize}
	
\paragraph{Common expressions}
	
	\begin{align}
	\partial^k \mathcal{F} &= \Box \partial^k \phi - (k+1) \partial^{k+1} \left( \partial \cdot \phi \right) + {{k+2} \choose 2} \partial^{k+2} \phi' \\
	\partial^k \mathcal{F}' &= 2 \Box \partial^k \varphi' - 2 \partial^k (\partial^2 \cdot \varphi) + (k+1) \partial^{k+1} (\partial \cdot \varphi') + {{k+2} \choose 2} \partial^{k+2} \varphi'' \\
	\partial^k \mathcal{F}'' &= 3 \Box \partial^k \varphi'' + 2(k+1) \partial^{k+1} (\partial \cdot \varphi'') + {{k+2} \choose 2} \partial^{k+2} \varphi''' \\
	\partial^k (\partial \cdot \mathcal{F}') &= 3 \Box \partial^k (\partial \cdot \varphi') - 2 \partial^k (\partial^3 \cdot \varphi) + (k+1) \partial^{k+1} (\partial^2 \cdot \varphi') \\ \nonumber &+ (k+1) \Box \partial^{k+1} \varphi'' + {{k+2} \choose 2} \partial^{k+2} (\partial \cdot \varphi'') \\
	\partial \cdot \mathcal{F} &= \Box \partial \varphi' - \partial (\partial^2 \cdot \varphi) + \partial^2 (\partial \cdot \varphi') \\
	\mathcal{F}' &= 2 \Box \varphi' - 2 \partial^2 \cdot \varphi + \partial (\partial \cdot \varphi') + \partial^2 \varphi'' \\
	\partial \cdot \mathcal{F}' &= 3 \Box \partial \cdot \varphi' - 2 \partial^3 \cdot \varphi + \partial (\partial^2 \cdot \varphi') + \Box \partial \varphi'' + \partial^2 (\partial \cdot \varphi'') \\
	\mathcal{F}'' &= 3 \Box \varphi'' + 3 \partial(\partial \cdot \varphi'') + \partial^2 \varphi''' \\
	\partial \cdot \mathcal{F}'' &= 6 \Box \partial \cdot \varphi'' + 3 \partial (\partial^2 \cdot \varphi'') + \Box \partial \varphi''' + \partial^2 (\partial \cdot \varphi''') \\
	\mathcal{F}''' &= 4 \Box \varphi''' + 6 \partial^2 \cdot \varphi'' + 5 \partial (\partial \cdot \varphi''') + \partial^2 \varphi^{[4]}
	\end{align}

\paragraph{Gauge variations}

	\begin{align}
	\delta_{\Lambda} \mathcal{F} &= 3 \partial^3 \Lambda' \\
	\delta_{\Lambda} (\partial \mathcal{F}) &= 12 \partial^{4} \Lambda' \\
	\delta_{\Lambda} (\partial \cdot \mathcal{F}) &= 3 \Box \partial^2 \Lambda' + 3 \partial^3 (\partial \cdot \Lambda') \\
	\delta_{\Lambda} (\partial \mathcal{F}') &= 6 \Box \partial^2 \Lambda' + 18 \partial^3 (\partial \cdot \Lambda') + 12 \partial^4 \Lambda'' \\
	\delta_{\Lambda} \mathcal{F}' &= 3 \Box \partial \Lambda' + 6 \partial^2 (\partial \cdot \Lambda') + 3 \partial^3 \Lambda'' \\	
	\delta_{\Lambda} (\partial \cdot \mathcal{F}') &= 3 \Box^2 \Lambda' + 9 \Box \partial (\partial \cdot \Lambda') + 6 \partial^2 (\partial^2 \cdot \Lambda') + 3 \Box \partial^2 \Lambda'' + 3 \partial^3 (\partial \cdot \Lambda'') \\
	\delta_{\Lambda} (\partial \mathcal{F}'') &= 12 \Box \partial (\partial \cdot \Lambda') + 12 \Box \partial^2 \Lambda'' + 24 \partial^2 (\partial^2 \cdot \Lambda') + 36 \partial^3 (\partial \cdot \Lambda'') + 12 \partial^4 \Lambda''' \\
	\delta_{\Lambda} \mathcal{F}'' &= 12 \Box \partial \cdot \Lambda' + 6 \Box \partial \Lambda'' + 12 \partial (\partial^2 \cdot \Lambda') + 12 \partial^2 (\partial \cdot \Lambda'') + 3 \partial^3 \Lambda''' \\
	\delta_{\Lambda} (\partial \cdot \mathcal{F}'') &= 24 \Box \partial^2 \cdot \Lambda' + 6 \Box^2 \Lambda'' + 18 \Box \partial (\partial \cdot \Lambda'') + 12 \partial (\partial^3 \cdot \Lambda') + 12 \partial^2 (\partial^2 \cdot \Lambda'') \\ \nonumber &+ 3 \Box \partial^2 \Lambda''' + 3 \partial^3 (\partial \cdot \Lambda''') \\
	\delta_{\Lambda} \mathcal{F}''' &= 36 \Box \partial \cdot \Lambda'' + 24 \partial^3 \cdot \Lambda' + 9 \Box \partial \Lambda''' +  36 \partial (\partial^2 \cdot \Lambda'') + 18 \partial^2 (\partial \cdot \Lambda''') \\ \nonumber & + 3 \partial^3 \Lambda^{[4]}
	\end{align}
\newpage
\subsubsection{$\mathcal{G}$-tensor}

\paragraph{Definition}

	\begin{equation}
	\boxed{\mathcal{G} = \mathcal{F} - \frac{1}{2} \eta \mathcal{F}'}
	\end{equation}
	
	\textbf{Important properties:}
	\begin{itemize}
		\item tensor of order $s$
		\item divergenceless for $s<3$
		\item $\Lambda$-gauge-invariant for $s<3$
	\end{itemize}
	
\paragraph{Common expressions in terms of $\mathcal{F}$}

	\begin{align}
	\partial \mathcal{G} &= \partial \mathcal{F} - \frac{1}{2} \eta \partial \mathcal{F}' \\
	\partial \cdot \mathcal{G} &= \partial \cdot \mathcal{F} - \frac{1}{2} \partial \mathcal{F}' - \frac{1}{2} \eta \partial \cdot \mathcal{F}' \\
	\partial \mathcal{G}' &= -\frac{1}{2} [ D + 2(s-3)] \partial \mathcal{F}' - \frac{1}{2} \eta \partial \mathcal{F}'' \\
	\mathcal{G}' &= -\frac{1}{2} [D + 2(s-3)] \mathcal{F}' - \frac{1}{2} \eta \mathcal{F}'' \\
	\partial \cdot \mathcal{G}' &= -\frac{1}{2} [D+2(s-3)] \partial \cdot \mathcal{F}' - \frac{1}{2} \partial \mathcal{F}'' - \frac{1}{2} \eta \partial \cdot \mathcal{F}'' \\
	\mathcal{G}'' &= -(D+2s-7) \mathcal{F}'' - \frac{1}{2} \eta \mathcal{F}'''
	\end{align}

\paragraph{Common expressions in terms of $\varphi$}
	
	\begin{align}
	\mathcal{G} &= \Box \varphi - \partial (\partial \cdot \varphi) + \partial^2 \varphi' - \eta \Box \varphi' + \eta \partial^2 \cdot \varphi - \frac{1}{2} \eta \partial (\partial \cdot \varphi') - \frac{1}{2} \eta \partial^2 \varphi'' \\
	\partial \mathcal{G} &= \Box \partial \phi - 2 \partial^{2} \left( \partial \cdot \phi \right) + 3 \partial^{3} \phi' - \eta \Box \partial \phi' + \eta \partial \left( \partial^{2} \cdot \phi \right) - \eta \partial^{2} \left( \partial \cdot \phi' \right) - \frac{3}{2} \eta \partial^{3} \phi'' \\
	\partial \cdot \mathcal{G} &=  - \frac{3}{2} \eta \Box \partial \cdot \phi' + \eta \partial^{3} \cdot \phi - \frac{1}{2} \eta \Box \partial \phi''- \frac{1}{2} \eta \partial \left( \partial^{2} \cdot \phi' \right)  \\ \nonumber &- \frac{3}{2} \partial^{3} \phi'' -\frac{1}{2} \eta \partial^{2} \left( \partial \cdot \phi'' \right) \\
	\partial \mathcal{G}' &= -[D+2(s-3)]\left(\Box \partial \varphi' - \partial(\partial^2 \cdot \varphi) + \partial^2(\partial \cdot \varphi') + \frac{3}{2} \partial^3 \varphi'' \right) \\ \nonumber &-\frac{3}{2} \eta \Box \partial \varphi'' - 2 \eta \partial^2 (\partial \cdot \varphi'') - \frac{3}{2} \eta \partial^3 \varphi''' \\
	\mathcal{G}' &= -[D+2(s-3)] \left( \Box \varphi' - \partial^2 \cdot \varphi + \frac{1}{2} \partial (\partial \cdot \varphi') + \frac{1}{2} \partial^2 \varphi'' \right) \\ \nonumber &-\frac{3}{2} \eta \Box \varphi'' - \eta \partial (\partial \cdot \varphi'') - \frac{1}{2} \eta \partial^2 \varphi'''\\
	\partial \cdot \mathcal{G}' &= -[D+2(s-3)] \left( \frac{3}{2} \Box \partial \cdot \varphi' - \partial^3 \cdot \varphi + \frac{1}{2} \partial (\partial^2 \cdot \varphi') \right) \\ \nonumber &-\frac{1}{2} (D+2s-3) \Box \partial \varphi'' - \frac{1}{2} [D + 2(s-1)] \partial^2 (\partial \cdot \varphi'')\\
	\nonumber &- \frac{5}{2} \eta \Box \partial \cdot \varphi'' -\frac{3}{2} \partial^3 \varphi'''  - \eta \partial (\partial^2 \cdot \varphi'') -\frac{1}{2} \eta \Box \partial \varphi''' - \frac{1}{2} \eta \partial^2 (\partial \cdot \varphi''') \\
	\mathcal{G}'' &= -(D+2s-7) \left( 3 \Box \varphi'' + 2 \partial (\partial \cdot \varphi'') + \partial^2 \varphi''' \right) -2\eta \Box \varphi''' - 2 \eta \partial^2 \cdot \varphi'' \\ \nonumber &- 2 \eta \partial (\partial \cdot \varphi''') - \frac{1}{2} \eta \partial^2 \varphi^{[4]}
	\end{align}

\paragraph{Gauge variations}

	\begin{align}
	\delta_{\Lambda} \mathcal{G} &= 3 \partial^3 \Lambda' - \frac{3}{2} \eta \Box \partial \Lambda' - 3 \eta \partial^2 (\partial \cdot \Lambda') - \frac{3}{2} \partial^3 \Lambda''\\
	\delta_{\Lambda} (\partial \mathcal{G}) &= 12 \partial^{4} \Lambda' - 3 \eta \Box \partial^{2} \Lambda' - 9 \eta \partial^{3} \left( \partial \cdot \Lambda' \right) - 6 \eta \partial^{4} \Lambda''\\
	\delta_{\Lambda} (\partial \cdot \mathcal{G}) &=  - \frac{3}{2} \eta \Box^{2} \Lambda' - \frac{9}{2} \eta \Box \partial \left( \partial \cdot \Lambda' \right) - 6 \partial^{3} \left( \partial \cdot \Lambda' \right) \\ \nonumber &- 3 \eta \partial^{2} \left( \partial^{2} \cdot \Lambda' \right)- 6 \partial^{4} \Lambda'' - \frac{3}{2} \eta \Box \partial^{2} \Lambda'' - \frac{3}{2} \eta \partial^{3} \left( \partial \cdot \Lambda'' \right)  
	\end{align}
\newpage
\subsubsection{$\mathcal{A}$-tensor}

\paragraph{Definition}

	\begin{equation}
	\boxed{\mathcal{A} = \mathcal{F} - 3\partial^3 \alpha}
	\end{equation}
	
	\textbf{Important properties:}
	\begin{itemize}
		\item tensor of order $s$
		\item divergenceless for $s<2$
		\item $\Lambda$-gauge-invariant
	\end{itemize}

\paragraph{Common expressions in terms of $\mathcal{F}$ and $\alpha$}
	
	\begin{align}
	\partial \cdot \mathcal{A} &= \partial \cdot \mathcal{F} - 3 \Box \partial^2 \alpha - 3 \partial^3 (\partial \cdot \alpha) \\
	\mathcal{A}' &= \mathcal{F}' - 3 \Box \partial \alpha - 3 \partial^2 (\partial \cdot \alpha) - 3 \partial^3 \alpha' \\
	\partial \mathcal{A}' &= \partial \mathcal{F}' - 6 \Box \partial^2 \alpha - 9 \partial^3 (\partial \cdot \alpha) - 12 \partial^4 \alpha' \\
	\partial \cdot \mathcal{A}' &= \partial \cdot \mathcal{F}' - 3 \Box^2 \alpha - 6 \Box \partial (\partial \cdot \alpha) - 3 \partial^2 (\partial^2 \cdot \alpha) - 3 \Box \partial^2 \alpha' - 3 \partial^3 (\partial \cdot \alpha')
	\end{align}
	
\paragraph{Common expressions in terms of $\varphi$ and $\alpha$}

	\begin{align}
	\mathcal{A} &=  \Box \varphi - \partial (\partial \cdot \varphi) + \partial^2 \varphi' - 3\partial^3 \alpha \\
	\partial \cdot \mathcal{A} &=  - \partial \left( \partial^{2} \cdot \varphi \right) + \Box \partial \varphi' + \partial^{2} \left( \partial \cdot \varphi' \right)  - 3 \Box \partial^{2} \alpha - 3 \partial^{3} \left( \partial \cdot \alpha \right)  \\
	\mathcal{A}' &= 2 \Box \varphi' - 2 \partial^{2} \cdot \varphi + \partial \left( \partial \cdot \varphi' \right) + \partial^{2} \varphi'' - 3 \Box \partial \alpha - 6 \partial^{2} \left( \partial \cdot \alpha \right) - 3 \partial^{3} \alpha' \\
	\partial \mathcal{A}' &= 2 \Box \partial \varphi' - 2 \partial \left( \partial^{2} \cdot \varphi \right) + 2 \partial^{2} \left( \partial \cdot \varphi' \right) \\ \nonumber & + 3 \partial^{3} \varphi'' - 6 \Box \partial^{2} \alpha - 18 \partial^{3} \left( \partial \cdot \alpha \right) - 12 \partial^{4} \alpha' \\
	\partial \cdot \mathcal{A}' &= 3 \Box \partial \cdot \varphi' - 2 \partial^{3} \cdot \varphi + \partial \left( \partial^{2} \cdot \varphi' \right) + \Box \partial \varphi'' + \partial^{2} \left( \partial \cdot \varphi'' \right)  \\ \nonumber & - 3 \Box^{2} \alpha  - 9 \Box \partial \left( \partial \cdot \alpha \right) - 6 \partial^{2} \left( \partial^{2} \cdot \alpha \right) - 3 \Box \partial^{2} \alpha' - 3 \partial^{3} \left( \partial \cdot \alpha' \right) 
	\end{align} 
\newpage
\subsubsection{$\mathcal{C}$-tensor}

\paragraph{Definition}

	\begin{equation}
	\boxed{\mathcal{C} = \varphi'' - 4 \partial \cdot \alpha - \partial \alpha'}
	\end{equation}

	\textbf{Important properties:}
	\begin{itemize}
		\item tensor of order $s-4$
		\item $\Lambda$-gauge-invariant
	\end{itemize}

	\paragraph{Common expressions in terms of $\varphi$ and $\alpha$}
	
	\begin{align}
	\mathcal{C} &= \varphi'' - 4 \partial \cdot \alpha - \partial \alpha' \\
	\partial \cdot \mathcal{C} &= \partial \cdot \varphi'' - 4 \partial^2 \cdot \alpha - \Box \alpha' - \partial (\partial \cdot \alpha') \\
	\partial^k \mathcal{C} &= \partial^k \varphi'' - 4 \partial^k (\partial \cdot \alpha) - (k+1) \partial^{k+1} \alpha' \\
	\partial^k (\partial \cdot \mathcal{C}) &= \partial^k (\partial \cdot \varphi'') - 4 \partial^k (\partial^2 \cdot \alpha) - \Box \partial^k \alpha' - (k+1) \partial^{k+1} (\partial \cdot \alpha')
	\end{align} 
	
\subsubsection{$\mathcal{B}$-tensor}

\paragraph{Definition}

	\begin{equation}
	\boxed{\mathcal{B} = \beta  + \Box \partial \cdot \alpha + \frac{1}{2}  \partial (\partial^2 \cdot \alpha) - \frac{1}{2} \partial^2 \cdot \varphi'}
	\end{equation}

	\textbf{Important properties:}
	\begin{itemize}
		\item tensor of order $s-4$
		\item $\Lambda$-gauge-invariant
	\end{itemize}
	
	\paragraph{Common expressions in terms of $\varphi$, $\alpha$ and $\beta$}
	
	\begin{align}
	\partial \mathcal{B} &= \partial \beta + \Box \partial (\partial \cdot \alpha) + \partial^2 (\partial^2 \cdot \alpha) - \frac{1}{2} \partial (\partial^2 \cdot \varphi') \\
	\partial \cdot \mathcal{B} &= \partial \cdot \beta + \frac{3}{2} \Box \partial^2 \cdot \alpha + \frac{1}{2} \partial (\partial^3 \cdot \alpha) - \frac{1}{2} \partial^3 \cdot \varphi'
	\end{align} 
\newpage
\subsection{Generalized Fronsdal tensor $\mathcal{F}_{n}$}

\paragraph{Definition}

	\begin{equation}
	\boxed{\mathcal{F}_{n} = \mathcal{F}_{n-1} - \frac{1}{n} \frac{\partial}{\Box} \left( \partial \cdot \mathcal{F}_{n-1} \right) +  \frac{1}{n(2n-1)} \frac{\partial^2}{\Box} \mathcal{F}_{n-1}{}' \, ,  \quad \mathcal{F}_{0}=\Box \varphi}\label{Fn}
	\end{equation}
	
	\textbf{Important properties:}
	\begin{itemize}
		\item for spin $s$, the minimal value of $n$ for which $\mathcal{F}_{n}$ becomes \\ $\Lambda$-gauge-invariant is $n=\left[ \frac{s-1}{2} \right]+1$
	\end{itemize}

	We list the generalized Fronsdal tensors for spins up to $s=8$ (i.e. up to $n=5$). As an exception, we do not include tensors of all orders, but only those that do not vanish for the specific spin(s) for which  $\mathcal{F}_{n}$ is minimally $\Lambda$-gauge invariant.

\paragraph{s=1,2}
	\begin{align}
	\mathcal{F}_{1} &= \Box \varphi - \partial \left( \partial \cdot \varphi \right) + \partial^{2} \varphi'
	\end{align}
\paragraph{s=3,4}
	\begin{align}
	\mathcal{F}_{2} &= \Box \varphi - \partial \left( \partial \cdot \varphi \right) + \frac{2}{3 \Box} \partial^{2} \left( \partial^{2} \cdot \varphi \right) + \frac{1}{3} \partial^{2} \varphi' - \frac{1}{\Box} \partial^{3} \left( \partial \cdot \varphi' \right) + \frac{1}{\Box} \partial^{4} \varphi''
	\end{align}
\paragraph{s=5,6}
	\begin{align}
	\mathcal{F}_{3} &= \Box \varphi - \partial \left( \partial \cdot \varphi \right) + \frac{4}{5 \Box} \partial^{2} \left( \partial^{2} \cdot \varphi \right) + \frac{1}{5} \partial^{2} \varphi' - \frac{2}{5 \Box^{2}} \partial^{3} \left( \partial^{3} \cdot \varphi \right) \\ \nonumber &  - \frac{3}{5 \Box} \partial^{3} \left( \partial \cdot \varphi' \right) + \frac{4}{5 \Box^{2}} \partial^{4} \left( \partial^{2} \cdot \varphi' \right) + \frac{1}{5 \Box} \partial^{4} \varphi'' - \frac{1}{\Box^{2}} \partial^{5} \left( \partial \cdot \varphi'' \right) + \frac{1}{\Box^{2}} \partial^{6} \varphi'''
	\end{align}
\paragraph{s=7,8}
	\begin{align}
	\mathcal{F}_{4} &= \Box \varphi - \partial \left( \partial \cdot \varphi \right) + \frac{6}{7 \Box} \partial^{2} \left( \partial^{2} \cdot \varphi \right) + \frac{1}{7} \partial^{2} \varphi' - \frac{4}{7 \Box^{2}} \partial^{3} \left( \partial^{3} \cdot \varphi \right) \\ \nonumber &  - \frac{3}{7 \Box} \partial^{3} \left( \partial \cdot \varphi' \right) + \frac{8}{35 \Box^{3}} \partial^{4} \left( \partial^{4} \cdot \varphi \right) + \frac{24}{35 \Box^{2}} \partial^{4} \left( \partial^{2} \cdot \varphi' \right) + \frac{3}{35 \Box} \partial^{4} \varphi'' \\ \nonumber &  - \frac{4}{7 \Box^{3}} \partial^{5} \left( \partial^{3} \cdot \varphi' \right) - \frac{3}{7 \Box^{2}} \partial^{5} \left( \partial \cdot \varphi'' \right) + \frac{6}{7 \Box^{3}} \partial^{6} \left( \partial^{2} \cdot \varphi'' \right) + \frac{1}{7 \Box^{2}} \partial^{6} \varphi''' \\ \nonumber &  - \frac{1}{\Box^{3}} \partial^{7} \left( \partial \cdot \varphi''' \right) + \frac{1}{\Box^{3}} \partial^{8} \varphi^{[4]}
	\end{align}

\subsection{Generalized Einstein tensor $\mathcal{G}_{n}$}

\paragraph{Definition}

	\begin{equation}
	\boxed{\mathcal{G}_{n} = \sum_{k=0}^n \frac{(-1)^k}{2^k n^{\underline{k}}} \eta^k \mathcal{F}_{n}^{[k]} \, ,  \quad \mathcal{G}_{0}=\Box \varphi} \label{Gn}
	\end{equation}

	\textbf{Important properties:}
	\begin{itemize}
		\item for spin $s$, the minimal value of $n$ for which $\mathcal{F}_{n}$ becomes $\Lambda$-gauge-invariant and divergenceless is $n=\left[ \frac{s-1}{2} \right]+1$
	\end{itemize}
	
	We list the generalized Einstein tensors for spins up to $s=8$ (i.e. up to $n=5$). As an exception, we do not include tensors of all orders, but only those that do not vanish for the specific spin(s) for which  $\mathcal{G}_{n}$ is minimally $\Lambda$-gauge invariant and divergenceless.
	
	\paragraph{s=1,2}
	\begin{align}
	\mathcal{G}_{1} &= \Box \varphi - \partial \left( \partial \cdot \varphi \right) + \partial^{2} \varphi' + \eta \partial^{2} \cdot \varphi - \eta \Box \varphi'
	\end{align}
	\paragraph{s=3,4}
	\begin{align}
	\mathcal{G}_{2} &= \Box \varphi - \partial \left( \partial \cdot \varphi \right) + \frac{2}{3\Box}\partial^{2} \left( \partial^{2} \cdot \varphi \right) + \frac{1}{3}\partial^{2} \varphi' + \frac{1}{3}\eta \partial^{2} \cdot \varphi \\ \nonumber &  - \frac{1}{3}\eta \Box \varphi' - \frac{1}{\Box}\partial^{3} \left( \partial \cdot \varphi' \right) - \frac{1}{3\Box}\eta \partial \left( \partial^{3} \cdot \varphi \right) + \frac{1}{3}\eta \partial \left( \partial \cdot \varphi' \right) \\ \nonumber &  + \frac{1}{\Box}\partial^{4} \varphi'' + \frac{1}{3\Box}\eta \partial^{2} \left( \partial^{2} \cdot \varphi' \right) - \frac{1}{3}\eta \partial^{2} \varphi'' + \frac{1}{3\Box}\eta^{2} \partial^{4} \cdot \varphi \\ \nonumber &  - \frac{2}{3}\eta^{2} \partial^{2} \cdot \varphi' + \frac{1}{3}\eta^{2} \Box \varphi''
	\end{align}
	\paragraph{s=5,6}
	\begin{align}
	\mathcal{G}_{3} &= \Box \varphi - \partial \left( \partial \cdot \varphi \right) + \frac{1}{5}\eta \partial^{2} \cdot \varphi - \frac{1}{5}\eta \Box \varphi' + \frac{1}{5}\partial^{2} \varphi' \\ \nonumber &  + \frac{4}{5\Box}\partial^{2} \left( \partial^{2} \cdot \varphi \right) - \frac{1}{5\Box}\eta \partial \left( \partial^{3} \cdot \varphi \right) + \frac{1}{5}\eta \partial \left( \partial \cdot \varphi' \right) - \frac{3}{5\Box}\partial^{3} \left( \partial \cdot \varphi' \right) \\ \nonumber &  - \frac{2}{5\Box^{2}}\partial^{3} \left( \partial^{3} \cdot \varphi \right) + \frac{2}{15\Box^{2}}\eta \partial^{2} \left( \partial^{4} \cdot \varphi \right) - \frac{1}{15\Box}\eta \partial^{2} \left( \partial^{2} \cdot \varphi' \right) + \frac{1}{5\Box}\partial^{4} \varphi'' \\ \nonumber &  + \frac{4}{5\Box^{2}}\partial^{4} \left( \partial^{2} \cdot \varphi' \right) + \frac{1}{15\Box}\eta^{2} \partial^{4} \cdot \varphi - \frac{1}{15}\eta \partial^{2} \varphi'' - \frac{2}{15}\eta^{2} \partial^{2} \cdot \varphi' \\ \nonumber &  + \frac{1}{15}\eta^{2} \Box \varphi'' - \frac{1}{5\Box^{2}}\eta \partial^{3} \left( \partial^{3} \cdot \varphi' \right) + \frac{1}{5\Box}\eta \partial^{3} \left( \partial \cdot \varphi'' \right) - \frac{1}{15\Box^{2}}\eta^{2} \partial \left( \partial^{5} \cdot \varphi \right) \\ \nonumber &  + \frac{2}{15\Box}\eta^{2} \partial \left( \partial^{3} \cdot \varphi' \right) - \frac{1}{\Box^{2}}\partial^{5} \left( \partial \cdot \varphi'' \right) - \frac{1}{15}\eta^{2} \partial \left( \partial \cdot \varphi'' \right) - \frac{2}{15\Box}\eta^{2} \partial^{2} \left( \partial^{2} \cdot \varphi'' \right) \\ \nonumber &  + \frac{1}{15\Box^{2}}\eta^{2} \partial^{2} \left( \partial^{4} \cdot \varphi' \right) + \frac{1}{15}\eta^{2} \partial^{2} \varphi''' + \frac{1}{5\Box^{2}}\eta \partial^{4} \left( \partial^{2} \cdot \varphi'' \right) - \frac{1}{5\Box}\eta \partial^{4} \varphi''' \\ \nonumber &  + \frac{1}{5}\eta^{3} \partial^{2} \cdot \varphi'' + \frac{1}{\Box^{2}}\partial^{6} \varphi''' - \frac{1}{15}\eta^{3} \Box \varphi''' + \frac{1}{15\Box^{2}}\eta^{3} \partial^{6} \cdot \varphi - \frac{1}{5\Box}\eta^{3} \partial^{4} \cdot \varphi'
	\end{align}
	\paragraph{s=7,8}
	\begin{align}
	\mathcal{G}_{4} &= \Box \varphi - \partial \left( \partial \cdot \varphi \right) + \frac{6}{7\Box}\partial^{2} \left( \partial^{2} \cdot \varphi \right) - \frac{1}{7}\eta \Box \varphi' + \frac{1}{7}\eta \partial^{2} \cdot \varphi \\ \nonumber &  + \frac{1}{7}\partial^{2} \varphi' - \frac{4}{7\Box^{2}}\partial^{3} \left( \partial^{3} \cdot \varphi \right) + \frac{1}{7}\eta \partial \left( \partial \cdot \varphi' \right) - \frac{3}{7\Box}\partial^{3} \left( \partial \cdot \varphi' \right) \\ \nonumber &  - \frac{1}{7\Box}\eta \partial \left( \partial^{3} \cdot \varphi \right) - \frac{2}{35}\eta^{2} \partial^{2} \cdot \varphi' + \frac{1}{35}\eta^{2} \Box \varphi'' + \frac{8}{35\Box^{3}}\partial^{4} \left( \partial^{4} \cdot \varphi \right) \\ \nonumber &  + \frac{1}{35\Box}\eta^{2} \partial^{4} \cdot \varphi + \frac{24}{35\Box^{2}}\partial^{4} \left( \partial^{2} \cdot \varphi' \right) + \frac{4}{35\Box^{2}}\eta \partial^{2} \left( \partial^{4} \cdot \varphi \right) + \frac{3}{35\Box}\partial^{4} \varphi'' \\ \nonumber &  - \frac{1}{35}\eta \partial^{2} \varphi'' - \frac{3}{35\Box}\eta \partial^{2} \left( \partial^{2} \cdot \varphi' \right) - \frac{4}{7\Box^{3}}\partial^{5} \left( \partial^{3} \cdot \varphi' \right) - \frac{3}{7\Box^{2}}\partial^{5} \left( \partial \cdot \varphi'' \right) \\ \nonumber &  + \frac{3}{35\Box}\eta \partial^{3} \left( \partial \cdot \varphi'' \right) - \frac{1}{35}\eta^{2} \partial \left( \partial \cdot \varphi'' \right) - \frac{2}{35\Box^{3}}\eta \partial^{3} \left( \partial^{5} \cdot \varphi \right) - \frac{1}{35\Box^{2}}\eta^{2} \partial \left( \partial^{5} \cdot \varphi \right) \\ \nonumber &  - \frac{1}{35\Box^{2}}\eta \partial^{3} \left( \partial^{3} \cdot \varphi' \right) + \frac{2}{35\Box}\eta^{2} \partial \left( \partial^{3} \cdot \varphi' \right) + \frac{4}{35\Box^{3}}\eta \partial^{4} \left( \partial^{4} \cdot \varphi' \right) + \frac{6}{7\Box^{3}}\partial^{6} \left( \partial^{2} \cdot \varphi'' \right) \\ \nonumber &  + \frac{1}{105}\eta^{2} \partial^{2} \varphi''' - \frac{1}{35\Box}\eta \partial^{4} \varphi''' + \frac{2}{105\Box^{3}}\eta^{2} \partial^{2} \left( \partial^{6} \cdot \varphi \right) - \frac{1}{35\Box^{2}}\eta^{2} \partial^{2} \left( \partial^{4} \cdot \varphi' \right) \\ \nonumber &  + \frac{1}{7\Box^{2}}\partial^{6} \varphi''' + \frac{1}{105\Box^{2}}\eta^{3} \partial^{6} \cdot \varphi - \frac{1}{105}\eta^{3} \Box \varphi''' + \frac{1}{35}\eta^{3} \partial^{2} \cdot \varphi'' \\ \nonumber &  - \frac{3}{35\Box^{2}}\eta \partial^{4} \left( \partial^{2} \cdot \varphi'' \right) - \frac{1}{35\Box}\eta^{3} \partial^{4} \cdot \varphi' + \frac{1}{7\Box^{2}}\eta \partial^{5} \left( \partial \cdot \varphi''' \right) - \frac{1}{7\Box^{3}}\eta \partial^{5} \left( \partial^{3} \cdot \varphi'' \right) \\ \nonumber &  - \frac{1}{\Box^{3}}\partial^{7} \left( \partial \cdot \varphi''' \right) - \frac{1}{35\Box}\eta^{2} \partial^{3} \left( \partial \cdot \varphi''' \right) + \frac{2}{35\Box^{2}}\eta^{2} \partial^{3} \left( \partial^{3} \cdot \varphi'' \right) - \frac{1}{35\Box^{3}}\eta^{2} \partial^{3} \left( \partial^{5} \cdot \varphi' \right) \\ \nonumber &  - \frac{1}{35\Box}\eta^{3} \partial \left( \partial^{3} \cdot \varphi'' \right) + \frac{1}{105}\eta^{3} \partial \left( \partial \cdot \varphi''' \right) + \frac{1}{35\Box^{2}}\eta^{3} \partial \left( \partial^{5} \cdot \varphi' \right) - \frac{1}{105\Box^{3}}\eta^{3} \partial \left( \partial^{7} \cdot \varphi \right) \\ \nonumber &  - \frac{1}{7\Box^{2}}\eta \partial^{6} \varphi^{[4]} + \frac{1}{35\Box}\eta^{2} \partial^{4} \varphi^{[4]} + \frac{1}{35\Box^{3}}\eta^{2} \partial^{4} \left( \partial^{4} \cdot \varphi'' \right) - \frac{1}{35\Box^{2}}\eta^{3} \partial^{2} \left( \partial^{4} \cdot \varphi'' \right) \\ \nonumber &  + \frac{1}{7\Box^{3}}\eta \partial^{6} \left( \partial^{2} \cdot \varphi''' \right) - \frac{2}{35\Box^{2}}\eta^{2} \partial^{4} \left( \partial^{2} \cdot \varphi''' \right) + \frac{1}{105\Box^{3}}\eta^{3} \partial^{2} \left( \partial^{6} \cdot \varphi' \right) + \frac{1}{35\Box}\eta^{3} \partial^{2} \left( \partial^{2} \cdot \varphi''' \right) \\ \nonumber &  - \frac{4}{105\Box^{2}}\eta^{4} \partial^{6} \cdot \varphi' - \frac{4}{105}\eta^{4} \partial^{2} \cdot \varphi''' + \frac{1}{105}\eta^{4} \Box \varphi^{[4]} + \frac{1}{\Box^{3}}\partial^{8} \varphi^{[4]} \\ \nonumber &  + \frac{2}{35\Box}\eta^{4} \partial^{4} \cdot \varphi'' + \frac{1}{105\Box^{3}}\eta^{4} \partial^{8} \cdot \varphi - \frac{1}{105}\eta^{3} \partial^{2} \varphi^{[4]}
	\end{align}

\subsection{Generalized Einstein tensor $\mathcal{E}$}

\paragraph{Definition}

	\begin{empheq}[box=\widefbox]{align}
	\mathcal{E_\varphi} &= \mathcal{A_\varphi} - \frac{1}{2} \eta \mathcal{A}_\varphi ' + \eta^2 \mathcal{B}_\varphi \\
	\mathcal{A}_\varphi &= \sum_{k=0}^{\left[ s/2 \right]}  (-1)^{k+1} (2k-1) \left\{ \prod_{j=0}^{k-1} \frac{n+j}{n-j+1} \right\} \frac{\partial^{2k}}{\Box^k} \mathcal{F}^{[k]}_{n+1} \\
	\mathcal{B}_\varphi &= \sum_{k=0}^{\left[ \frac{s-4}{2}  \right]} \frac{1}{2^{k-1} (k+2)!} \eta^k \mathcal{B}_0^{[k]} \\
	\mathcal{B}_0 &= \sum_{k=0}^{\left[ s/2 \right]} \frac{(-1)^{k} (1-4n^2)}{4(n-k)(n-k+1)(2k+1)} \left\{ \prod_{j=0}^{k-1} \frac{n+j}{n-j+1} \right\} \frac{\partial^{2k}}{\Box^k} \mathcal{F}^{[k+2]}_{n+1} \\
	n &= \left[ \frac{s-1}{2} \right]
	\end{empheq}
	
	\textbf{Important properties:}
	\begin{itemize}
		\item gauge-invariant
		\item divergenceless
		\item doubly-traceless
	\end{itemize}
	
	We list the generalized non-local Einstein tensor $\mathcal{E}$ and its building blocks $\mathcal{A_\varphi}$ and $\mathcal{B}_\varphi$  for spins up to $s=8$.
	
\subsubsection{$\mathcal{E}$ in terms of $\varphi$}\label{E}
	\begin{align} 
	\mathcal{E}_{0} &= \Box \varphi
	\end{align}
	\begin{align} 
	\mathcal{E}_{1} = \mathcal{E}_{0}& - \partial \left( \partial \cdot \varphi \right)
	\end{align}
	\begin{align} 
	\mathcal{E}_{2} = \mathcal{E}_{1}& + \partial^{2} \varphi' + \eta \partial^{2} \cdot \varphi - \eta \Box \varphi'
	\end{align}
	\begin{align} 
	\mathcal{E}_{3} = \mathcal{E}_{2}& + \frac{2}{\Box^{2}}\partial^{3} \left( \partial^{3} \cdot \varphi \right) - \frac{3}{\Box}\partial^{3} \left( \partial \cdot \varphi' \right) - \frac{1}{\Box}\eta \partial \left( \partial^{3} \cdot \varphi \right) + \eta \partial \left( \partial \cdot \varphi' \right)
	\end{align}
	\begin{align} 
	\mathcal{E}_{4} = \mathcal{E}_{3}& - \frac{8}{\Box^{3}}\partial^{4} \left( \partial^{4} \cdot \varphi \right) + \frac{12}{\Box^{2}}\partial^{4} \left( \partial^{2} \cdot \varphi' \right) - \frac{3}{\Box}\partial^{4} \varphi'' + \frac{2}{\Box^{2}}\eta \partial^{2} \left( \partial^{4} \cdot \varphi \right) \\ \nonumber &  - \frac{3}{\Box}\eta \partial^{2} \left( \partial^{2} \cdot \varphi' \right) + \eta \partial^{2} \varphi'' - \frac{1}{\Box}\eta^{2} \partial^{4} \cdot \varphi   + 2\eta^{2} \partial^{2} \cdot \varphi' - \eta^{2} \Box \varphi''
	\end{align}
	\begin{align} 
	\mathcal{E}_{5} = \mathcal{E}_{4}& + \frac{24}{\Box^{4}}\partial^{5} \left( \partial^{5} \cdot \varphi \right) - \frac{40}{\Box^{3}}\partial^{5} \left( \partial^{3} \cdot \varphi' \right) + \frac{15}{\Box^{2}}\partial^{5} \left( \partial \cdot \varphi'' \right) - \frac{4}{\Box^{3}}\eta \partial^{3} \left( \partial^{5} \cdot \varphi \right) \\ \nonumber &  + \frac{7}{\Box^{2}}\eta \partial^{3} \left( \partial^{3} \cdot \varphi' \right) - \frac{3}{\Box}\eta \partial^{3} \left( \partial \cdot \varphi'' \right) + \frac{1}{\Box^{2}}\eta^{2} \partial \left( \partial^{5} \cdot \varphi \right)  \\ \nonumber &  - \frac{2}{\Box}\eta^{2} \partial \left( \partial^{3} \cdot \varphi' \right) + \eta^{2} \partial \left( \partial \cdot \varphi'' \right)
	\end{align}
	\begin{align} 
	\mathcal{E}_{6} = \mathcal{E}_{5}& - \frac{64}{\Box^{5}}\partial^{6} \left( \partial^{6} \cdot \varphi \right) + \frac{120}{\Box^{4}}\partial^{6} \left( \partial^{4} \cdot \varphi' \right) - \frac{60}{\Box^{3}}\partial^{6} \left( \partial^{2} \cdot \varphi'' \right) + \frac{5}{\Box^{2}}\partial^{6} \varphi''' \\ \nonumber &  + \frac{8}{\Box^{4}}\eta \partial^{4} \left( \partial^{6} \cdot \varphi \right) - \frac{16}{\Box^{3}}\eta \partial^{4} \left( \partial^{4} \cdot \varphi' \right) + \frac{9}{\Box^{2}}\eta \partial^{4} \left( \partial^{2} \cdot \varphi'' \right)   - \frac{1}{\Box}\eta \partial^{4} \varphi''' \\ \nonumber &- \frac{4}{3\Box^{3}}\eta^{2} \partial^{2} \left( \partial^{6} \cdot \varphi \right) + \frac{3}{\Box^{2}}\eta^{2} \partial^{2} \left( \partial^{4} \cdot \varphi' \right)  - \frac{2}{\Box}\eta^{2} \partial^{2} \left( \partial^{2} \cdot \varphi'' \right) \\ \nonumber & + \frac{1}{3}\eta^{2} \partial^{2} \varphi''' + \frac{1}{3\Box^{2}}\eta^{3} \partial^{6} \cdot \varphi  - \frac{1}{\Box}\eta^{3} \partial^{4} \cdot \varphi' + \eta^{3} \partial^{2} \cdot \varphi'' - \frac{1}{3}\eta^{3} \Box \varphi'''
	\end{align}
	\begin{align} 
	\mathcal{E}_{7} = \mathcal{E}_{6}& + \frac{160}{\Box^{6}}\partial^{7} \left( \partial^{7} \cdot \varphi \right) - \frac{336}{\Box^{5}}\partial^{7} \left( \partial^{5} \cdot \varphi' \right) + \frac{210}{\Box^{4}}\partial^{7} \left( \partial^{3} \cdot \varphi'' \right) - \frac{35}{\Box^{3}}\partial^{7} \left( \partial \cdot \varphi''' \right)  \\ \nonumber & - \frac{16}{\Box^{5}}\eta \partial^{5} \left( \partial^{7} \cdot \varphi \right) + \frac{36}{\Box^{4}}\eta \partial^{5} \left( \partial^{5} \cdot \varphi' \right) - \frac{25}{\Box^{3}}\eta \partial^{5} \left( \partial^{3} \cdot \varphi'' \right)   + \frac{5}{\Box^{2}}\eta \partial^{5} \left( \partial \cdot \varphi''' \right) \\ \nonumber &+ \frac{2}{\Box^{4}}\eta^{2} \partial^{3} \left( \partial^{7} \cdot \varphi \right) - \frac{5}{\Box^{3}}\eta^{2} \partial^{3} \left( \partial^{5} \cdot \varphi' \right)   + \frac{4}{\Box^{2}}\eta^{2} \partial^{3} \left( \partial^{3} \cdot \varphi'' \right) - \frac{1}{\Box}\eta^{2} \partial^{3} \left( \partial \cdot \varphi''' \right) \\ \nonumber &- \frac{1}{3\Box^{3}}\eta^{3} \partial \left( \partial^{7} \cdot \varphi \right)   + \frac{1}{\Box^{2}}\eta^{3} \partial \left( \partial^{5} \cdot \varphi' \right) - \frac{1}{\Box}\eta^{3} \partial \left( \partial^{3} \cdot \varphi'' \right) + \frac{1}{3}\eta^{3} \partial \left( \partial \cdot \varphi''' \right)
	\end{align}
	\begin{align} 
	\mathcal{E}_{8} = \mathcal{E}_{7}& + \frac{44}{5\Box^{4}}\eta^{2} \partial^{4} \left( \partial^{6} \cdot \varphi' \right) - \frac{384}{\Box^{7}}\partial^{8} \left( \partial^{8} \cdot \varphi \right) + \frac{896}{\Box^{6}}\partial^{8} \left( \partial^{6} \cdot \varphi' \right) - \frac{7}{\Box^{3}}\partial^{8} \varphi^{[4]} \\ \nonumber &  - \frac{672}{\Box^{5}}\partial^{8} \left( \partial^{4} \cdot \varphi'' \right) + \frac{1}{\Box^{2}}\eta \partial^{6} \varphi^{[4]} + \frac{32}{\Box^{6}}\eta \partial^{6} \left( \partial^{8} \cdot \varphi \right)   - \frac{19}{\Box^{3}}\eta \partial^{6} \left( \partial^{2} \cdot \varphi''' \right) \\ \nonumber & - \frac{80}{\Box^{5}}\eta \partial^{6} \left( \partial^{6} \cdot \varphi' \right) + \frac{66}{\Box^{4}}\eta \partial^{6} \left( \partial^{4} \cdot \varphi'' \right)   - \frac{1}{5\Box}\eta^{2} \partial^{4} \varphi^{[4]} + \frac{14}{5\Box^{2}}\eta^{2} \partial^{4} \left( \partial^{2} \cdot \varphi''' \right) \\ \nonumber &+ \frac{168}{\Box^{4}}\partial^{8} \left( \partial^{2} \cdot \varphi''' \right)   - \frac{41}{5\Box^{3}}\eta^{2} \partial^{4} \left( \partial^{4} \cdot \varphi'' \right) - \frac{16}{5\Box^{5}}\eta^{2} \partial^{4} \left( \partial^{8} \cdot \varphi \right) \\ \nonumber & + \frac{7}{5\Box^{2}}\eta^{3} \partial^{2} \left( \partial^{4} \cdot \varphi'' \right)   + \frac{1}{15}\eta^{3} \partial^{2} \varphi^{[4]}- \frac{19}{15\Box^{3}}\eta^{3} \partial^{2} \left( \partial^{6} \cdot \varphi' \right)  \\ \nonumber &- \frac{3}{5\Box}\eta^{3} \partial^{2} \left( \partial^{2} \cdot \varphi''' \right)   + \frac{2}{5\Box^{4}}\eta^{3} \partial^{2} \left( \partial^{8} \cdot \varphi \right) - \frac{1}{15}\eta^{4} \Box \varphi^{[4]} \\ \nonumber & + \frac{4}{15}\eta^{4} \partial^{2} \cdot \varphi'''  - \frac{2}{5\Box}\eta^{4} \partial^{4} \cdot \varphi'' + \frac{4}{15\Box^{2}}\eta^{4} \partial^{6} \cdot \varphi' - \frac{1}{15\Box^{3}}\eta^{4} \partial^{8} \cdot \varphi
	\end{align}
\newpage
\subsubsection{$\mathcal{A}_\varphi$ and $\mathcal{B}_\varphi$ in terms of $\mathcal{F}_{n+1}$}
	
	\begin{align} 
	\mathcal{A}_{0} &= \mathcal{F}_{0}\\
	\mathcal{B}_{0} &= 0
	\end{align}
	\begin{align} 
	\mathcal{A}_{1} &= \mathcal{F}_{1}\\
	\mathcal{B}_{1} &= 0
	\end{align}
	\begin{align} 
	\mathcal{A}_{2} &= \mathcal{F}_{1}\\
	\mathcal{B}_{2} &= 0
	\end{align}
	\begin{align} 
	\mathcal{A}_{3} &= \mathcal{F}_{2} + \frac{1}{2\Box}\partial^{2} \mathcal{F}_{2}'\\
	\mathcal{B}_{3} &= 0
	\end{align}
	\begin{align} 
	\mathcal{A}_{4} &= \mathcal{F}_{2} + \frac{1}{2\Box}\partial^{2} \mathcal{F}_{2}' - \frac{3}{\Box^{2}}\partial^{4} \mathcal{F}_{2}''\\
	\mathcal{B}_{4} &=  - \frac{3}{8}\mathcal{F}_{2}''
	\end{align}
	\begin{align} 
	\mathcal{A}_{5} &= \mathcal{F}_{3} + \frac{2}{3\Box}\partial^{2} \mathcal{F}_{3}' - \frac{3}{\Box^{2}}\partial^{4} \mathcal{F}_{3}''\\
	\mathcal{B}_{5} &=  - \frac{5}{8}\mathcal{F}_{3}''
	\end{align}
	\begin{align} 
	\mathcal{A}_{6} &= \mathcal{F}_{3} + \frac{2}{3\Box}\partial^{2} \mathcal{F}_{3}' - \frac{3}{\Box^{2}}\partial^{4} \mathcal{F}_{3}'' + \frac{20}{\Box^{3}}\partial^{6} \mathcal{F}_{3}'''\\
	\mathcal{B}_{6} &=  - \frac{5}{8}\mathcal{F}_{3}'' + \frac{5}{12\Box}\partial^{2} \mathcal{F}_{3}''' - \frac{5}{144}\eta \mathcal{F}_{3}'''
	\end{align}
	\begin{align} 
	\mathcal{A}_{7} &= \mathcal{F}_{4} + \frac{3}{4\Box}\partial^{2} \mathcal{F}_{4}' - \frac{3}{\Box^{2}}\partial^{4} \mathcal{F}_{4}'' + \frac{25}{2\Box^{3}}\partial^{6} \mathcal{F}_{4}'''\\
	\mathcal{B}_{7} &=  - \frac{35}{48}\mathcal{F}_{4}'' + \frac{35}{96\Box}\partial^{2} \mathcal{F}_{4}''' - \frac{35}{576}\eta \mathcal{F}_{4}''' + \frac{35}{288\Box}\eta \partial \left( \partial \cdot \mathcal{F}_{4}''' \right)
	\end{align}
	\begin{align} 
	\mathcal{A}_{8} &= \mathcal{F}_{4} + \frac{3}{4\Box}\partial^{2} \mathcal{F}_{4}' - \frac{3}{\Box^{2}}\partial^{4} \mathcal{F}_{4}'' + \frac{25}{2\Box^{3}}\partial^{6} \mathcal{F}_{4}''' - \frac{105}{\Box^{4}}\partial^{8} \mathcal{F}_{4}^{[4]}\\
	\mathcal{B}_{8} &=  - \frac{35}{48}\mathcal{F}_{4}'' + \frac{35}{96\Box}\partial^{2} \mathcal{F}_{4}''' - \frac{35}{576}\eta \mathcal{F}_{4}''' + \frac{35}{288\Box}\eta \partial \left( \partial \cdot \mathcal{F}_{4}''' \right) \\ \nonumber &  - \frac{7}{8\Box^{2}}\partial^{4} \mathcal{F}_{4}^{[4]} - \frac{49}{576\Box}\eta \partial^{2} \mathcal{F}_{4}^{[4]} + \frac{35}{1152\Box}\eta^{2} \partial^{2} \cdot \mathcal{F}_{4}''' - \frac{7}{384}\eta^{2} \mathcal{F}_{4}^{[4]}
	\end{align}
\newpage	
\subsubsection{$\mathcal{A}_\varphi$ and $\mathcal{B}_\varphi$ in terms of $\mathcal{R}$}
	\begin{align} 
	\mathcal{A}_{0} &= \mathcal{R}\\
	\mathcal{B}_{0} &= 0
	\end{align}
	\begin{align} 
	\mathcal{A}_{1} &= \partial \cdot \mathcal{R}\\
	\mathcal{B}_{1} &= 0
	\end{align}
	\begin{align} 
	\mathcal{A}_{2} &= \mathcal{R}'\\
	\mathcal{B}_{2} &= 0
	\end{align}
	\begin{align} 
	\mathcal{A}_{3} &= \frac{1}{\Box}\partial \cdot \mathcal{R}' + \frac{1}{2\Box^{2}}\partial^{2} \left( \partial \cdot \mathcal{R}'' \right)\\
	\mathcal{B}_{3} &= 0
	\end{align}
	\begin{align} 
	\mathcal{A}_{4} &= \frac{1}{\Box}\mathcal{R}'' + \frac{1}{2\Box^{2}}\partial^{2} \mathcal{R}''' - \frac{3}{\Box^{3}}\partial^{4} \mathcal{R}^{[4]}\\
	\mathcal{B}_{4} &=  - \frac{3}{8\Box}\mathcal{R}^{[4]}
	\end{align}
	\begin{align} 
	\mathcal{A}_{5} &= \frac{1}{\Box^{2}}\partial \cdot \mathcal{R}'' + \frac{2}{3\Box^{3}}\partial^{2} \left( \partial \cdot \mathcal{R}''' \right) - \frac{3}{\Box^{4}}\partial^{4} \left( \partial \cdot \mathcal{R}^{[4]} \right)\\
	\mathcal{B}_{5} &=  - \frac{5}{8\Box^{2}}\partial \cdot \mathcal{R}^{[4]}
	\end{align}
	\begin{align} 
	\mathcal{A}_{6} &= \frac{1}{\Box^{2}}\mathcal{R}''' + \frac{2}{3\Box^{3}}\partial^{2} \mathcal{R}^{[4]} - \frac{3}{\Box^{4}}\partial^{4} \mathcal{R}^{[5]} + \frac{20}{\Box^{5}}\partial^{6} \mathcal{R}^{[6]}\\
	\mathcal{B}_{6} &=  - \frac{5}{8\Box^{2}}\mathcal{R}^{[5]} + \frac{5}{12\Box^{3}}\partial^{2} \mathcal{R}^{[6]} - \frac{5}{144\Box^{2}}\eta \mathcal{R}^{[6]}
	\end{align}
	\begin{align} 
	\mathcal{A}_{7} &= \frac{1}{\Box^{3}}\partial \cdot \mathcal{R}''' + \frac{3}{4\Box^{4}}\partial^{2} \left( \partial \cdot \mathcal{R}^{[4]} \right) - \frac{3}{\Box^{5}}\partial^{4} \left( \partial \cdot \mathcal{R}^{[5]} \right) + \frac{25}{2\Box^{6}}\partial^{6} \left( \partial \cdot \mathcal{R}^{[6]} \right)\\
	\mathcal{B}_{7} &=  - \frac{35}{48\Box^{3}}\partial \cdot \mathcal{R}^{[5]} + \frac{35}{96\Box^{4}}\partial^{2} \left( \partial \cdot \mathcal{R}^{[6]} \right) - \frac{35}{576\Box^{3}}\eta \partial \cdot \mathcal{R}^{[6]} \\ \nonumber &+ \frac{35}{288\Box^{4}}\eta \partial \left( \partial^{2} \cdot \mathcal{R}^{[6]} \right)
	\end{align}
	\begin{align} 
	\mathcal{A}_{8} &= \frac{1}{\Box^{3}}\mathcal{R}^{[4]} + \frac{3}{4\Box^{4}}\partial^{2} \mathcal{R}^{[5]} - \frac{3}{\Box^{5}}\partial^{4} \mathcal{R}^{[6]} + \frac{25}{2\Box^{6}}\partial^{6} \mathcal{R}^{[7]} - \frac{105}{\Box^{7}}\partial^{8} \mathcal{R}^{[8]}\\
	\mathcal{B}_{8} &=  - \frac{35}{48\Box^{3}}\mathcal{R}^{[6]} + \frac{35}{96\Box^{4}}\partial^{2} \mathcal{R}^{[7]} - \frac{35}{576\Box^{3}}\eta \mathcal{R}^{[7]} + \frac{35}{288\Box^{4}}\eta \partial \left( \partial \cdot \mathcal{R}^{[7]} \right) \\ \nonumber &  - \frac{7}{8\Box^{5}}\partial^{4} \mathcal{R}^{[8]} - \frac{49}{576\Box^{4}}\eta \partial^{2} \mathcal{R}^{[8]} + \frac{35}{1152\Box^{4}}\eta^{2} \partial^{2} \cdot \mathcal{R}^{[7]} - \frac{7}{384\Box^{3}}\eta^{2} \mathcal{R}^{[8]}
	\end{align}
\newpage
\section{HS computer algebra}\label{program}\label{a2}

	There exist a number of excellent computer programs for symbolic algebra, such as Wolfram Mathematica \cite{mathematica} and Cadabra \cite{cadabra1}\cite{cadabra2}, but we have decided to write a C\texttt{++} code from scratch. The main reason for choosing to do this was the need to implement a simple working code for the Francia-Sagnotti HS formalism. While it could surely be done in the aforementioned programs or similar ones, writing the code from scratch seemed like a particularly clean and straightforward solution. In addition to that, this allowed us to implement the option of exporting the results directly to  \TeX\, code.	
	The interested reader is free to use and modify the code and is encouraged to implement it in any computer algebra system.\footnotemark 	
	One thing that is lacking in the code is its optimization, which was sacrificed for human-readability.\footnotetext{Any \href{mailto:vukovic.ac@gmail.com}{feedback} from anyone who chooses to do so would be greatly appreciated!}
	
	The code contains three fundamental objects, \textbf{terms}, \textbf{expressions} and \textbf{symbols}.	
	A \textbf{term} represents a generic \textit{linear} term one would find in the Francia-Sagnotti formalism, i.e.
	\begin{equation}
	c \eta^j \Box^k \partial^l \left( \partial^m \cdot \varphi^{[n]}_{(s)} \right)
	\end{equation}
	and is defined by eight parameters, the numerical coefficient $c$, the exponents $j,k,l$ and $m$, the number of traces $n$, the spin (technically, the tensor order) of the field $s$ and the field name, which is "$\varphi$" in this particular case.	
	A \textbf{symbol} (for example $c$) is defined as an ordered pair of integers that form a maximally reduced fraction.	
	An \textbf{expression} is simply a linear combination of \textbf{terms}.
	
	Only the standard C/C\texttt{++} libraries are used.\footnotemark \footnotetext{To be specific, \texttt{iostream}, \texttt{iomanip}, \texttt{algorithm}, \texttt{cmath} , \texttt{vector} and \texttt{assert.h} .}
\newpage
\subsection{Fraction algebra}
	
	We define \texttt{pint} as a data type alias for elements in $\mathbb{N}_0$  and \texttt{lint} for elements in $\mathbb{Z}$.
	\begin{lstlisting}
	typedef unsigned long long	pint;
	typedef long long int		lint;
	\end{lstlisting}
	Next, we define a structure \texttt{symb} to represent \textbf{symbols} and the common binary and unary operations on them.
\begin{lstlisting}
struct symb{
	lint numerator;			// numerator	defined
	pint denominator;			// denominator	defined
	
	symb(){						// default	constructor
		numerator = 1;
		denominator= 1;
	}
	symb (lint x){				// integer	constructor
		numerator = x;
		denominator = 1;
	}
	symb (lint x, lint y){	// fraction	constructor
        numerator = x;
		if(y<0){
			denominator = pint(-y);
			numerator = -numerator;
		}
		else
			denominator = pint(y);
	}
	
	bool operator == (symb y){	// equal
		return (this->numerator * y.denominator == y.numerator * this->denominator);}
	bool operator == (lint x){	// equal
		return (symb(x) == symb(this->numerator, this->denominator));}
	bool operator != (symb y){	// not equal
		return !(this->numerator * y.denominator == y.numerator * this->denominator);}
	bool operator != (lint x){	// not equal
		return (symb(x) != symb(this->numerator, this->denominator));}
	bool operator > (symb y){	// greater than
		return (this->numerator * lint(y.denominator) > y.numerator * lint(this->denominator));}
	bool operator < (symb y){	// less than
		return (this->numerator * lint(y.denominator) < y.numerator * lint(this->denominator));}
	symb operator + (symb y){	// addition
		return symb(lint(this->numerator * y.denominator + y.numerator * this->denominator), pint(this->denominator*y.denominator));}
	symb operator + (symb y){	// subtraction
		return symb(lint(this->numerator * y.denominator - y.numerator * this->denominator), pint(this->denominator*y.denominator));}
	symb operator * (symb y){	// multiplication
		return symb(this->numerator*y.numerator, this->denominator*y.denominator);}
	symb operator / (symb y){	// division
		return (symb(this->numerator, this->denominator) * y.inverse());}
	symb inverse (){				// inversion
		if (this->numerator > 0)
			return symb(this->denominator,this->numerator);
		return symb(-1*this->denominator, -1*this->numerator);
	}
	void Simplify (){				// reduction to an irreducible fraction
		pint g = gcd(abs(numerator), denominator);	
		numerator = numerator / lint(g);
		denominator = denominator / lint(g);
	}
};
\end{lstlisting}
We define the \textbf{greatest common divisor} function $\gcd(a,b)$.
\begin{lstlisting}
pint gcd (pint a, pint b){
	if (b==0)
		return a;
	else
		return gcd(b, a%b);
}
\end{lstlisting}
We define the \textbf{falling factorial} function $n^{\underline{k}} = n (n-1) \dots (n-k+1)$.
\begin{lstlisting}
pint FallingFactorial(pint n, pint k){
	if(k==0)
		return 1;
	if(k>n)
		return 0;
	pint out = n;
	for(pint i=1; i<k; i++)
		out *= n-i;
	return out;
}
\end{lstlisting}
We define the \textbf{factorial} function $n!$.
\begin{lstlisting}
pint Factorial(pint n){
	if(n==0)
		return 1;
	return FallingFactorial(n,n);
}
\end{lstlisting}
We define the \textbf{binomial coefficient} ${n \choose k}$.
\begin{lstlisting}
symb Choose(pint n, pint k){
	if(k>n)
		return symb(0);
	return Simplify(symb(FallingFactorial(n,k),Factorial(k)));
}
\end{lstlisting}

\subsection{HS objects and operations}

	\textbf{Expressions} are defined as data-vectors of \textbf{terms} using a data type alias \texttt{expression}.
\begin{lstlisting}
typedef vector<term> expression;
\end{lstlisting}
	We define a structure \texttt{term} to represent \textbf{terms} and functions to extract their algebraic properties.
\begin{lstlisting}
struct term{
	string field;         // field name (in LaTeX) defined
	symb fact;            // multiplicative factor defined
	pint ord;             // tensor order defined
	pint eta;             // number of metric tensors defined
	lint box;             // number of D'Alambertians defined
	pint grad;            // number of gradients defined
	pint div;             // number of divergences defined
	pint trace;           // number of traces defined
	
	term (){					// default constructor
		field = "";
		fact = symb(0);
		ord = 0;
		eta = 0;
		box = 0;
		grad = 0;
		div = 0;
		trace = 0;
	}
	term (string in_field, symb in_fact, pint in_ord,
	pint in_eta, lint in_box, pint in_grad, pint in_div, pint in_trace){
		field = in_field;
		fact = Simplify(in_fact);
		ord = in_ord;
		eta = in_eta;
		box = in_box;
		grad = in_grad;
		div = in_div;
		trace = in_trace;
	}
	
	int DerivativeOrder(){	// calculates the derivative order
		return grad + div + 2*box;}
	int TermOrder(){			// calculates the tensorial order
		return ord + 2*eta + grad - div - 2*trace;}
	int MinimalOrder(){		// calculates the minimal tensorial order
		return 2*trace + div;// for which the term does not vanish
	}
	void Reduce(){				// cancels algebraically inconsistent (trivially vanishing) terms
		if(ord - 2*trace - div < 0 || fact == 0){
			field = "";
			fact = symb(0);
			ord = 0;
			eta = 0;
			box = 0;
			grad = 0;
			div = 0;
			trace = 0;
		}
		fact.Simplify();
	}
};
\end{lstlisting}
	We define a structure \texttt{parameters} that holds the \textbf{parameters of a term}, needed for some more complicated operations.
\begin{lstlisting}
struct parameters{
	pint trace;
	pint metric;
	pint div;
	pint grad;
	lint box;
	symb factor;
	string field;
	
	bool operator == (parameters x){
		return (this->trace == x.trace && this->metric == x.metric && this->div == x.div && this->grad == x.grad && this->box == x.box && this->factor == x.factor);
	}
};
\end{lstlisting}
	We define a structure \texttt{transformation}, which holds two sets of \texttt{parameters}, first defining a \textbf{condition} on the term, the second defining the \textbf{operation} to be performed if the conditions are satisfied.
\begin{lstlisting}
struct transformation{
	parameters condition;
	parameters operation;
};
\end{lstlisting}
	We define a structure \texttt{GaugeTransformation} for \textbf{gauge transformations} that transform a single term into a single term, specifying the transformation law.
\begin{lstlisting}
struct GaugeTransformation{
	term original;
	term transformed;
};
\end{lstlisting}
	We define a function that takes a \textbf{gradient of a term}, i.e. multiplies it by $\partial$.
\begin{lstlisting}
term TakeGradient(term interm){
	term outterm = interm;
	outterm.grad += 1;
	outterm.fact = outterm.fact * symb(outterm.grad);
	outterm.Reduce();
	return outterm;
}
\end{lstlisting}
	We define a function that takes \textbf{$n$ consecutive gradients of a term}, which is \textbf{\underline{not}} the same as multiplying by $\partial^n$.
\begin{lstlisting}
term TakeGradient(term interm, pint n){
	term outterm = interm;
	for(pint i=0; i<n; i++)
		outterm = TakeGradient(outterm);
	return outterm;
}
\end{lstlisting}
	We define a function that takes a \textbf{gradient of a term} \textit{without} the combinatorial factors, for example $\partial^3 \varphi \mapsto \partial^4 \varphi$, needed for some more complicated operations.
\begin{lstlisting}
term TakeGradientNoNorm(term interm){
	term outterm = interm;
	outterm.grad += 1;
	outterm.Reduce();
	return outterm;
}
\end{lstlisting}
	We define a function that takes \textbf{$n$ consecutive gradients of a term} \textit{without} the combinatorial factors, needed for some more complicated operations.
\begin{lstlisting}
term TakeGradientNoNorm(term interm, pint number){
	term outterm = interm;
	for(pint i=0; i<number; i++)
		outterm = TakeGradientNoNorm(outterm);
	return outterm;
}
\end{lstlisting}
	We define a function that takes a \textbf{multi-gradient of a term}, i.e. multiplies it by $\partial^n$.
\begin{lstlisting}
term TakeMultiGradient(term interm, pint n){
	symb temp = Choose(interm.grad + exponent, n);
	term outterm = TakeGradientNoNorm(interm, n);
	outterm.fact = outterm.fact * temp;
	return outterm;
}
\end{lstlisting}
	We define a function that \textbf{multiplies a term by the metric tensor} $\eta$.
\begin{lstlisting}
term MultiplyByMetric(term interm){
	term outterm = interm;
	outterm.fact = outterm.fact * (outterm.eta + 1);
	outterm.eta++;
	outterm.Reduce();
	return outterm;
}
\end{lstlisting}
\newpage
	We define a function that, if a term satisfies a condition given by a set of \textbf{parameters}, modifies the term into one that minimally satisfies it. This is needed for some more complicated operations.
\begin{lstlisting}[escapechar=@]
term TrimmedToCondition(term interm, parameters cond){
	if (interm.box < cond.box || interm.div < cond.div
		|| interm.eta < cond.metric || interm.grad < cond.grad
		|| interm.trace < cond.trace || cond.field != interm.field)
	return NULLTERM;@\footnotemark@

	term outterm = interm;
	outterm.box = cond.box;
	outterm.div = cond.div;
	outterm.eta = cond.metric;
	outterm.grad = cond.grad;
	outterm.trace = cond.trace;
	outterm.fact = cond.factor;
	return outterm;
}
\end{lstlisting}
	We define a function that transforms a \textbf{term} into an \textbf{expression} containing that one term.\footnotetext{\texttt{term NULLTERM("",0,0,0,0,0,0,0);}}
\begin{lstlisting}
expression ToExpression(term infield){
	expression outexp;
	infield.Reduce();
	outexp.push_back(infield);
	return Simplify(outexp);
}
\end{lstlisting}
	We define a function that takes a \textbf{trace of a term}. Note the dependence on \texttt{dim}, i.e. the spacetime dimension $D$.
\begin{lstlisting}
expression TakeTrace(pint dim, term interm){
	expression outexp;
	outexp.clear();

	if(ToExpression(interm).size()==0)
		return outexp;

	pint multiplier;
	term temp = interm;
	temp.Reduce();
	if(temp.eta >= 1 && !IsNull(temp)){
		multiplier = dim + 2*(temp.ord + temp.grad - temp.div - 2*temp.trace + temp.eta -1 );
		temp.fact = temp.fact * symb(multiplier);
		temp.eta--;
		temp.Reduce();
		if(!IsNull(temp))
			outexp = ToExpression(temp);
	}

	temp = interm;
	temp.Reduce();
	if(temp.grad >= 2 && !IsNull(temp)){
		temp.box++;
		temp.grad-=2;
		temp.Reduce();
		if(!IsNull(temp))
			outexp = Add(outexp, temp);
	}

	temp = interm;
	temp.Reduce();
	if(temp.grad >= 1 && temp.ord - 2*temp.trace - temp.div >= 1 && !IsNull(temp)){
		multiplier = 2;
		temp.fact = temp.fact * symb(multiplier);
		temp.grad--;
		temp.div++;
		temp.Reduce();
		if(!IsNull(temp))
			outexp = Add(outexp, temp);
	}

	temp = interm;
	temp.Reduce();
	if(temp.ord - 2*temp.trace - temp.div >= 2 && !IsNull(temp)){
		temp.trace++;
		temp.Reduce();
		if(!IsNull(temp))
			outexp = Add(outexp, temp);
	}
	return Simplify(outexp);
}
\end{lstlisting}
	We define a function that takes a \textbf{divergence of a term}, i.e. contracts it with $\partial$.
\begin{lstlisting}
expression TakeDivergence(term interm){
	if(ToExpression(interm).size()==0)
		return ToExpression(interm);

	term temp_term = interm;
	temp_term.div++;
	expression outexp = ToExpression(temp_term);

	if(interm.eta > 0){
		temp_term = interm;
		temp_term.eta--;
		temp_term.Reduce();
		outexp = Add(outexp, TakeGradient(temp_term));
	}

	if(interm.grad > 0){
		temp_term = interm;
		temp_term.box += 1;
		temp_term.grad = temp_term.grad - 1;
		temp_term.Reduce();
		outexp.push_back(temp_term);
	}
	return Simplify(outexp);
}
\end{lstlisting}
	We define a function that \textbf{takes a "box" of an expression}, i.e. operates on it with a D'Alambertian $\Box$.
\begin{lstlisting}
expression TakeBox(expression inexp){
	if(inexp.size()==0)
		return inexp;

	expression outexp = inexp;
	for(pint i=0; i<inexp.size(); i++)
		outexp[i].box++;
	return outexp;
}
\end{lstlisting}
	We define a function that \textbf{simplifies an expression}, i.e. checks if all terms are of the same tensorial order and cancels those that do not make algebraic sense (for example $\partial^2 \cdot \varphi$ if $s=1$). After checking, it gathers all the terms.
\begin{lstlisting}
expression Simplify(expression inexp){
	if(inexp.size()==0)
		return inexp;

	assert(CheckConsistency(inexp));
	return SortTerms(Gather(inexp));
}
\end{lstlisting}
	We define a function that \textbf{gathers} all the terms in an expression that are equal up to a multiplicative constant into a single term. 
\begin{lstlisting}
expression Gather(expression inexp){
	if(inexp.size()==0)
		return inexp;

	expression outexp = inexp;
	for (pint i=0; i<inexp.size(); i++){
		for (pint j=i+1; j<inexp.size(); j++){
			if(CompareUpToFactor(inexp[i],inexp[j])){
				outexp[i].fact = outexp[i].fact + outexp[j].fact;
				outexp[j].fact = 0;
			}
		}
	}

	for(pint i=0; i<outexp.size(); i++)
		outexp[i].Reduce();

	bool termpowers_OK;
	for (pint i=inexp.size(); i>0; i--){
		termpowers_OK = inexp[i-1].fact != 0;
		termpowers_OK = termpowers_OK && inexp[i-1].eta >= 0;
		termpowers_OK = termpowers_OK && inexp[i-1].div >= 0;
		termpowers_OK = termpowers_OK && inexp[i-1].grad >= 0;
		termpowers_OK = termpowers_OK && inexp[i-1].trace >= 0;
		termpowers_OK = termpowers_OK && 2*inexp[i-1].trace + inexp[i-1].div <= inexp[i-1].ord;

		if (!termpowers_OK || inexp[i-1].TermOrder() < 0 || IsNull(inexp[i-1]))
			outexp.erase(outexp.begin() + i - 1);
	}
	return outexp;
}
\end{lstlisting}
	We define a function that \textbf{adds a term to an expression}.
\begin{lstlisting}
expression Add(expression inexp, term interm){
	if(inexp.size()==0)
		return ToExpression(interm);

	expression outexp = inexp;
	for(pint i=0; i<inexp.size(); i++){
		if (CompareUpToFactor(outexp[i],interm)){
			outexp[i].fact = outexp[i].fact + interm.fact;
			return Simplify(outexp);
		}
	}
	outexp.push_back(interm);
	return Simplify(outexp);
}
\end{lstlisting}
	We define a function that \textbf{adds two terms} into an expression.
\begin{lstlisting}
expression Add(term interm1, term interm2){
	return Add(ToExpression(interm1), interm2);
}
\end{lstlisting}
	We define a function that \textbf{multiplies an expression with a symbol}.
\begin{lstlisting}
expression Multiply(expression inexp, symb innumber){
	if(inexp.size()==0)
		return inexp;
	expression outexp = inexp;
	for (pint i=0; i<outexp.size(); i++)
		outexp[i].fact = outexp[i].fact * innumber;
	return Simplify(outexp);
}
\end{lstlisting}
	We define a function that extracts the \textbf{traceless part of an expression}.
\begin{lstlisting}
expression MakeTraceless(pint dim, expression inexp){
	if(inexp.size()==0 || TakeTrace(dim,inexp).size()==0)
		return inexp;

	if(IsTraceful(inexp))
		return inexp;

	expression outexp = inexp;
	expression newterm;
	int s = inexp[0].TermOrder();
	int sumlimit = s/2;
	int sign = 1;
	lint multiplier;

	for(int n=1; n<=sumlimit; n++){
		multiplier = 1;
		sign *= -1;
		for(int j=1; j<=n; j++)
			multiplier = multiplier * (dim + 2*(s-j-1));
		newterm = TakeTrace(dim, inexp, n);
		newterm = MultiplyByMetric(newterm,n);
		newterm = Multiply(newterm,symb(sign,multiplier));
		outexp = Add(outexp,newterm);
	}
	return outexp;
}
\end{lstlisting}
	We define a function that extracts the \textbf{traceful} part of an expression.
\begin{lstlisting}
expression TracefulPart(pint dim, expression inexp){
	if(IsTraceful(inexp))
		return inexp;

	expression traceless;
	traceless = MakeTraceless(dim, inexp);
	if(traceless.size()==0)
		return inexp;

	traceless = Multiply(traceless,symb(-1));
	return Add(inexp, traceless);
}
\end{lstlisting}
	We define a function that performs a \textbf{gauge variation} according to a single gauge variation law.
\begin{lstlisting}
expression GaugeVariation(pint dim, expression original, GaugeTransformation transform){
	expression outexp = Substitute(dim, original, transform.original, transform.transformed);
	return outexp;
}
\end{lstlisting}
	We define a function that \textbf{transforms a term} according to a given set of parameters.
\begin{lstlisting}
expression Transform(pint dim, term interm, parameters param){
	term tempterm = interm;
	expression outexp = ToExpression(tempterm);

	if (param.trace > 0)
		outexp = TakeTrace(dim, outexp, param.trace);
	if (param.div > 0)
		outexp = TakeDivergence(outexp, param.div);
	if (param.grad > 0)
		outexp = TakeGradient(outexp, param.grad);
		
	if (param.metric > 0)
		outexp = MultiplyByMetric(outexp, param.metric);
		
	outexp = TakeBox(outexp, param.box);
	outexp = Multiply(outexp, param.factor);
	return outexp;
}
\end{lstlisting}
	We define a function that \textbf{transforms a term} according to a given set of parameters, with the option of \textbf{renaming the field variable}. 
\begin{lstlisting}
expression NamedTransform(pint dim, term interm, parameters param, string target, string replacement){
	term outterm = interm;
	if(interm.field == target){
		outterm.field = replacement;
		expression outexp = Transform(dim, outterm, param);
		return outexp;
	}
	return ToExpression(interm);
}
\end{lstlisting}
	We define a function that \textbf{transforms an expression} according to a given set of parameters.
\begin{lstlisting}
expression Transform(pint dim, expression inexp, parameters param){
	expression outexp = inexp;

	if (param.trace > 0)
		outexp = TakeTrace(dim, outexp, param.trace);
	if (param.div > 0)
		outexp = TakeDivergence(outexp, param.div);
	if (param.grad > 0)
		outexp = TakeMultiGradient(outexp, param.grad);
	if (param.metric > 0)
		outexp = MultiplyByMetric(outexp, param.metric);
	
	outexp = TakeBox(outexp, param.box);
	outexp = Multiply(outexp, param.factor);
	return outexp;
}
\end{lstlisting}
	We define a function that \textbf{transforms an expression} according to a given \textbf{conditional transformation law}.
\begin{lstlisting}
expression ConditionalTransform(pint dim, expression inexp, transformation trans){
	expression outexp = inexp;
	if(ConditionSatisfied(inexp, trans.condition))
		outexp = Transform(dim, outexp, trans.operation);
	return outexp;
}
\end{lstlisting}
	We define a function that \textbf{transforms an expression} according to a given \textbf{conditional transformation law}, with the option of \textbf{renaming the field variable}.
\begin{lstlisting}
expression ConditionalNamedTransform(pint dim, expression inexp, transformation trans, string target, string replacement){
	expression outexp = inexp;
	outexp = NamedTransform(dim, outexp, trans.operation, target, replacement);
	return outexp;
}
\end{lstlisting}
	We define a function that \textbf{substitutes a term} in an expression with another term.
\begin{lstlisting}
expression Substitute(pint dim, expression original, term old, term new_term){
	expression outexp;
	pint length = original.size();
	if(length==0)
		return original;
		
	// removes non-linear operators
	lint box_min = original[0].box;
	for(pint i=0; i<length; i++){
		if(original[i].box < box_min)
			box_min = original[i].box;
	}
	if(box_min < 0)
		original = TakeBox(original,-box_min);

	outexp.clear();
	transformation trans;
	trans.condition = GetParameters(new_term);
	parameters posttrans;
	parameters targetparams = GetParameters(old);
	targetparams.field = old.field;
	expression tempexp;
	term tempterm;

	for (pint i=0; i<length; i++){
		tempterm = original[i];
		tempterm.ord = new_term.ord;
		tempexp = ToExpression(TrimmedToCondition(tempterm,targetparams));
		if(!IsNull(TrimmedToCondition(tempterm,targetparams))){
			trans.operation = ConditionDifference(tempexp, trans.condition);
			tempexp = ConditionalNamedTransform(dim, tempexp, trans, old.field, new_term.field);

			posttrans = TrimToCondition(original[i], targetparams);
			tempexp = Transform(dim,tempexp,posttrans);

			outexp = Add(outexp, tempexp);
		}
	else{
		tempterm = original[i];
		tempexp = ToExpression(tempterm);
		outexp = Add(outexp, tempexp);
		}
	}
	// restores non-linear operators
	if(box_min < 0)
		outexp = TakeBox(outexp,box_min);
	return outexp;
}
\end{lstlisting}
	We define a function that \textbf{cancels a given term} in an expression.
\begin{lstlisting}
expression Nullify(pint dim, expression original, term constraint){
	term zero = constraint;
	zero.fact = symb(0);
	zero.field = " ";
	return Substitute(dim, original, constraint, zero);
}

\end{lstlisting}
	We define a function that outputs the difference between a parametrized condition and the parameters of an expression.
\begin{lstlisting}
parameters ConditionDifference(expression inexp, parameters condition){
	parameters outparam;
	parameters inparams = GetParameters(inexp);
	outparam.trace = 0;
	outparam.metric = 0;
	outparam.grad = 0;
	outparam.div = 0;
	outparam.box = 0;

	if(inexp.size()==0){
		outparam.factor = 1;
		return outparam;
	}

	if(condition.trace > inparams.trace)
		outparam.trace = condition.trace - inparams.trace;
	if(condition.metric > inparams.metric)
		outparam.metric = condition.metric - inparams.metric;
	if(condition.grad > inparams.grad)
		outparam.grad = condition.grad - inparams.grad;
	if(condition.div > inparams.div)
		outparam.div = condition.div - inparams.div;
	if(condition.box >= inparams.box)
		outparam.box = condition.box - inparams.box;

	outparam.factor = condition.factor / inparams.factor;
	return outparam;
}
\end{lstlisting}
\newpage
	We define a function that outputs the \textbf{maximal set of parameters} shared by all the terms.
\begin{lstlisting}
parameters GetParameters(expression inexp){
	pint length = inexp.size();
	parameters output;

	if(length == 0){
		output.trace = 0;
		output.div = 0;
		output.grad = 0;
		output.metric = 0;
		output.box = 0;
		output.factor = symb(0);
		return output;
	}

	pint trace_count = 0;
	pint div_count = 0;
	pint grad_count = 0;
	pint box_count = 0;
	pint metric_count = 0;

	lint multiplier = 1;
	while(multiplier!=0){
		for(pint i=0; i<length; i++)
			multiplier *= inexp[i].trace;
		if(multiplier !=0){
			for(pint i=0; i<length; i++){
				inexp[i].trace--;
				trace_count++;
			}
		}
	}

	multiplier = 1;
	while(multiplier!=0){
		for(pint i=0; i<length; i++)
			multiplier *= inexp[i].div;
			if(multiplier !=0){
				for(pint i=0; i<length; i++){
					inexp[i].div--;
					div_count++;
				}
			}
		}

	multiplier = 1;
	for(pint i=0; i<length; i++)
	multiplier *= inexp[i].grad;

	while(multiplier!=0){
		for(pint i=0; i<length; i++)
			multiplier *= inexp[i].grad;
		if(multiplier !=0){
			for(pint i=0; i<length; i++){
				inexp[i].grad--;
				grad_count++;
			}
		}
	}

	multiplier = 1;
	while(multiplier!=0){
		for(pint i=0; i<length; i++)
			multiplier *= inexp[i].box;
				if(multiplier !=0){
					for(pint i=0; i<length; i++){
						inexp[i].box--;
						box_count++;
					}
				}
			}

	multiplier = 1;
	while(multiplier!=0){
		for(pint i=0; i<length; i++)
			multiplier *= inexp[i].eta;
			if(multiplier !=0){
				for(pint i=0; i<length; i++){
					inexp[i].eta--;
					metric_count++;
				}
			}
		}

	output.trace = trace_count;
	output.div = div_count;
	output.grad = grad_count;
	output.box = box_count;
	output.metric = metric_count;
	output.factor = inexp[0].fact;
	return output;
}
\end{lstlisting}
	We define a function that, if a term satisfies a condition given by a set of \textbf{parameters}, outputs the difference between a parametrized condition and the parameters of that term. This is needed for some more complicated operations.
\begin{lstlisting}
parameters TrimToCondition(term interm, parameters cond){
	parameters outparam;

	if(Compare(interm,NULLTERM)){
		outparam.box = 0;
		outparam.div = 0;
		outparam.metric = 0;
		outparam.grad = 0;
		outparam.trace = 0;
		outparam.factor = 1;
	}

	else{
		outparam.box = interm.box - cond.box;
		outparam.div = interm.div - cond.div;
		outparam.metric = interm.eta - cond.metric;
		outparam.grad = interm.grad - cond.grad;
		outparam.trace = interm.trace - cond.trace;
		outparam.factor = interm.fact / cond.factor;
	}
	return outparam;
}
\end{lstlisting}
	We define a boolean function that \textbf{checks algebraic consistency} of a given \textbf{expression}.
\begin{lstlisting}
bool CheckConsistency(expression inexp){
	if (inexp.size()==0)
		return true;

	for (pint i=0; i<inexp.size()-1; i++){
		if(!CheckConsistency(inexp[i]))
			return false;

		for (pint j=i+1; j<inexp.size(); j++)
			if (inexp[i].TermOrder() != inexp[j].TermOrder())
				return false;
	}
	return true;
}
\end{lstlisting}
	We define a boolean function that \textbf{checks algebraic consistency} of a given \textbf{term}.
\begin{lstlisting}
bool CheckConsistency(term interm){
	return (interm.TermOrder()>=0);
}
\end{lstlisting}
	We define a boolean function that \textbf{compares two expressions} to see if they are equal.
\begin{lstlisting}
bool Compare(expression inexp1, expression inexp2){
	expression tempexp1 = Simplify(inexp1);
	expression tempexp2 = Simplify(inexp2);

	if (tempexp1.size() != tempexp2.size())
		return false;

	pint inexp_size = inexp1.size();
	pint counter = 0;

	for(pint i=0; i<inexp_size; i++)
		for(pint j=0; j<inexp_size; j++)
			if(CompareUpToFactor(tempexp1[i],tempexp2[j]) && tempexp1[i].fact == tempexp2[j].fact)
				counter++;
	return (counter == inexp_size);
}
\end{lstlisting}
	We define a boolean function that \textbf{checks the mutual algebraic consistency} of two expressions, i.e. if we are permitted to add them.
\begin{lstlisting}
bool CheckConsistency(expression inexp1, expression inexp2){
	if(!CheckConsistency(inexp1))
		return false;
	if(!CheckConsistency(inexp2))
		return false;
	if(inexp1.size()==0 || inexp2.size()==0)
		return true;
	if(inexp1[0].TermOrder() != inexp2[0].TermOrder())
		return false;
	return true;
}
\end{lstlisting}
	We define a boolean function that \textbf{compares two terms} to see if they are equal up to a multiplicative factor.
\begin{lstlisting}
bool CompareUpToFactor(term interm1, term interm2){
	return (
		interm1.field == interm2.field &&
		interm1.ord == interm2.ord &&
		interm1.eta == interm2.eta &&
		interm1.box == interm2.box &&
		interm1.grad == interm2.grad &&
		interm1.div == interm2.div &&
		interm1.trace == interm2.trace
	);
}
\end{lstlisting}
	We define a boolean function that \textbf{compares two terms} to see if they are of the same form, i.e. if they are equal up to a multiplicative factor \textbf{and} the field variable itself.
\begin{lstlisting}
bool CompareUpToFactorAndName(term interm1, term interm2){
	return (
		interm1.ord == interm2.ord &&
		interm1.eta == interm2.eta &&
		interm1.box == interm2.box &&
		interm1.grad == interm2.grad &&
		interm1.div == interm2.div &&
		interm1.trace == interm2.trace
	);
}
\end{lstlisting}
\newpage
	We define a boolean function that \textbf{compares two terms} to see if they are exactly equal.
\begin{lstlisting}
bool Compare(term interm1, term interm2){
	return(CompareUpToFactor(interm1,interm2) && interm1.fact == interm2.fact);
}
\end{lstlisting}
	We define a boolean function that \textbf{checks if a term is identical to zero}.
\begin{lstlisting}
bool IsNull(term interm){
	interm.Reduce();
	return (Compare(interm,NULLTERM));
}
\end{lstlisting}
	We define a boolean function that \textbf{checks if an expression has non-vanishing trace}.
\begin{lstlisting}
bool IsTraceful(expression inexp){
	if(inexp.size()==0)
		return false;

	pint metric_counter = 0;
		for(pint i=0; i<inexp.size(); i++)
			if(inexp[i].eta > 0)
				metric_counter++;

	if(metric_counter == inexp.size())
		return true;
	return false;
}
\end{lstlisting}
	We define a boolean function that \textbf{checks if an expression exactly satisfies a condition} given by a set of parameters.
\begin{lstlisting}
bool ConditionSatisfied(expression inexp, parameters condition){
	parameters param = GetParameters(inexp);
	if(condition.trace != param.trace)
		return false;
	if(condition.div != param.div)
		return false;
	if(condition.grad != param.grad)
		return false;
	if(condition.box != param.box)
		return false;
	if(condition.metric != param.metric)
		return false;
	if(condition.factor != param.factor)
		return false;
	return true;
}
\end{lstlisting}
\newpage
	We define a boolean function that \textbf{checks if an expression sufficiently satisfies a condition} given by a set of parameters.
\begin{lstlisting}
bool MinConditionSatisfied(expression inexp, parameters condition){
	parameters param = GetParameters(inexp);
	if(condition.trace >= param.trace)
		return false;
	if(condition.div >= param.div)
		return false;
	if(condition.grad >= param.grad)
		return false;
	if(condition.box >= param.box)
		return false;
	if(condition.metric >= param.metric)
		return false;
	return true;
}
\end{lstlisting}
	We define a boolean function that \textbf{checks if a term sufficiently satisfies a condition} given by a set of parameters.
\begin{lstlisting}
bool MinConditionSatisfied(term interm, parameters condition){
	if(condition.trace > interm.trace)
		return false;
	if(condition.div > interm.div)
		return false;
	if(condition.grad > interm.grad)
		return false;
	if(condition.box > interm.box)
		return false;
	if(condition.metric > interm.eta)
		return false;
	return true;
}
\end{lstlisting}
	We define a function (along with a boolean condition) that \textbf{sorts terms in an expression} by the minimal order for which they do not vanish.
\begin{lstlisting}
bool SortByMinOrder(term a, term b){
	return a.MinimalOrder() < b.MinimalOrder();
}
expression SortTerms(expression inexp){
	expression outexp = inexp;
	sort(outexp.begin(), outexp.end(), SortByMinOrder);
	return outexp;
}
\end{lstlisting}
\newpage
	Some functions we use are not explicitly defined here for the sake of brevity. Those are either obvious duplicates of a function for some other operation, straightforward generalizations that repeat a function a certain number of times, \textbf{term} functions generalized to \textbf{expression} functions using a simple \texttt{for} loop or \textbf{expression} functions generalized to \textbf{term} functions using the \texttt{ToExpression} function.	
	Since their structure should be quite simple to figure out from previously defined functions, we list here only their names (i.e. declarations).
	
\begin{lstlisting}
term MultiplyByMetric(term interm, pint n);
term MultiplyByMetricNoNorm(term interm);
term MultiplyByMetricNoNorm(term interm, pint number);
term MultiplyByMultiMetric(term interm, pint n);
expression TakeTrace(pint dim, expression inexp);
expression TakeTrace(pint dim, expression inexp, pint number);
expression TakeTrace(pint dim, term interm, pint number);
expression TakeGradient(expression inexp);
expression TakeGradient(expression inexp, pint number);
expression TakeGradientNoNorm(expression inexp);
expression TakeGradientNoNorm(expression inexp, pint number);
expression TakeMultiGradient(expression inexp, pint exponent);
expression TakeDivergence(expression inexp);
expression TakeDivergence(expression inexp, pint number);
expression TakeDivergence(term interm, pint number);
expression MultiplyByMetric(expression inexp);
expression MultiplyByMetric(expression inexp, pint number);
expression MultiplyByMultiMetric(expression inexp, pint exponent);
expression TakeBox(expression inexp, lint number);
expression TakeInverseBox(expression inexp);
expression Add(expression inexp1, expression inexp2);
expression Subtract(expression first, expression second);
expression GaugeVariation(pint dim, expression original, GaugeTransformation gauge1, GaugeTransformation gauge2);
expression GaugeVariation(pint dim, expression original, GaugeTransformation gauge1, GaugeTransformation gauge2, GaugeTransformation gauge3);
expression NamedTransform(pint dim, expression inexp, parameters param, string target, string replacement);
parameters ConditionDifference(term interm, parameters condition);
parameters GetParameters(term interm);
bool ConditionSatisfied(term interm, parameters condition);
\end{lstlisting}
\newpage
\subsection{Geometric objects and important examples}

	We define several functions that use the machinery of HS computer algebra to calculate some quantities of interest. The reader is encouraged to examine these functions in detail as they serve as instructive examples of the previously defined code.

	We define a function that constructs the generalized Fronsdal tensor $\mathcal{F}_{n}$ as defined in \eqref{Fn}.
\begin{lstlisting}
expression GeneralizedFronsdal(pint dim, pint spin, pint n){
	term f = term("\\varphi", 1, spin, 0, 1, 0, 0, 0);

	expression temp = ToExpression(f);
	
	expression temp1;
	expression temp2;
	expression temp3;
	
	for(pint i=1; i<=n; i++){
		temp1 = temp;

		temp2 = TakeDivergence(temp);
		temp2 = TakeGradient(temp2);
		temp2 = Multiply(temp2, symb(-1,i));
		temp2 = TakeInverseBox(temp2);
		
		temp3 = TakeTrace(dim, temp);
		temp3 = TakeMultiGradient(temp3, 2);
		temp3 = Multiply(temp3, symb(1, i*(2*i - 1)));
		temp3 = TakeInverseBox(temp3);
		
		temp = Add(Add(temp1,temp2),temp3);
	}
	return Simplify(temp);
}
\end{lstlisting}
	We define functions that calculate $a_k$ and $b_k$ as defined in \eqref{ak} and \eqref{bk}.
\begin{lstlisting}
symb A(lint k, lint n){
	symb result = symb(1);
	for(lint i=1; i<=k; i++)
		result = result * symb((1-2*i)*(n+i-1),(2*i-3)*(n-i+2));

	return result;
}

symb B(lint k, lint n){
	symb factor = symb(1 - 4*n*n, 1 - 4*k*k);
	factor = factor * symb(1,4*(n-k)*(n-k+1));
	factor = factor*A(k,n);
	return factor;
}
\end{lstlisting}
\newpage
	We define a function that constructs the $\mathcal{B}_0$ tensor as defined in \eqref{D}.
\begin{lstlisting}
expression NonLocalB0(pint dim, pint s){
	expression result;
	result.clear();
	symb factor;
	lint n = (lint(s)-1)/2;
	lint M = (lint(s))/2;
	expression F = GeneralizedFronsdal(dim,s,n+1);
	expression temp;

	for(lint k=0; k<=M; k++){
		temp = TakeTrace(dim,F,k+2);
		temp = TakeMultiGradient(temp,2*k);
		temp = TakeBox(temp, -k);
		temp = Multiply(temp,B(k,n));
		result = Add(result,temp);
	}
	return Simplify(result);
}
\end{lstlisting}
	We define a function that constructs the $\mathcal{B}$ tensor as defined in \eqref{B}.
\begin{lstlisting}
expression NonLocalB(pint dim, pint s){
	expression result;
	expression temp;
	result.clear();

	for(lint k=0; k <= (lint(s)-4)/2; k++){
		temp = TakeTrace(dim,NonLocalB0(dim,s),k);
		temp = Multiply(temp, symb(2, pow(2,k)));
		temp = Multiply(temp, symb(1, Factorial(k+2)));
		temp = MultiplyByMultiMetric(temp,k);
		result = Add(result, temp);
	}
	return Simplify(result);
}
\end{lstlisting}
	We define a function that constructs the generalized Einstein tensor $\mathcal{E}$ as defined in \eqref{E}.
\begin{lstlisting}
expression NonLocalE(pint dim, pint s){
	expression result;
	expression temp;
	result = NonLocalB(dim,s);
	result = MultiplyByMultiMetric(result,2);

	temp = NonLocalA(dim,s);
	result = Add(result, temp);
	temp = TakeTrace(dim, temp);
	temp = MultiplyByMetric(temp);
	temp = Multiply(temp, symb(-1,2));
	result = Add(result, temp);

	return Simplify(result);
}
\end{lstlisting}
	We define a function that constructs the generalized Einstein tensor $\mathcal{G}_{n}$ as defined in \eqref{Gn}.
\begin{lstlisting}
expression GeneralizedEinstein(pint dim, pint spin, pint n){
	expression result;
	result.clear();
	expression temp;

	for(pint k=0; k<=n; k++){
		temp = GeneralizedFronsdal(dim,spin,n);
		temp = TakeTrace(dim,temp,k);
		temp = MultiplyByMultiMetric(temp,k);
		temp = Multiply(temp, symb(1,pow(2,k)));
		temp = Multiply(temp, symb(1,FallingFactorial(n,k)));
		if(k%2==1)
			temp = Multiply(temp,-1);
		result = Add(result, temp);
	}
	return result;
}
\end{lstlisting}
	We define a function that constructs the $E_\varphi (k)$ function as defined in \eqref{Evarphi}.
\begin{lstlisting}
expression kPhi(pint dim, pint spin, int k){
	// Fronsdal
	term fronsdal_1 = term("\\varphi", 1, spin, 0, 1, 0, 0, 0);
	term fronsdal_2 = term("\\varphi",-1, spin, 0, 0, 1, 1, 0);
	term fronsdal_3 = term("\\varphi", 1, spin, 0, 0, 2, 0, 1);
	expression fronsdal = Add(Add(fronsdal_1, fronsdal_2), fronsdal_3);
	
	// C-tensor
	term C_1 = term("\\varphi"	, 1, spin	, 0, 0, 0, 0, 2);
	term C_2 = term("\\alpha"	,-4, spin-3	, 0, 0, 0, 1, 0);
	term C_3 = term("\\alpha"	,-1, spin-3	, 0, 0, 1, 0, 1);
	
	expression C 		= Add(Add(C_1, C_2), C_3);
	expression BoxC		= TakeBox(C);
	expression DivC		= TakeDivergence(C);
	expression GradDivC	= TakeGradient(DivC);
	expression TrC		= TakeTrace(dim, C);
	expression Grad2TrC	= TakeMultiGradient(TrC, 2);
	
	// A-tensor
	expression A	= Add(fronsdal, ToExpression(term("\\alpha",-3, spin-3, 0, 0, 3, 0, 0)));
	expression Tr2A	= TakeTrace(dim, A, 2);
	
	// B-tensor
	term b = term("\\beta", 1, spin-4, 0, 0, 0, 0, 0);
	expression B = Add(Add(Add(
		ToExpression(b),
		ToExpression(term("\\varphi",symb(-1,2)	,spin, 0, 0, 0, 2, 1))),
		ToExpression(term("\\alpha"	,1,spin-3, 0, 1, 0, 1, 0))),
		ToExpression(term("\\alpha"	,symb(1,2),spin-3, 0, 0, 1, 2, 0)));
	
	expression result;
	
	expression t1 = A;
	
	expression t2 = TakeTrace(dim, A);
	t2 = MultiplyByMetric(t2);
	t2 = Multiply(t2, symb(-1,2));
	
	expression t3 = C;
	t3 = TakeMultiGradient(C,2);
	t3 = MultiplyByMetric(t3);
	t3 = Multiply(t3, symb(1+k, 4));
	
	expression t4 = B;
	t4 = MultiplyByMultiMetric(B,2);
	t4 = Multiply(t4, 1-k);
	
	result = Add(Add(Add(t1,t2),t3),t4);
	return result;
}
\end{lstlisting}
	We define a function that constructs the $E_\alpha (k)$ function as defined in \eqref{Ealpha}.
\begin{lstlisting}
expression kAlpha(pint dim, pint spin, int k){
	// Fronsdal
	term fronsdal_1 = term("\\varphi", 1, spin, 0, 1, 0, 0, 0);
	term fronsdal_2 = term("\\varphi",-1, spin, 0, 0, 1, 1, 0);
	term fronsdal_3 = term("\\varphi", 1, spin, 0, 0, 2, 0, 1);
	expression fronsdal = Add(Add(fronsdal_1, fronsdal_2), fronsdal_3);
	
	// C-tensor
	term C_1 = term("\\varphi"	, 1, spin	, 0, 0, 0, 0, 2);
	term C_2 = term("\\alpha"	,-4, spin-3	, 0, 0, 0, 1, 0);
	term C_3 = term("\\alpha"	,-1, spin-3	, 0, 0, 1, 0, 1);
	
	expression C 		= Add(Add(C_1, C_2), C_3);
	expression BoxC		= TakeBox(C);
	expression DivC		= TakeDivergence(C);
	expression GradDivC	= TakeGradient(DivC);
	expression TrC		= TakeTrace(dim, C);
	expression Grad2TrC	= TakeMultiGradient(TrC, 2);
	
	// A-tensor
	expression A	= Add(fronsdal, ToExpression(term("\\alpha",-3, spin-3, 0, 0, 3, 0, 0)));
	expression Tr2A	= TakeTrace(dim, A, 2);
	
	// B-tensor
	term b = term("\\beta", 1, spin-4, 0, 0, 0, 0, 0);
	expression B = Add(Add(Add(
		ToExpression(b),
		ToExpression(term("\\varphi",symb(-1,2)	,spin, 0, 0, 0, 2, 1))),
		ToExpression(term("\\alpha"	,1,spin-3, 0, 1, 0, 1, 0))),
		ToExpression(term("\\alpha"	,symb(1,2),spin-3, 0, 0, 1, 2, 0)));
	
	expression result;
	
	expression t1 = TakeDivergence(TakeTrace(dim,A));
	
	expression t2 = TakeGradient(C);
	t2 = TakeBox(t2);
	t2 = Multiply(t2, symb(-(k+1),2));
	
	expression t3 = TakeMultiGradient(DivC, 2);
	t3 = Multiply(t3, symb(-(k+1),2));
	
	expression t4 = TakeGradient(B);
	t4 = Multiply(t4, 2*(k - 1));
	
	expression t5 = TakeDivergence(B);
	t5 = MultiplyByMetric(t5);
	t5 = Multiply(t5, symb(k-1,1));
	
	result = Add(Add(Add(Add(t1,t2),t3),t4),t5);
	result = Multiply(result, symb(-3,2));
	result = Multiply(result, Choose(spin, 3));
	return result;
}
\end{lstlisting}
	We define a function that constructs the $E_\beta (k)$ function as defined in \eqref{Ebeta}.
\begin{lstlisting}
expression kBeta(pint dim, pint spin, int k){
	// C-tensor
	term C_1 = term("\\varphi"	, 1, spin	, 0, 0, 0, 0, 2);
	term C_2 = term("\\alpha"	,-4, spin-3	, 0, 0, 0, 1, 0);
	term C_3 = term("\\alpha"	,-1, spin-3	, 0, 0, 1, 0, 1);
	
	expression C 		= Add(Add(C_1, C_2), C_3);
	expression result;
	result = Multiply(C, 3*(1-k));
	result = Multiply(result, Choose(spin,4));
	return result;
}
\end{lstlisting}
\newpage
\subsection{Outputting results in \TeX}
	
	We define functions that convert \textbf{symbols}, \textbf{terms}, and \textbf{expressions} to a \TeX\, string ready to be used with any \TeX\, engine.	
	All of these functions are simply called with \texttt{TeX(...)} and are of type \texttt{string}.	
	Some expressions will inevitably have a lot of terms that won't always render properly. For that purpose, we introduce two global parameters that take care of automatic aligned line breaking in the 
	
	\verb|\begin{align} ... \end{align}| 
		
	\LaTeX\, environment.
\begin{lstlisting}
const bool break_line = false;	// set TRUE for long expressions
const pint break_line_after = 5;	// set number of terms before a new line
\end{lstlisting}
The\quad \texttt{symb} $\longrightarrow$ \TeX \quad function:
\begin{lstlisting}
string TeX (symb fraction){
	if (fraction==1)
		return "";
	return "\\frac{" + to_string(abs(fraction.numerator)) + "}{" + to_string(fraction.denominator) + "}";
}
\end{lstlisting}
The\quad \texttt{term} $\longrightarrow$ \TeX \quad function:
\begin{lstlisting}	
string TeX (term in){
	string texout = in.field;
	string temp = "";
	bool parentheses = in.div * in.grad != 0;

	// TRACES
	if (in.trace > 0 && in.trace <= 3){   // trace primes for traces from 1 to 3
		for (pint i=1; i<=in.trace; i++)
			temp += "\'";
		texout += temp;
	}
	if (in.trace > 3){
		temp = to_string(in.trace);
		texout += "^{[" + temp + "]}";
	}
	if (parentheses)
		texout = texout + " \\right)";

	// DIVERGENCES
	if (in.div == 1)
		texout = "\\partial \\cdot " + texout;
	if (in.div > 1)
		texout = "\\partial^{" + to_string(in.div) + "} \\cdot " + texout;
	if (parentheses)
		texout = "\\left( " + texout;

	// GRADIENTS
	if (in.grad == 1)
		texout = "\\partial " + texout;
	if (in.grad > 1)
		texout = "\\partial^{" + to_string(in.grad) + "} " + texout;

	// BOXES
	if (in.box == 1)
		texout = "\\Box " + texout;
	if (in.box > 1)
		texout = "\\Box^{" + to_string(in.box) + "} " + texout;

	// METRICS
	if (in.eta == 1)
		texout = "\\eta " + texout;
	if (in.eta > 1)
		texout = "\\eta^{" + to_string(in.eta) + "} " + texout;

    // MULTIPLICATIVE FACTOR AND INVERSE BOX
	if (in.fact == 0)
		texout = "0";
	if (in.fact.denominator == 1 && in.fact.numerator!= 1 && in.box >= 0)
		texout = to_string(abs(in.fact.numerator)) + texout;
	else{
		if(in.fact.denominator == 1){
			if(in.box == -1)
				texout = "\\frac{" + to_string(abs(in.fact.numerator)) 
					+ "}{\\Box}" + texout;
			if(in.box < -1)
				texout = "\\frac{" + to_string(abs(in.fact.numerator)) 
					+ "}{\\Box^{" + to_string(abs(in.box)) + "}}" + texout;
		}
	else{
		if(in.box == -1)
			texout = "\\frac{" + to_string(abs(in.fact.numerator)) 
				+ "}{" + to_string(in.fact.denominator) + "\\Box}" + texout;
		if(in.box < -1)
			texout = "\\frac{" + to_string(abs(in.fact.numerator)) + "}{" 
				+ to_string(in.fact.denominator) + "\\Box^{" 
				+ to_string(abs(in.box)) + "}}" + texout;
		if(in.box >= 0)
			texout = "\\frac{" + to_string(abs(in.fact.numerator)) + "}{" 
				+ to_string(in.fact.denominator) + "}" + texout;
		}
	}
	if (in.fact.numerator < 0)
		texout = "-" + texout;

	return texout;
}
\end{lstlisting}
\newpage
The\quad \texttt{expression} $\longrightarrow$ \TeX \quad function:
\begin{lstlisting}
string TeX (expression in){
	expression simplified = Simplify(in);
	string texout = "";
	pint NumberOfTerms = simplified.size();
	if (NumberOfTerms == 0)
		texout = "0";

	for(pint i=0; i<NumberOfTerms; i++){
		if(i==0)
			if(simplified[i].fact.numerator > 0)
				texout = texout + TeX(simplified[i]);
			else{
				simplified[i].fact.numerator *= -1;
				texout = texout + " - " + TeX(simplified[i]);
			}
		else{
			if(simplified[i].fact.numerator > 0)
				texout = texout + " + " + TeX(simplified[i]);
			else{
				simplified[i].fact.numerator *= -1;
				texout = texout + " - " + TeX(simplified[i]);
			}
		}
		if(break_line && i%break_line_after==0 && i!=0 && i<NumberOfTerms-2)
			texout = texout + " \\\\ \\nonumber & ";
	}
	return texout;
}
\end{lstlisting}

\end{document}